\documentclass[aps,prd,twocolumn,floatfix,superscriptaddress,showpacs,amsmath,notitlepage,nofootinbib]{revtex4-1}
\usepackage{graphicx}
\usepackage{bm}
\usepackage{slashed}
\usepackage{amsmath}
\usepackage{enumitem}
\usepackage{hhline}
\usepackage{multirow}
\usepackage[utf8]{inputenc} 
\usepackage[dvipsnames]{xcolor}
\usepackage[colorlinks,citecolor=blue]{hyperref}

\newcommand\physrep{{Phys. Rept.}}
\newcommand\aj{{Astron. J.}}
\newcommand\aap{{Astron. Astrophys.}}
\newcommand\apjs{{Astrophys. J., Suppl. Ser.}}
\newcommand\apjl{{Astrophys. J. Lett.}} 
\newcommand\mnras{{Mon. Not. Roy. Astron. Soc.}} 
\newcommand\pasj{{Publ. Astron. Soc. Japan}}
\newcommand\araa{{Annu. Rev. Astron. Astrophys.}}

\begin{document}
\title{The Production of Actinides in Neutron Star Mergers}

\author{Meng-Ru Wu}
\email{mwu@gate.sinica.edu.tw}
\affiliation{Institute of Physics, Academia Sinica, Taipei, 11529, Taiwan}
\affiliation{Institute of Astronomy and Astrophysics, Academia Sinica, Taipei, 10617, Taiwan}

\author{Projjwal Banerjee}
\email{projjwal.banerjee@gmail.com}
\affiliation{Department of Physics, Indian Institute of Technology Palakkad, Kerala 678558, India}

\date{\today}

\begin{abstract}
Although the multimessenger detection of the neutron star merger event GW170817 confirmed that mergers are promising sites producing the majority of nature's heavy elements via the rapid neutron-capture process ($r$-process), 
a number of issues related to the production of translead nuclei -- the actinides -- remain to be answered. 
In this short review paper, we summarize the general requirements for actinide production in $r$-process and the impact of nuclear physics inputs. 
We also discuss recent efforts addressing the actinide production in neutron star mergers from different perspectives, 
including signatures that may be probed by future kilonova and $\gamma$-ray observations, 
the abundance scattering in metal-poor stars, and constraints put by the presence of short-lived radioactive actinides in the Solar system.
\end{abstract}
\maketitle
\section{Introduction}\label{intro}

The detection of the kilonova accompanying the
gravitational wave emissions from the binary neutron star merger (BNSM) event GW170817~\cite{abbott2017a,abbott2017b} provided a first direct evidence that the rapid neutron-capture process 
($r$-process) takes place in BNSMs (see Refs.~\cite{Metzger:2019zeh,Nakar:2019fza,Shibata:2019wef,Margutti:2020xbo,Perego:2021dpw} for recent reviews and references therein).
However, the comparison of the observed spectroscopic and photometric data to theory models
only confirmed the existence of heavy elements with high atomic opacity~\cite{Kasen+17,Drout:2017ijr,Cowperthwaite+17,Villar+17,Kawaguchi+18}, likely lanthanides, 
as well as the potential signature of lighter $r$-process elements such as strontium~\cite{Watson:2019xjv} (see also \cite{Domoto:2021xfq,Gillanders:2021qvx,Perego:2020evn}).
Whether or not even heavier nuclear isotopes -- the actinides -- can be produced in BNSM events remain unclear (cf.~\cite{Kasliwal:2018fwk}).

On the other hand, quite a number of metal-poor stars in our galaxy enriched by the $r$-process also contain thorium 
whose relative abundance to the lanthanides are similar to those in our Solar system (see e.g., Refs.~\citep{Sneden2008,Cowan2019} for reviews).
This indicates that the major $r$-process(es) producing astrophysical sites should also produce certain amount of actinides.
Attempts to use the actinide abundances in metal-poor stars to date their ages and therefore connect to the age of the Universe through the so-called ``nuclear cosmochronometry'' have
been made in \cite{Fowler:1960,Schramm:1970,Cowan:1991,Meyer:2000,Panov:2017lbx,Wu:2021jzv}.
Moreover, short-lived radioactive actinide isotopes whose life times are on the order of few million years (Myrs), have been shown to be present currently in the Earth's deep-sea crusts. Additionally, isotopic anomalies in meteorites clearly show that such radioactive isotopes were present   at the time of formation of the Solar system. These detection have been used as direct evidence of the production of actinides by BNSMs or to constrain the $r$-process conditions in potential production sites~\cite{Wallner_2015,hotokezaka2015,Bartos:2019cec,Cote2021,Wallner:2021,bwj2022}.

In this article, we review recent efforts that addressed the issue of actinide production by $r$-process in BNSMs.
In Sec.~\ref{sec:req}, we discuss the astrophysical conditions allowing for actinide production in BNSMs and the impact of uncertain nuclear physics inputs.
In Sec.~\ref{sec:kn}, we highlight recently proposed features associated with actinide production that may be detected by future kilonova observations or the next-generation $\gamma$-ray telescopes.
Sec.~\ref{sec:mpstars} focuses on the observation of actinide abundances in metal-poor stars and their implications.
Sec.~\ref{sec:slr} discusses the implications of measurements of the short-lived radioactive isotopes on actinide production in BNSMs.
We conclude this review in Sec.~\ref{sec:conclu}.

\section{Theory requirements and nuclear physics inputs}
\label{sec:req}

Earlier $r$-process nucleosynthesis studies that adopted either the parametrized BNSM ejecta properties or the outflow trajectories extracted from numerical simulations modeling the post-merger evolution of BNSMs, revealed that 
a significant amount of actinides with mass fraction $X_{\rm act}\gtrsim 10^{-3}$ is only produced in ejecta with $Y_e \lesssim 0.2$~\cite{Goriely2011,Korobkin2012,Wanajo:2014wha,Just:2014fka,Lippuner:2015gwa,Wu:2016pnw,Siegel:2017nub,Holmbeck:2018xet,Holmbeck:2019xnd,Eichler:2019rzj,Fernandez:2020oow,Nedora:2020hxc,Kullmann:2021gvo,Fujibayashi:2022ftg}.
This can be qualitatively understood from relating the averaged nuclear mass number at the end of the $r$-process ($\langle A\rangle_f$) to the
averaged initial seed nuclear mass number ($\langle A\rangle_s$) and the abundance ratio of free neutrons to seed nuclei ($R_{n/s}$)
before the onset of $r$-process by
\begin{equation}\label{eq:af}
\langle A\rangle_f = \langle A\rangle_s + R_{n/s}.
\end{equation}

\begin{figure}[t]
    \centering
    \includegraphics[width=\linewidth]{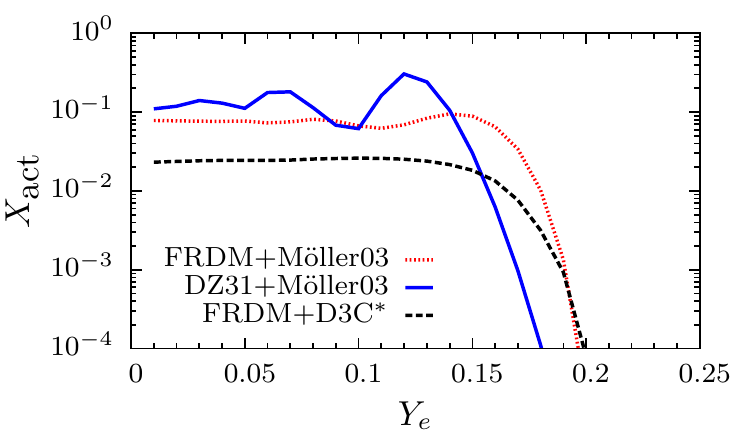}
    \caption{The actinide mass fraction $X_{\rm act}$
    as a function of $Y_e$ for a trajectory with an initial entropy per nucleon $s=10$~$k_B$ and a dynamical timescale $\tau=10$~ms. 
    The red dotted, blue solid, and black dashed lines are obtained with different combinations of theoretical nuclear mass models (FRDM or DZ31) and (``+'') $\beta$-decay rates (M\"oller03 or D3C$^*$).
    }
    \label{fig:xact}
\end{figure}

Assuming that right before the onset of $r$-process (usually around $T\simeq 3-5$~GK), the nuclear composition is dominated by free neutrons, with the presence of a small amount of seed nuclei formed via charged-particle reactions (or the $\alpha$-process)~\cite{Woosley:1992ApJ},
the neutron-to-seed ratio $R_{n/s}$ can be related to the electron fraction $Y_e$ of the ejecta by
\begin{equation}\label{eq:rns}
R_{n/s} = \frac{\langle Z \rangle_s - Y_e \langle A \rangle_s}{Y_e},
\end{equation}
where $\langle Z\rangle_s$ is the averaged initial seed nuclear charge number. 
To produce a significant amount of actinides, the $r$-process
nucleosynthesis needs to proceed beyond the third peak at $A\simeq 195$, which corresponds to $\langle A \rangle_f \gtrsim 150$~\cite{Lippuner:2015gwa}.
Using this condition together with Eqs.~\eqref{eq:af} and \eqref{eq:rns}, and considering typical values of $\langle Z \rangle_s \simeq 30$ and $\langle A \rangle_s \simeq 80$,
one gets $Y_e\lesssim 0.2$ as the rough threshold value for actinide production in BNSM ejecta.
Fig.~\ref{fig:xact} shows the actinide mass fractions $X_{\rm act}$ as a function of $Y_e$ computed after a time evolution of one day (24 hrs) for a typical parametrized BNSM ejecta trajectory, with an initial entropy per nucleon $s=10$~$k_B$ and a dynamical timescale $\tau=10$~ms used in~\cite{Wu+19} (see \cite{Lippuner:2015gwa} for the form of parametrization), but supplied with different nuclear physics inputs.
It clearly shows that although different nuclear physics inputs give rise to quantitatively different amount of actinides for a given value of $Y_e$, the actinide mass fractions all drop to less than $10^{-4}$ at values of $Y_e\gtrsim 0.2$.
Current BNSM and post-merger simulations suggest that actinides with $X_{\rm act} \gtrsim 10^{-3}$ can be produced in outflows containing a non-negligible component of $Y_e\lesssim 0.2$, ejected dynamically during the merging phase (due to tidal disruption or BNS collision) as well as from post-merger accretion disk outflows with a central black hole or short-lived hyper-massive neutron star that collapses to a black hole within $\sim$ tens of ms~\cite{Goriely2011,Korobkin2012,Wanajo:2014wha,Just:2014fka,Wu:2016pnw,Siegel:2017nub,Fernandez:2020oow,Nedora:2020hxc,Kullmann:2021gvo,Fujibayashi:2022ftg}.

Let us now discuss the impact of nuclear physics inputs on actinide production. 
As studied in Refs.~\cite{Goriely:1999jq,Panov:2013tfa,Eichler:2014kma,Mendoza2015,Goriely:2016gfe,Holmbeck:2018xet,Vassh:2018wcf,Zhu:2018oay,Wu+19,Giuliani:2019oot,Eichler:2019rzj,Zhu:2020eyk,Lemaitre:2021yje}, the theoretical nuclear physics inputs such as the nuclear mass predictions, the $\beta$-decay half-lives of translead nuclei, and their fission properties can all greatly impact the production of actinides. 
This is clearly demonstrated by Fig.~\ref{fig:xact}, where variation of up to a factor of $\sim 10$ in $X_{\rm act}$ is found for the same $Y_e$ when using different nuclear physics inputs based on two nuclear mass models -- the Finite
Range Droplet Model (FRDM)~\cite{Moller:1993ed} and the  Duflo-Zuker mass formula with
31 parameters (DZ31)~\cite{Duflo:1995ep} -- used in~\cite{Mendoza2015,Wu:2016pnw}, combined with two different $\beta$-decay rates from \cite{Moller:2003fn} (labeled by M\"oller03) and \cite{Marketin:2015gya} (labeled by D3C$^*$).

The effect of nuclear masses comes in mainly through the change of the $r$-process path, which can be defined by a set of nuclei, each of which is the most abundant one in an isotopic chain.  
For $r$-process that operates at temperature that is high enough ($T\gtrsim 0.7$~GK) so that the balance between the neutron captures and photo-dissociations i.e, the $(n,\gamma)\leftrightarrow (\gamma,n)$ equilibrium, can be maintained, the $r$-process path locates at nuclei that have two-neutron separation energy $S_{2n}$ just above~\cite{arnould2007}
\begin{equation}\label{eq:sn0}
  S_{2n}^0 = \frac{T_9}{2.52}\left( 34.075 - \log n_n + \frac{3}{2} \log T_9 \right),
\end{equation}
where $T_9$ is the temperature in GK and $n_n$ is the free neutron number density.
Thus, changes in nuclear mass prediction affect how far away from stability the $r$-process path locates on the nuclear chart. 
Then, due to the typical condition close to the steady $\beta$ flow during the $r$-process, the abundances of neighboring isotopic chains satisfy~\cite{Kratz:1993ApJ}
\begin{equation}\label{eq:betaflow}
  Y(Z)\langle \lambda_\beta(Z) \rangle = Y(Z+1)\langle \lambda_\beta(Z+1) \rangle,
\end{equation}
where $Y(Z)=\sum_A Y(Z,A)$ is the total nuclear abundance of an isotopic chain with proton number $Z$, and $\langle\lambda_\beta(Z) \rangle = \sum_A \lambda_\beta(Z,A) Y(Z,A) / Y(Z)$ is the averaged $\beta$-decay rate of the same chain. 
Because $\langle\lambda_\beta(Z) \rangle$ may be well approximated by $\lambda_\beta$ of the nuclei on the $r$-process path, Eq.~\eqref{eq:betaflow} implies the following: 
for two different $r$-process calculations that have the same $\beta$-decay inputs, the case where the $r$-process path locates further away from the stability in a certain region will have lower nuclear abundances in that region when compared to the other case.

\begin{figure}[t]
    \centering
    \includegraphics[width=\linewidth]{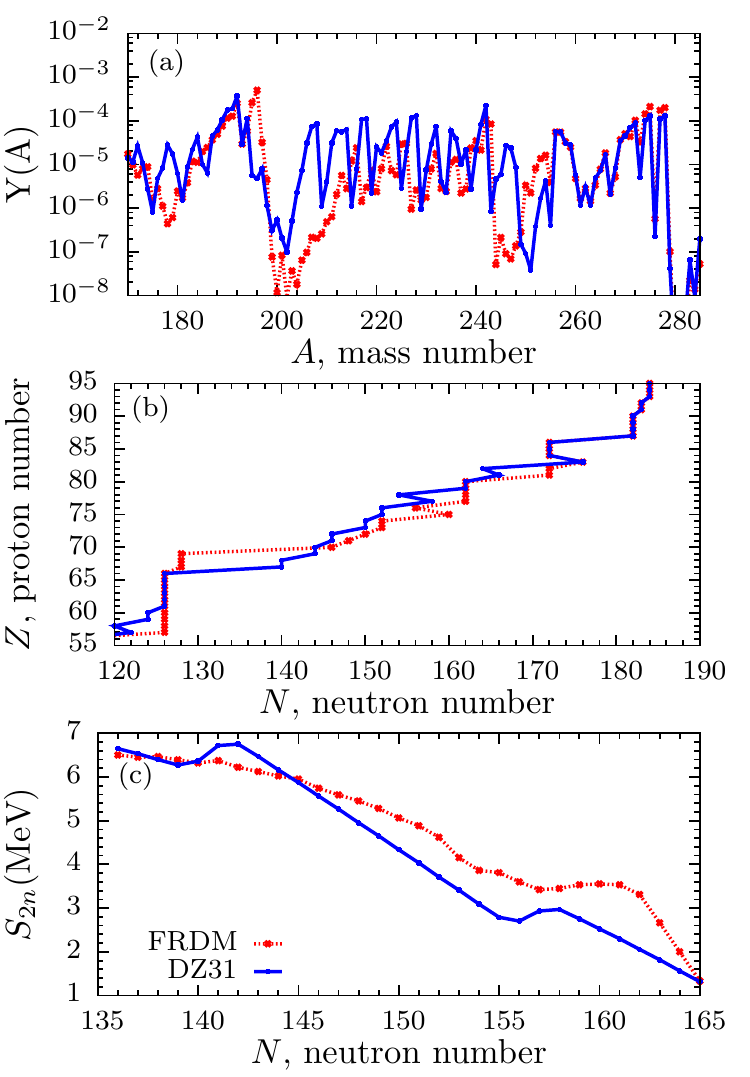}
    \caption{Comparison of the abundance $Y(A)$ [panel (a)],
    $r$-process path [panel (b)], and the two-neutron separation energy $S_{2n}(N)$ for $Z=78$ from calculations
    with $Y_e=0.07$, for trajectory with an initial entropy per nucleon $s=10$~$k_B$ and a dynamical timescale $\tau=10$~ms. 
    The red dotted and blue solid lines are obtained with the FRDM and the DZ31 mass model, respectively. For both cases, $S_{2n}^0\simeq 3$~MeV.
    }
    \label{fig:mass}
\end{figure}
In Fig.~\ref{fig:mass}, we show in panel (a) the comparison of the abundance $Y(A)$ at the time when $R_{n/s}=1$ for $Y_e=0.07$ for cases with FRDM+M\"oller03 and DZ31+M\"oller03, in panel (b) the corresponding $r$-process paths, and in panel (c) the $S_{2n}(N)$ for $Z=78$ around $A=230$.
Panel (b) clearly shows that the FRDM mass model results in a path further away from the stability than the DZ31 case. 
The corresponding $Y(A)$ around the same mass number range are also a factor of a few smaller than the DZ31 case as shown in panel (a). 
This difference originates from different evolution of $S_{2n}$ in mass models shown in panel (c). 
The DZ31 model predicts a faster decline of $S_{2n}$ than FRDM in the actinide region, which results in an $r$-process path closer to stability and higher actinide abundances as discussed above.

\begin{figure}[t]
    \centering
    \includegraphics[width=\linewidth]{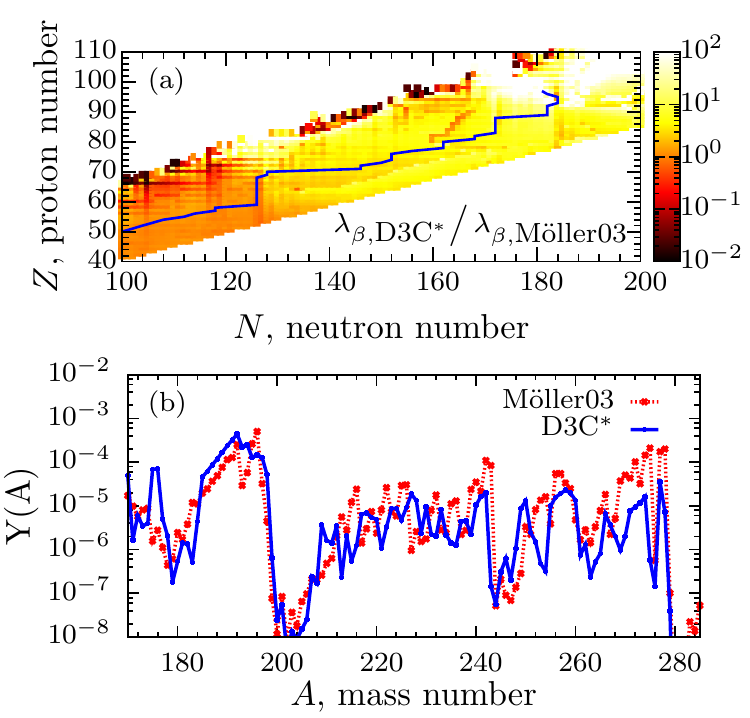}
    \caption{Panel (a): The ratio of $\beta$-decay rates from D3C$^*$~\cite{Marketin:2015gya} and M\"oller03~\cite{Moller:2003fn} models.
    The blue line shows the $r$-process path obtained from a calculation with $Y_e = 0.07$, an initial entropy per nucleon $s = 10$~$k_B$, and a dynamical timescale $\tau=10$~ms, using the FRDM mass model and the D3C$^*$ $\beta$-decay rates at the time when $R_{n/s}=1$.
    Panel (b): The comparison of abundances for $A\geq 170$ using the same trajectory.
    Both lines are with the FRDM mass model, while the D3C$^*$ (M\"oller03) $\beta$-decay rates are used for the blue (red) line.
    }
    \label{fig:beta}
\end{figure}
For $\beta$-decays, their impact on the actinide abundances comes in directly through Eq.~\eqref{eq:betaflow}.
Models that predict relatively faster decay rates in region of actinides compared to other parts of the nuclear chart lead to generally smaller amount of actinide production~\cite{Holmbeck:2018xet}. 
Comparing the two example $\beta$-decay models, M\"oller03 and D3C$^*$, Fig.~\ref{fig:beta}(a) shows that D3C$^*$ predicts shorter half-lives than M\"oller03 by $\sim$ a factor of $\mathcal{O}(10)$ for $A\gtrsim 200$.
This then results in smaller actinide abundances during the $r$-process as shown in panel (b) for cases with $Y_e=0.07$, and is the primary reason for the generally lower $X_{\rm act}$ shown in Fig.~\ref{fig:xact} for D3C$^*$.

The properties of nuclear fission can also substantially affect the actinide abundances in $r$-process with low enough initial $Y_e \lesssim 0.1$ obtained in e.g., the tidally disrupted material in BNSMs or NS -- black hole mergers~\cite{Vassh:2018wcf,Giuliani:2019oot,Lemaitre:2021yje}. 
These impacts can be generally classified as \emph{direct} and \emph{indirect} ones~\cite{Giuliani:2019oot}.
The theory dependent prediction of fission barrier and fission path for neutron-rich nuclei of $A\gtrsim 250$ directly determines the relevant fission rates. 
After the $r$-process freeze-out, when all decay channels (including fission) start to compete, higher fission barriers (or lower fission rates) along relevant isobaric chains can lead to a higher amount of actinides surviving against fission~\cite{Vassh:2018wcf,Giuliani:2019oot}.
Consequently, this can result in a larger amount of actinides with $A\gtrsim 250$ at times relevant for kilonova observation (see next section for the particular importance of $^{254}$Cf).

The fission rate predictions, together with other rates, also determine the free neutron abundance post the $r$-process freeze-out via a quasi-equilibrium condition found in \cite{Giuliani:2019oot}.
This can lead to more than one order of magnitude difference in free neutron abundance.
Cases with larger post freeze-out free neutron abundances undergo more neutron-induced fission, which then results in  less $A\gtrsim 250$ actinides that survive against fission.
Moreover, a larger amount of free neutrons can also transport non-fissioning actinides of $220\lesssim A \lesssim 230$ to higher masses, which then affects the $\alpha$-decay heating relevant for kilonovae (see also next section).

The above discussions outline the major impact of uncertain nuclear physics inputs on the actinide production in merger ejecta. 
In the next section, we will discuss how they can possibly leave imprints on kilonova emission of BNSMs.

\section{Potential electromagnetic observables}
\label{sec:kn}

The unstable nuclei produced by $r$-process in the BNSM ejecta within the timescale of $\sim \mathcal{O}(1)$~s gradually decay back to the stability while the ejecta expands.
The released nuclear energy from all possible decay channels, including $\beta$-decays, $\alpha$-decays, and different kinds of fission (neutron-induced, $\beta$-induced, spontaneous, $\gamma$-induced, etc.) 
directly powers the electromagnetic emission -- kilonova -- at epochs when thermalized photons can escape, which happens roughly after $\mathcal{O}(1)$~days post the merger~\cite{Li:1998bw,Metzger:2010sy,Barnes:2013wka,Tanaka:2013ana,Metzger:2019zeh}.
Modeling the kilonova lightcurve and spectral evolution plays a decisive role in extracting the underlying properties of the expanding ejecta from observation~\cite{Barnes:2016umi,Wollaeger:2017ahm,Kasen+17,Tanaka:2017qxj,Waxman:2017sqv,Coughlin:2018miv,Kawaguchi+18,Hotokezaka:2019uwo,Kawaguchi:2020,Ristic:2021ksz,Breschi:2021tbm}.

\begin{figure}[t]
    \centering
    \includegraphics[width=\linewidth]{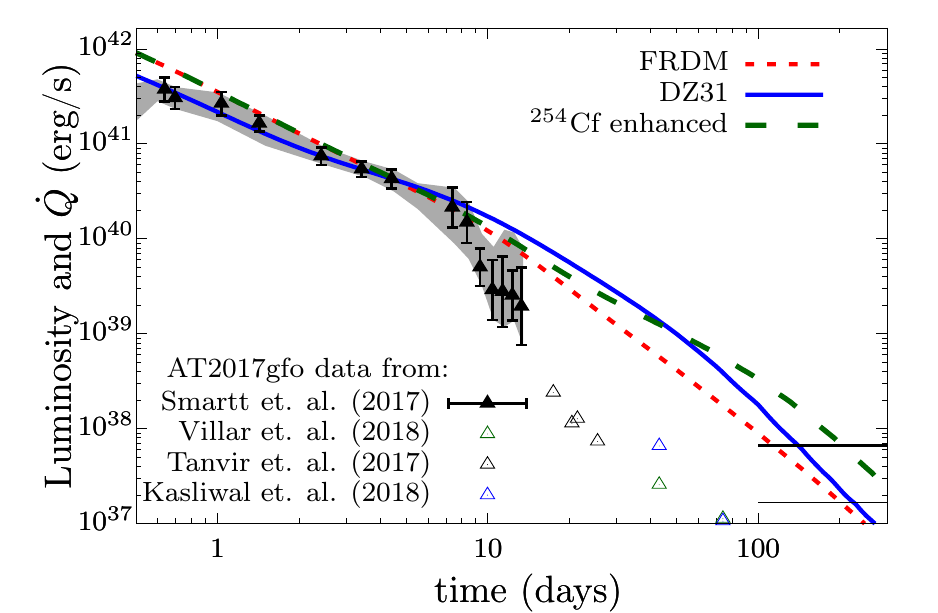}
    \caption{Impact of the existence of actinides on the bolometric lightcurve of kilonovae. 
    The dotted red, solid blue, and dashed green lines (labeled by FRDM, DZ31, and $^{254}$Cf enhanced) represent cases where the lightcurves are dominated by the energy release from $\beta$-decays, $\alpha$-decay actinides, and spontaneous fission of $^{254}$Cf, normalized to $t=6$~days.
    The filled and the hollow triangles are the observed and the lower bound of the brightness of the GW170817 kilonova.
    Data for this plot are taken from \cite{Wu+19}.
    }
    \label{fig:lcQ}
\end{figure}

For this purpose, it was realized in recent years that the exact amount of actinide present in BNSM ejecta can critically affect the amount of the radioactive heating relevant for kilonovae~\cite{Barnes:2016umi,Rosswog:2016dhy,Wanajo2018,Zhu:2018oay,Wu+19,Vassh:2018wcf,Giuliani:2019oot,Zhu:2020eyk,Barnes:2020nfi}. 
For instance, Refs.~\cite{Barnes:2016umi} and \cite{Rosswog:2016dhy} first found that the radioactive heating rate (after taking into account the effect of particle thermalization) differ by a factor of 2--6 during $\sim 1-10$~days when adopting different nuclear mass models shown in Fig.~\ref{fig:xact}, for very neutron-rich ejecta with $Y_e\lesssim 0.1$. The origin of this difference can be related to the produced amount of nuclei that $\alpha$ decay at relevant times. 
Later, Ref.~\cite{Wu+19} further identified that specifically the $\alpha$-decay nuclei $^{222}$Rn, $^{224}$Ra, $^{223}$Ra and $^{225}$Ac with half-lives $t_{1/2}$ of 3.8, 3.6, 11.43 and 10.0~days\footnote{In $r$-process, $^{225}$Ac is actually produced via the $\beta$-decay of $^{225}$Ra, which has a similar half-life of $14.9$~days.}, as well as the spontaneous fissioning nuclei $^{254}$Cf ($t_{1/2}=60.5$~days), can potentially leave interesting imprints on the bolometric kilonova lightcurve at $t\gtrsim 10$~days (see Fig.~\ref{fig:lcQ}). 
Around the same time, Ref.~\cite{Zhu:2018oay} studied independently the impact of $^{254}$Cf on certain bands of the lightcurves and reached similar conclusion. 
These findings suggest that a precise measurement of future kilonova late-time lightcurves from BNSM events can possibly be used to infer the presence of $\alpha$-decay actinides to the abundance level of $\sim 10^{-5}$ and $^{254}$Cf to $\sim 10^{-6}$.  
More recently, Ref.~\cite{Zhu:2020eyk,Barnes:2020nfi} suggested that the uncertainty from nuclear physics inputs on the kilonova heating rate can be as large as one order of magnitude even at the peak time, based on results that explored outflow conditions with single $Y_e$ values.

The primary reason behind the possibility of the above discoveries is rather simple: these nuclei generate much larger energy per decay ($\alpha$ or fission) than that from the $\beta$-decay of non-actinides.
For instance, the $\alpha$ decays of aforementioned actinides with half-lives of days are followed by a sequence of fast $\alpha$ and $\beta$ decays with much shorter half-lives before reaching stable isotopes. 
The total released energy in each of these decay chains is $\sim 30$~MeV. 
For the case of the fission of $^{254}$Cf, it releases $\sim 180$~MeV per decay. 
These values are much larger than the typical $\beta$-decay energy release of $\lesssim \mathcal{O}(1)$~MeV per nucleus.
Moreover, the thermalization efficiencies of products from the $\alpha$-decay sequence and fission are typically higher than those from $\beta$-decay, where neutrinos and $\gamma$-rays can carry a significant part of the energy away without thermalization~\cite{Barnes:2016umi,Hotokezaka:2015cma}.  
Consequently, these decay channels can dominate the energy deposition into the ejecta even with much smaller abundances than the $\beta$-decay non-actinide nuclei and leave imprints on kilonova observations.
A caution here is that the detailed effects of the excessive heating on the precise spectral evolution of kilonovae at such late times require improved radiation emission modeling during the nebula phase of kilonovae~\cite{Hotokezaka:2021ofe,Pognan:2021wpy}.
Future work along this direction is crucial if one would like to convincingly use the late-time kilonova lightcurve to obtain information regarding the presence of actinides or to even constrain the associated theoretical nuclear physics inputs.

Besides the impact of actinide production on kilonova lightcurves, their presence might also contribute to the $\gamma$-ray emission from individual BNSM event~\cite{Hotokezaka:2015cma,Li:2018wee,Korobkin:2019uxw,Wang:2020qkn,Chen:2021tob,Chen:2022nsj} or from old BNSM remnants ($\sim 10^4$--$10^6$~yrs old) residing inside the Milky Way~\cite{Wu:2019xrq,Korobkin:2019uxw,Terada:2022hut}.
In particular, \cite{Wu:2019xrq} suggested that if future sub-MeV $\gamma$-ray missions can reach a line sensitivity of $\sim 10^{-8}$~$\gamma$~cm$^{-2}$~s$^{-1}$, then there is a non-negligible chance ($\sim 10-30\%$) to detect lines from the decay of $^{230}$Th with a number fraction of $\sim 10^{-5}-10^{-6}$ per merger in Milky Way's old BNSM remnants.
If such lines are detected and the remnant is confirmed to be produced by a past BNSM, it will provide a direct evidence of actinide production in BNSMs.
Meanwhile, Ref.~\cite{Wang:2020qkn} found that the $\gamma$ emissions from fissioning actinides may dominate the smeared spectra above $\gtrsim 3$~MeV during $\sim 10-1000$~days post merger.
This would require a Galactic event even for future missions, but can in principle be a potential signature of actinide production in mergers.

\section{Actinides in metal-poor stars}
\label{sec:mpstars}
\begin{figure}[t]
    \centering
    \includegraphics[width=\linewidth]{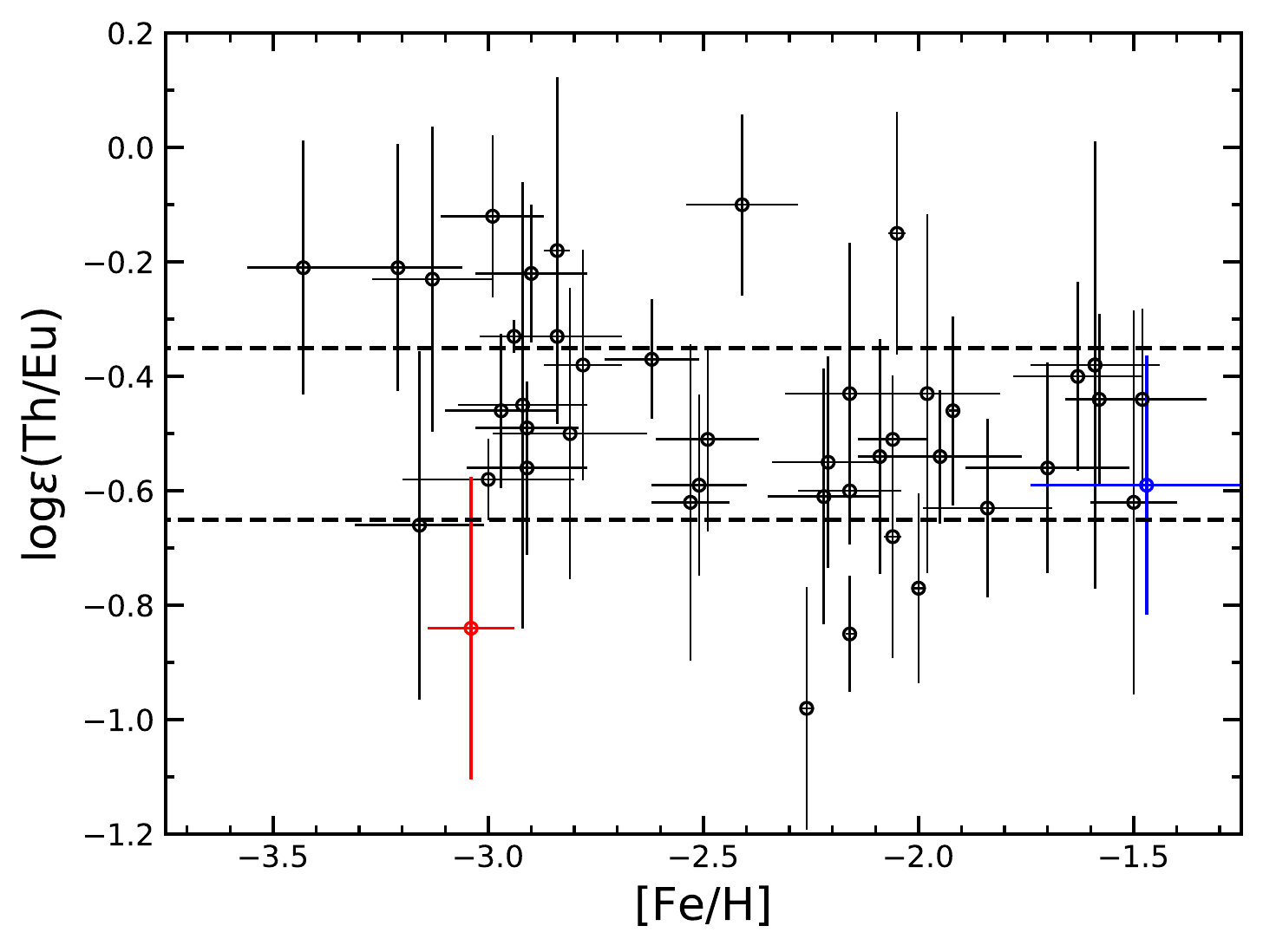}
    \caption{Observed abundance of Th/Eu in metal-poor stars in the Milky Way (black), Reticulum II (red) and Ursa Minor (blue). Data is generated using the SAGA database \cite{SAGA} and adapted from Refs.~\cite{johnson2001,cowan2002,hill2002,honda2004,ivans2006,frebel2007,aoki2007,lai2008,roederer2009,roederer2010,mashonkina2010,roederer2012,roederer2014,siqueira2014,mashokina2014,placco2017,Holmbeck2018obs,sakari2018,roederer2018,cain2018,gull2018,sakari2018b,jiap2018}.  }
    \label{fig:theu}
\end{figure}

Stars with [Fe/H]$\lesssim -2$ are known as very metal-poor (VMP) stars. Low mass VMP stars of $\lesssim 0.9\, M_\odot$ are thought to have formed within $\sim 1$--$2$ Gyr from the time of Big Bang and are still alive today. 
The surface composition of VMP stars provide a direct snapshot of the interstellar medium (ISM) in the early Galaxy. 
This is particularly important because the abundance patterns of elements in the early Galaxy are thought to be from a few nucleosynthetic events. 
With regard to $r$-process, VMP stars that are formed from ISM enriched in $r$-process (${\rm [Ba/Eu]<0}$), provide a direct measurement of the nucleosynthesis from individual $r$-process sources. 
Depending on the level of $r$-process enrichment, they are classified as $r$-I ($0.3\leq {\rm [Eu/Fe]} \leq 1$) and $r$-II (${\rm [Eu/Fe]}>1$) stars \cite{beers2005}. 
Starting with the discovery of the famous $r$-II star  CS 22892-052 \cite{beers1992,sneden2003}, a number of $r$-I and $r$-II stars have been discovered that have detailed abundance patterns for heavy elements. 
Remarkably, these stars show an almost identical abundance pattern that agrees well with the Solar $r$-process pattern from Ba to Pb~\cite{Sneden2008,Cowan2019}). 
However, some of the $r$-II stars show an enhanced Th abundance (relative to Eu) when compared to the rest of the $r$-II stars \cite{mashokina2014}. 
These are referred to as actinide-boost stars. Fig.~\ref{fig:theu} shows the observed Th/Eu ratio vs [Fe/H] as observed in VMP as well as metal-poor stars. 
As can be seen for the figure, there is a considerable scatter of up to an order of magnitude in the observed Th/Eu ratio. Despite the large scatter, the definition of stars with ``normal'' Th/Eu ratio is somewhat subjective due to the considerable $1 \sigma $ uncertainty of $\gtrsim 0.15$ dex in the observed Th/Eu ratio in individual stars. 
The horizontal lines shown in Fig.~\ref{fig:theu} shows the band corresponding to $\log \epsilon({\rm Th/Eu})=-0.5 \pm 0.15$\footnote{$\log\epsilon({\rm X})\equiv\log(N_{\rm X}/N_{\rm H})+12$, where $N_{\rm X}$ and $N_{\rm H}$ are the number of atoms of element X and hydrogen, respectively. 
For elements X and Y, $\log\epsilon({\rm X/Y})=\log\epsilon{\rm (X)}-\log\epsilon{\rm (Y)}=\log(N_{\rm X}/N_{\rm Y})$.} as adopted recently by Ref.~\cite{Holmbeck2018obs} to define ``normal'' Th/Eu stars. 
With this definition, there are $3$ stars with Th/Eu values that are clearly above the band and 5 more stars that have mean values above $-0.35$ but still lies within the band when the $1\sigma$ error is taken into account. 
Similarly, there are two stars whose Th/Eu values are clearly below the band and 3 more stars whose mean values are below $-0.65$ but lie within the band when $1\sigma$ uncertainty is taken into account. 
Regardless of the exact definition of what a normal $r$-process Th/Eu ratio is, it is clear that the difference in the Th/Eu ratio in stars above and below the band is statistically significant  and points to variations in the actinide abundance that is in sharp contrast with the robust pattern seen in elements between the second and third $r$-process peak. 

Ref.~\cite{Holmbeck2018obs,Holmbeck:2018xet} investigated whether BNSMs can account for the scattering of the observed $\log \epsilon({\rm Th/Eu})$ discussed above, and found that a proper mixture of low $Y_e\lesssim 0.15$ and high $Y_e$ components can easily reproduce the observed range of $\log \epsilon({\rm Th/Eu})$.
In such a scenario, the low $Y_e\lesssim 0.15$ component may produce $\log \epsilon({\rm Th/Eu}) \gtrsim 0$ and the high $Y_e\gtrsim 0.15$ component plays the role of diluting $\log \epsilon({\rm Th/Eu})$.
For example, both the FRDM+M\"oeller03 model and DZ31+M\"oeller03 model shown in Fig.~\ref{fig:mass} produce $0\lesssim \log \epsilon({\rm Th/Eu}) \lesssim 1.0$ for $Y_e\lesssim 0.15$ (taken at $t=13$~Gyr).
Since this value decreases rapidly with increasing $Y_e\gtrsim 0.15$, a dilution factor of $\sim 1-10$ for both cases can lower $\log \epsilon({\rm Th/Eu})$ to the observed range shown in Fig.~\ref{fig:theu}.
On the other hand, the model FRDM+D3C$^*$ only obtains $\log \epsilon({\rm Th/Eu}) \lesssim -0.7$, which appears to be inconsistent with the observations.  
This may point to the possibility of constraining the $\beta$-decay half-lives of exotic neutron-rich heavy nuclei with metal-poor stars.
However, given the large uncertainties involved in model predictions and  observations, it is too early to draw any definite conclusions. 

In addition to the Th (and U) abundances that can be directly used to infer the actinide production in BNSMs, a precise measurement of the abundance of Pb can also provide useful information. This is because  most of the non-fissioning actinides produced by $r$-process ultimately decay to Pb isotopes (other than those blocked by Th). 
In particular, it may be used to constrain the amount of $\alpha$-decay nuclei relevant for kilonova discussed in Sec.~\ref{sec:kn} assuming BNSMs are indeed the sites enriching these stars~\cite{Hotokezaka:2019uwo}.
Notwithstanding the large theoretical and even larger observational uncertainties associated with the Pb abundance  that precludes a definite answer now, it may be worthwhile to examine this in future with improved theoretical understanding as well as better observational data (e.g.,~\cite{Roederer2022}).

\section{Short-lived radioactive Actinides}
\label{sec:slr}
Radioactive isotopes that have half-lives of a few Myr are called short-lived radioactive isotopes (SLRIs). They provide crucial information about the local enrichment history over the relatively short timescales of their half-lives. 
Broadly, abundance of SLRIs can be measured in three distinct regimes. 
The first is the measurement of live SLRIs present in the ISM using $\gamma$-ray telescopes that detect the emitted $\gamma$-rays following radioactive decay of SLRIs such as $^{26}$Al and $^{60}$Fe. 
Such measurements have been used to directly probe ongoing star formation in the Galaxy~\cite{Diehl:2006cf,Wang:2020xx}, and can in principle be used to probe the BNSM history as well (see Sec.~\ref{sec:kn}).
On the other hand, live SLRIs that are incident on Earth are eventually deposited at the Earth's deep sea floor which grows over time. Measuring SLRIs in different layers of such sea crusts gives the enrichment history of the local ISM over the last $\sim 10$--25 Myrs. 
Lastly, and remarkably, abundance of SLRIs that were present at the time of the formation of the Solar system can be measured by analyzing the isotopic anomalies in meteorites resulting from the SLRIs decaying to stable daughter isotopes. This can probe the nucleosynthetic events that occurred within a few Myr before the formation of the Solar system. 

\begin{figure}[t]
    \centering
    \includegraphics[width=\linewidth]{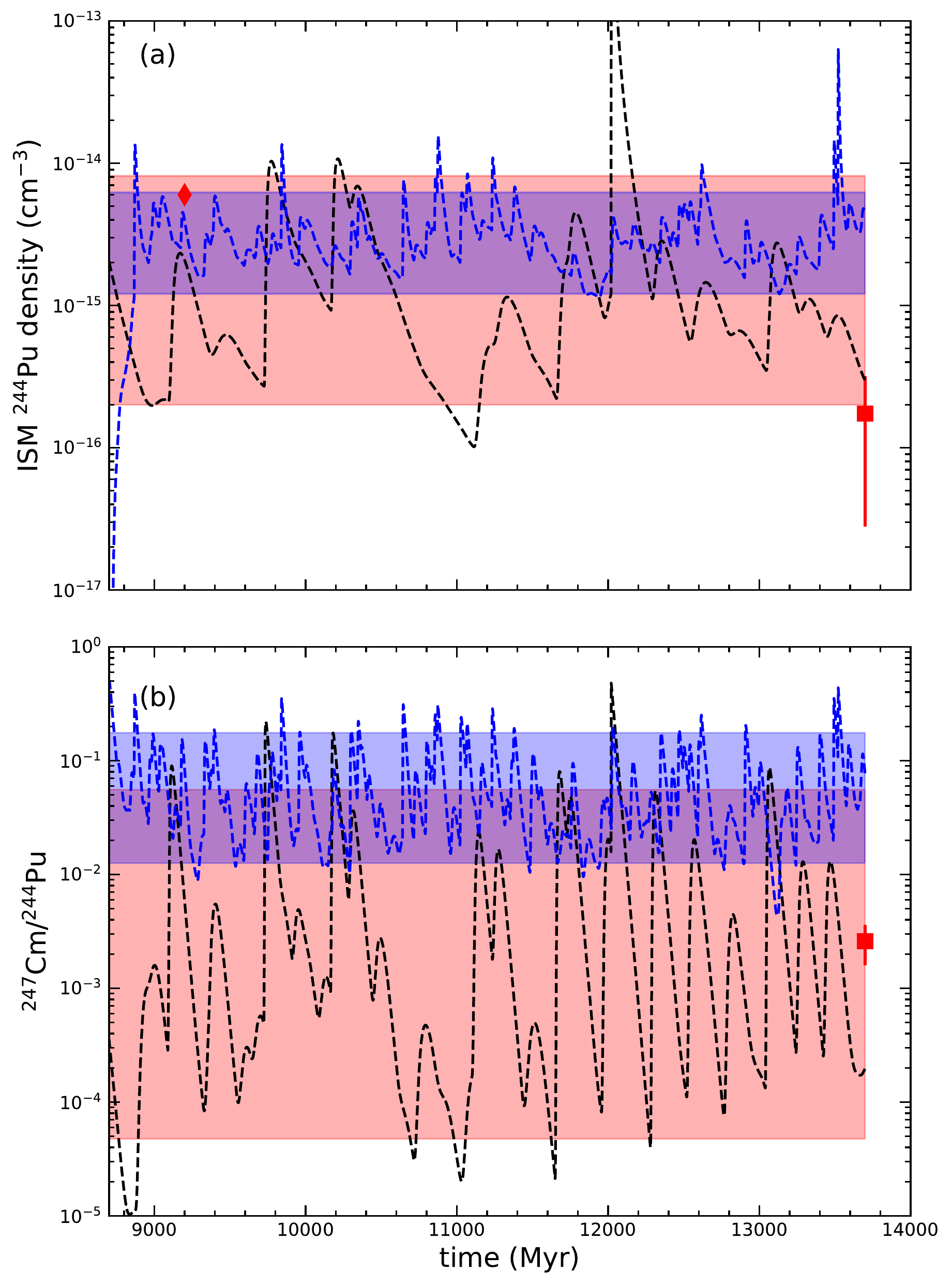}
    \caption{(a) Evolution of $^{244}$Pu ISM density for $r$-process sources with rates of 10 Myr$^{-1}$ (red) and 500 Myr$^{-1}$ (blue). The shaded regions show the  corresponding 90 percentile range. The estimated $^{244}$Pu ISM in the ESS (red diamond) is shown along with the estimated local ISM density over the last 25 Myr estimated from earth's crust (red square) \cite{Wallner:2021}.  (b) Evolution of the corresponding $^{247}$Cm/$^{244}$Pu ratio along with the  measured ratio in the ESS (red square). The shaded regions correspond to the 90 percentile range. }
    \label{fig:PuCm}
\end{figure}
With regard to $r$-process, the important SLRIs are $^{129}$I (half-life of 15.6 Myr) and actinides $^{244}$Pu (half-life of 81 Myr) and $^{247}$Cm (half-life of 15.7 Myr). 
In terms of live SLRIs measured in deep sea crusts, $^{244}$Pu is particularly important because of its relatively longer half-life. Ref.~\cite{Wallner_2015} have shown that the amount of live $^{244}$Pu accumulated in deep sea crust over the last 25 Myr can be used to estimate the local ISM density of $^{244}$Pu which turns out to be about two orders of magnitude lower than the amount expected from a frequent $r$-process source associated with a regular core-collapse SN. 
More recent measurements of deep sea crust that correspond to the last $\sim 10$ Myr also point towards a rare $r$-process source \cite{Wallner:2021}. 
Interestingly, $^{244}$Pu has also been measured in meteorites and has been shown to be present in the early Solar system (ESS) with an abundance of $^{244}$Pu/$^{238}$U$= (7\pm 1)\times 10^{-3}$ \citep{Hudson1989}. 
Surprisingly, the abundance measured in the ESS corresponds to an ISM density that is $\sim$ two orders of magnitude higher than the value inferred from deep sea measurement. 
Fig.~\ref{fig:PuCm}a shows the evolution of local ISM $^{244}$Pu density with time for two different  $r$-process sources with a frequencies of $500\,{\rm Myr}^{-1}$ and $10\,{\rm Myr}^{-1}$ using the turbulent gas diffusion prescription used in Ref.~\cite{Hotokezaka:2015zea}. 
As can bee clearly seen from the figure, an $r$-process source that has a lower frequency is required to account for the $\sim 2$ orders of magnitude variation in local $^{244}$Pu density. 
This essentially rules out the possibility of a frequent $r$-process source associated with core-collapse SN and points to 
a rare but prolific source with an upper limit of the inferred occurrence rate of $\lesssim 10\,{\rm Myr}^{-1}$ \cite{Hotokezaka:2015zea}, which is consistent with
BNSMs or some rare supernovae (SNe) such as the magneto-rotational SNe (MRSNe)~\cite{Nishimura2006,Winteler+12,Mosta+17}.  
In addition to the constraint put by $^{244}$Pu ISM density, Ref.~\cite{Bartos:2019cec} showed that the abundance ratio of $^{247}$Cm/$^{244}$Pu in the ESS can also be used to directly constrain the $r$-process event rate during the formation of the Solar system with a inferred rate of $1$--100 Myr$^{-1}$. Fig.~\ref{fig:PuCm}b shows the corresponding evolution of the $^{247}$Cm/$^{244}$Pu for the models plotted in Fig.~\ref{fig:PuCm}a. As can be seen clearly from the figure, an $r$-process source with higher frequency results in  $^{247}$Cm/$^{244}$Pu that is always higher than the measured value in the ESS. On the other hand, a rare $r$-process source of BNSM-type can easily account for the observed value.

The ESS measurement of $^{247}$Cm is also  particularly interesting when combined with the measurement of $^{129}$I. In a recent study by Ref.~\cite{Cote2021}, it was shown that due to the fortuitous coincidence of nearly identical half-lives of these two isotopes that are also almost exclusively produced by $r$-process, the $^{129}$I/$^{247}$Cm ratio stays constant over time  and is not sensitive to the typical uncertainties associated with Galactic chemical evolution. 
Assuming a rare $r$-process source, it was shown that the value of $438\pm 184$ measured in the ESS corresponds to the ``last'' $r$-process event that contributed to the Solar system inventory. Because the ratio $^{129}$I/$^{247}$Cm is particularly sensitive to the neutron richness, this can directly constrain neutron richness of the ``last'' $r$-process event. 
It was shown that the measured value in the ESS is consistent with moderately neutron-rich ejecta that are associated with disk winds following BNSMs. 
On the other hand, the value is inconsistent with both very neutron-rich ejecta that is usually associated with tidally-disrupted ejecta, and mildly neutron-rich ejecta that associated with MRSNe. 
\begin{figure}[t]
    \centering
    \includegraphics[width=\linewidth]{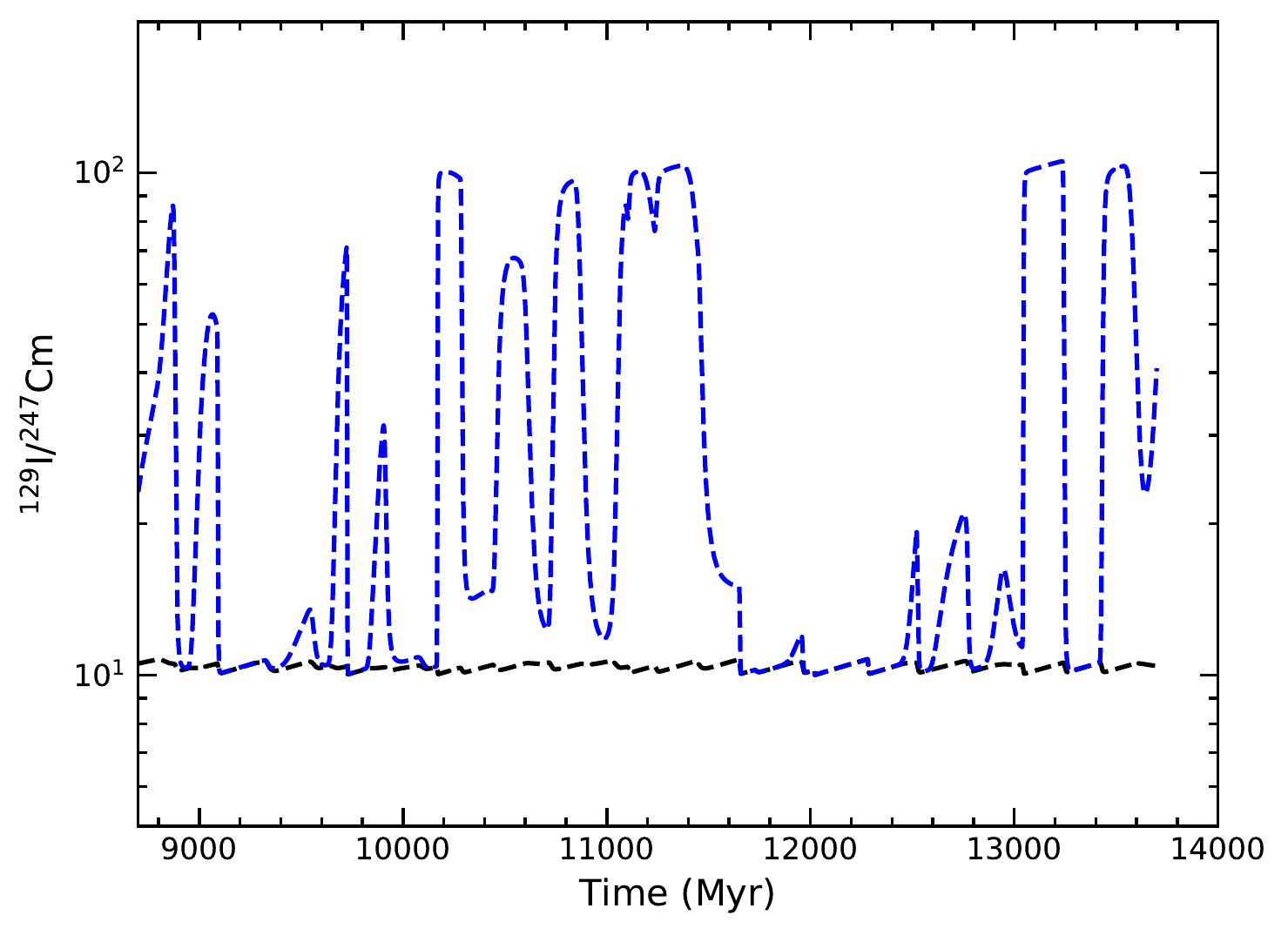}
    \caption{Evolution of $^{129}$I/$^{247}$Cm: One $r$-process source with $^{129}$I/$^{247}$Cm production ratio of 10 is plotted in black. The case where there are two equally frequent sources with $^{129}$I/$^{247}$Cm production ratios of 10 and 100, respectively, is shown in blue. In both cases the total rate of $r$-process events is 10 Myr$^{-1}$.}
    \label{fig:ICm}
\end{figure}
A recent study by Ref.~\cite{bwj2022} has found that the origin of $^{247}$Cm and $^{129}$I in the ESS is likely more complicated. In particular, when the turbulent gas diffusion prescription is used to model the $^{129}$I and $^{247}$Cm abundance, it was found that in addition to a major contributing event, at least two more events contribute substantially to the ESS inventory of $^{129}$I/$^{247}$Cm. As an illustration, Fig.~\ref{fig:ICm} shows the comparison between evolution of the
$^{129}$I/$^{247}$Cm ratio for one (black) and two (blue) $r$-process sources. In the case with two sources, it is assumed that they are equally frequent and have distinct $^{129}$I/$^{247}$Cm production ratios of 10 and 100. 
As can be seen clearly from the figure, with two sources, the value of $^{129}$I/$^{247}$Cm not only takes 10 and 100 but also intermediate values for substantial intervals.  
Thus, the measured value of the $^{129}$I/$^{247}$Cm ratio may not correspond to the production ratios of the individual $r$-process event. Furthermore, it was shown that when the measured values of $^{129}$I/$^{127}$I and $^{247}$Cm/$^{235}$U are taken in to account, the effect is even more dramatic with  the major contributor accounting for only $\sim 50\%$ of the $^{129}$I in the ESS. 
Overall, it was found that it is not possible to put straightforward constraint on the ``last" $r$-process source using the measured ESS ratio of $^{129}$I/$^{247}$Cm. Nevertheless, 
$^{129}$I/$^{247}$Cm can still be used to put important constraints on the nature of $r$-process sources that were operating  during the formation of the Solar system. In particular, an $r$-process source that is primarily composed of low $Y_e$ ejecta as a major source of $r$-process  is highly disfavored.

\section{Conclusion}\label{sec:conclu}
In this short review article, we have discussed 
several aspects related to actinide production in $r$-process sites and particularly in BNSMs.
First, we reviewed the general condition ($Y_e\lesssim 0.2$) for actinide production via $r$-process in typical neutron star merger outflows. 
We then demonstrated with simple examples the impact of yet-uncertain nuclear physics inputs (including masses, $\beta$-decay half-lives, and fission) on actinide production relevant for kilonova observations. 
Mass models that predict steeper drops of neutron separation energies around $Z\sim 78$ and $N\sim 150$ generally lead to an $r$-process path closer to the stability and permit higher actinide yields around $A\sim 230$.
For $\beta$-decays, a mass model that predicts longer half-lives for actinides directly leads to a smaller amount of actinide production.
Nuclear fission predictions directly affect the amount of actinides that survive against fission for $A\gtrsim 250$, and indirectly influence $A\sim 220-230$ through their impact on post freeze-out abundance of neutrons.

We further reviewed the recent findings regarding the impact of actinide production in BNSM outflows on kilonova lightcurves. 
Both the $\alpha$-decay nuclei ($^{222}$Rn, $^{224}$Ra,
$^{223}$Ra and $^{225}$Ac) and the spontaneous fission nucleus $^{254}$Cf can dominate the radioactive heating for kilonovae at late times ($\gtrsim 10$~days) even with subdominant yields, due to their larger energy release per nucleus than $\beta$ decays.
Detailed late-time observations of future kilonovae, together with improved theoretical modeling of radiation at relevant times can possibly probe the presence of actinides in BNSM outflows. 
Moreover, the amount of actinides produced in BNSMs may also leave imprints on $\gamma$-ray emission for live events or for older remnants ($\sim 10^4$--$10^6$~yr old) in our Milky Way, which may be probed by the next generation MeV $\gamma$-ray missions.

Beyond the detection of electromagnetic signatures that may be realized in future, we also discussed the implications of actinide production in mergers from clues obtained with abundances observed in VMP stars as well as with the SLRIs measured in deep-sea crusts and meteorites. 
The observed VMP stars show a large scatter of Th abundances by $\sim$ a factor of 10 relative to the Eu abundances (classified as actinide-boost, normal, and actinide-poor).
This scatter can be explained by a mixture of high and low $Y_e$ components of BNSM ejecta. 
However, nuclear physics models that predict low Th yields are in tension with actinide boost stars, and may be constrained with improved observational uncertainties, if these stars were indeed enriched by BNSMs.
Besides, better determination of the abundance of Pb in VMP stars in future can also offer insights into the amount of $\alpha$-decay actinides production during BNSMs.

With regard to SLRIs, we reviewed the important conclusions derived in literature that measurements of $^{244}$Pu and $^{247}$Cm in ESS and in Earth's deep-sea crust suggest they were dominantly produced by a rare and prolific BNSM-like source with an occurrence frequency of $\sim 10$~Myr in Milky Way, instead of a more frequent source like typical core-collapse supernovae.
Furthermore, we discussed recent findings on the implication of the measured abundance ratio of SLRIs $^{247}$Cm to $^{129}$I (both have almost identical half-lives) in the ESS, as inferred from meteorites, on the $r$-process condition in BNSMs. 
If the ratio measured in the ESS ratio was from a single ``last'' polluting $r$-process event, then a moderately neutron-rich condition (similar to the BNSM accretion disk outflow) is preferred. 
However, it remains likely that multiple events contributed to this ratio such that a direct inference deduced from single event assumption is not valid.
In this case, this measurement still suggests that an $r$-process source dominantly consisting of low $Y_e$ material as the major source of $r$-process is disfavored.

In summary, in recent years, tremendous observational efforts in several directions have helped to gain crucial insights in to $r$-process and indicates BNSMs as the likely source. This is also consistent with current theoretical predictions although a clear answer, particularly regarding the exact level of neutron richness and associated actinide production, is still lacking. In order to reach a definitive conclusion, further improvements are required in theoretical predictions combined with additional observational data.
It is hopeful that with all the probes summarized in the review, and potentially other new ideas, this question will be fully answered in the coming decades.

\begin{acknowledgments}
We are grateful for the fruitful collaboration with Jennifer Barnes, Samuel Giuliani, Gabriel Mart\`inez-Pinedo, and Brian Metzger leading to works relevant to this paper. We also thank Benoit C\^ote, Kenta Hotokezaka, Yong-Zhong Qian, Friedel Thielemann, and Zhen Yuan for insightful discussions and exchanges over the past few years. M.-R.~W. acknowledges supports from the Ministry of Science and Technology, Taiwan under Grant No.~110-2112-M-001-050, the Academia Sinica under Project No.~AS-CDA-109-M11, and the Physics Division, National Center for Theoretical Sciences, Taiwan.
\end{acknowledgments}


\begin{thebibliography}{124}%
\makeatletter
\providecommand \@ifxundefined [1]{%
 \@ifx{#1\undefined}
}%
\providecommand \@ifnum [1]{%
 \ifnum #1\expandafter \@firstoftwo
 \else \expandafter \@secondoftwo
 \fi
}%
\providecommand \@ifx [1]{%
 \ifx #1\expandafter \@firstoftwo
 \else \expandafter \@secondoftwo
 \fi
}%
\providecommand \natexlab [1]{#1}%
\providecommand \enquote  [1]{``#1''}%
\providecommand \bibnamefont  [1]{#1}%
\providecommand \bibfnamefont [1]{#1}%
\providecommand \citenamefont [1]{#1}%
\providecommand \href@noop [0]{\@secondoftwo}%
\providecommand \href [0]{\begingroup \@sanitize@url \@href}%
\providecommand \@href[1]{\@@startlink{#1}\@@href}%
\providecommand \@@href[1]{\endgroup#1\@@endlink}%
\providecommand \@sanitize@url [0]{\catcode `\\12\catcode `\$12\catcode
  `\&12\catcode `\#12\catcode `\^12\catcode `\_12\catcode `\%12\relax}%
\providecommand \@@startlink[1]{}%
\providecommand \@@endlink[0]{}%
\providecommand \url  [0]{\begingroup\@sanitize@url \@url }%
\providecommand \@url [1]{\endgroup\@href {#1}{\urlprefix }}%
\providecommand \urlprefix  [0]{URL }%
\providecommand \Eprint [0]{\href }%
\providecommand \doibase [0]{http://dx.doi.org/}%
\providecommand \selectlanguage [0]{\@gobble}%
\providecommand \bibinfo  [0]{\@secondoftwo}%
\providecommand \bibfield  [0]{\@secondoftwo}%
\providecommand \translation [1]{[#1]}%
\providecommand \BibitemOpen [0]{}%
\providecommand \bibitemStop [0]{}%
\providecommand \bibitemNoStop [0]{.\EOS\space}%
\providecommand \EOS [0]{\spacefactor3000\relax}%
\providecommand \BibitemShut  [1]{\csname bibitem#1\endcsname}%
\let\auto@bib@innerbib\@empty
\bibitem [{\citenamefont {{Abbott}}\ \emph
  {et~al.}(2017{\natexlab{a}})\citenamefont {{Abbott}}, \citenamefont
  {{Abbott}}, \citenamefont {{Abbott}}, \citenamefont {{Acernese}},
  \citenamefont {{Ackley}}, \citenamefont {{Adams}}, \citenamefont {{Adams}},
  \citenamefont {{Addesso}}, \citenamefont {{Adhikari}}, \citenamefont
  {{Adya}}, \citenamefont {{Affeldt}}, \citenamefont {{Afrough}}, \citenamefont
  {{Agarwal}}, \citenamefont {{Agathos}}, \citenamefont {{Agatsuma}},
  \citenamefont {{Aggarwal}}, \citenamefont {{Aguiar}}, \citenamefont
  {{Aiello}}, \citenamefont {{Ain}}, \citenamefont {{Bailes}}, \citenamefont
  {{Baker}}, \citenamefont {{Baldaccini}}, \citenamefont {{Ballardin}},
  \citenamefont {{Ballmer}}, \citenamefont {{Banagiri}}, \citenamefont
  {{Barayoga}}, \citenamefont {{Barclay}}, \citenamefont {{Barish}},
  \citenamefont {{Barker}}, \citenamefont {{Barkett}}, \citenamefont
  {{Barone}}, \citenamefont {{Barr}}, \citenamefont {{Barsotti}}, \citenamefont
  {{Barsuglia}}, \citenamefont {{Barta}}, \citenamefont {{Barthelmy}},
  \citenamefont {{Bartlett}}, \citenamefont {{Bartos}}, \citenamefont
  {{Bassiri}}, \citenamefont {{Basti}}, \citenamefont {{Batch}},\ and\
  \citenamefont {{Bawaj}}}]{abbott2017a}%
  \BibitemOpen
  \bibfield  {author} {\bibinfo {author} {\bibfnamefont {B.~P.}\ \bibnamefont
  {{Abbott}}}, \bibinfo {author} {\bibfnamefont {R.}~\bibnamefont {{Abbott}}},
  \bibinfo {author} {\bibfnamefont {T.~D.}\ \bibnamefont {{Abbott}}}, \bibinfo
  {author} {\bibfnamefont {F.}~\bibnamefont {{Acernese}}}, \bibinfo {author}
  {\bibfnamefont {K.}~\bibnamefont {{Ackley}}}, \bibinfo {author}
  {\bibfnamefont {C.}~\bibnamefont {{Adams}}}, \bibinfo {author} {\bibfnamefont
  {T.}~\bibnamefont {{Adams}}}, \bibinfo {author} {\bibfnamefont
  {P.}~\bibnamefont {{Addesso}}}, \bibinfo {author} {\bibfnamefont {R.~X.}\
  \bibnamefont {{Adhikari}}}, \bibinfo {author} {\bibfnamefont {V.~B.}\
  \bibnamefont {{Adya}}}, \bibinfo {author} {\bibfnamefont {C.}~\bibnamefont
  {{Affeldt}}}, \bibinfo {author} {\bibfnamefont {M.}~\bibnamefont
  {{Afrough}}}, \bibinfo {author} {\bibfnamefont {B.}~\bibnamefont
  {{Agarwal}}}, \bibinfo {author} {\bibfnamefont {M.}~\bibnamefont
  {{Agathos}}}, \bibinfo {author} {\bibfnamefont {K.}~\bibnamefont
  {{Agatsuma}}}, \bibinfo {author} {\bibfnamefont {N.}~\bibnamefont
  {{Aggarwal}}}, \bibinfo {author} {\bibfnamefont {O.~D.}\ \bibnamefont
  {{Aguiar}}}, \bibinfo {author} {\bibfnamefont {L.}~\bibnamefont {{Aiello}}},
  \bibinfo {author} {\bibfnamefont {A.}~\bibnamefont {{Ain}}}, \bibinfo
  {author} {\bibfnamefont {M.}~\bibnamefont {{Bailes}}}, \bibinfo {author}
  {\bibfnamefont {P.~T.}\ \bibnamefont {{Baker}}}, \bibinfo {author}
  {\bibfnamefont {F.}~\bibnamefont {{Baldaccini}}}, \bibinfo {author}
  {\bibfnamefont {G.}~\bibnamefont {{Ballardin}}}, \bibinfo {author}
  {\bibfnamefont {S.~W.}\ \bibnamefont {{Ballmer}}}, \bibinfo {author}
  {\bibfnamefont {S.}~\bibnamefont {{Banagiri}}}, \bibinfo {author}
  {\bibfnamefont {J.~C.}\ \bibnamefont {{Barayoga}}}, \bibinfo {author}
  {\bibfnamefont {S.~E.}\ \bibnamefont {{Barclay}}}, \bibinfo {author}
  {\bibfnamefont {B.~C.}\ \bibnamefont {{Barish}}}, \bibinfo {author}
  {\bibfnamefont {D.}~\bibnamefont {{Barker}}}, \bibinfo {author}
  {\bibfnamefont {K.}~\bibnamefont {{Barkett}}}, \bibinfo {author}
  {\bibfnamefont {F.}~\bibnamefont {{Barone}}}, \bibinfo {author}
  {\bibfnamefont {B.}~\bibnamefont {{Barr}}}, \bibinfo {author} {\bibfnamefont
  {L.}~\bibnamefont {{Barsotti}}}, \bibinfo {author} {\bibfnamefont
  {M.}~\bibnamefont {{Barsuglia}}}, \bibinfo {author} {\bibfnamefont
  {D.}~\bibnamefont {{Barta}}}, \bibinfo {author} {\bibfnamefont {S.~D.}\
  \bibnamefont {{Barthelmy}}}, \bibinfo {author} {\bibfnamefont
  {J.}~\bibnamefont {{Bartlett}}}, \bibinfo {author} {\bibfnamefont
  {I.}~\bibnamefont {{Bartos}}}, \bibinfo {author} {\bibfnamefont
  {R.}~\bibnamefont {{Bassiri}}}, \bibinfo {author} {\bibfnamefont
  {A.}~\bibnamefont {{Basti}}}, \bibinfo {author} {\bibfnamefont {J.~C.}\
  \bibnamefont {{Batch}}}, \ and\ \bibinfo {author} {\bibfnamefont
  {M.}~\bibnamefont {{Bawaj}}},\ }\href {\doibase
  10.1103/PhysRevLett.119.161101} {\bibfield  {journal} {\bibinfo  {journal}
  {\prl}\ }\textbf {\bibinfo {volume} {119}},\ \bibinfo {eid} {161101}
  (\bibinfo {year} {2017}{\natexlab{a}})},\ \Eprint
  {http://arxiv.org/abs/1710.05832} {arXiv:1710.05832 [gr-qc]} \BibitemShut
  {NoStop}%
\bibitem [{\citenamefont {{Abbott}}\ \emph
  {et~al.}(2017{\natexlab{b}})\citenamefont {{Abbott}}, \citenamefont
  {{Abbott}}, \citenamefont {{Abbott}}, \citenamefont {{Acernese}},
  \citenamefont {{Ackley}}, \citenamefont {{Adams}}, \citenamefont {{Adams}},
  \citenamefont {{Addesso}}, \citenamefont {{Adhikari}}, \citenamefont
  {{Adya}}, \citenamefont {{Affeldt}}, \citenamefont {{Afrough}}, \citenamefont
  {{Agarwal}}, \citenamefont {{Agathos}}, \citenamefont {{Agatsuma}},
  \citenamefont {{Aggarwal}}, \citenamefont {{Aguiar}}, \citenamefont
  {{Aiello}}, \citenamefont {{Ain}}, \citenamefont {{Ajith}}, \citenamefont
  {{Allen}}, \citenamefont {{Allen}}, \citenamefont {{Allocca}}, \citenamefont
  {{Altin}}, \citenamefont {{Amato}}, \citenamefont {{Ananyeva}}, \citenamefont
  {{Anderson}}, \citenamefont {{Anderson}}, \citenamefont {{Angelova}},
  \citenamefont {{Antier}}, \citenamefont {{Appert}}, \citenamefont {{Arai}},
  \citenamefont {{Araya}}, \citenamefont {{Areeda}}, \citenamefont {{Arnaud}},
  \citenamefont {{Arun}}, \citenamefont {{Ascenzi}}, \citenamefont {{Ashton}},
  \citenamefont {{Ast}}, \citenamefont {{Aston}},\ and\ \citenamefont
  {{Astone}}}]{abbott2017b}%
  \BibitemOpen
  \bibfield  {author} {\bibinfo {author} {\bibfnamefont {B.~P.}\ \bibnamefont
  {{Abbott}}}, \bibinfo {author} {\bibfnamefont {R.}~\bibnamefont {{Abbott}}},
  \bibinfo {author} {\bibfnamefont {T.~D.}\ \bibnamefont {{Abbott}}}, \bibinfo
  {author} {\bibfnamefont {F.}~\bibnamefont {{Acernese}}}, \bibinfo {author}
  {\bibfnamefont {K.}~\bibnamefont {{Ackley}}}, \bibinfo {author}
  {\bibfnamefont {C.}~\bibnamefont {{Adams}}}, \bibinfo {author} {\bibfnamefont
  {T.}~\bibnamefont {{Adams}}}, \bibinfo {author} {\bibfnamefont
  {P.}~\bibnamefont {{Addesso}}}, \bibinfo {author} {\bibfnamefont {R.~X.}\
  \bibnamefont {{Adhikari}}}, \bibinfo {author} {\bibfnamefont {V.~B.}\
  \bibnamefont {{Adya}}}, \bibinfo {author} {\bibfnamefont {C.}~\bibnamefont
  {{Affeldt}}}, \bibinfo {author} {\bibfnamefont {M.}~\bibnamefont
  {{Afrough}}}, \bibinfo {author} {\bibfnamefont {B.}~\bibnamefont
  {{Agarwal}}}, \bibinfo {author} {\bibfnamefont {M.}~\bibnamefont
  {{Agathos}}}, \bibinfo {author} {\bibfnamefont {K.}~\bibnamefont
  {{Agatsuma}}}, \bibinfo {author} {\bibfnamefont {N.}~\bibnamefont
  {{Aggarwal}}}, \bibinfo {author} {\bibfnamefont {O.~D.}\ \bibnamefont
  {{Aguiar}}}, \bibinfo {author} {\bibfnamefont {L.}~\bibnamefont {{Aiello}}},
  \bibinfo {author} {\bibfnamefont {A.}~\bibnamefont {{Ain}}}, \bibinfo
  {author} {\bibfnamefont {P.}~\bibnamefont {{Ajith}}}, \bibinfo {author}
  {\bibfnamefont {B.}~\bibnamefont {{Allen}}}, \bibinfo {author} {\bibfnamefont
  {G.}~\bibnamefont {{Allen}}}, \bibinfo {author} {\bibfnamefont
  {A.}~\bibnamefont {{Allocca}}}, \bibinfo {author} {\bibfnamefont {P.~A.}\
  \bibnamefont {{Altin}}}, \bibinfo {author} {\bibfnamefont {A.}~\bibnamefont
  {{Amato}}}, \bibinfo {author} {\bibfnamefont {A.}~\bibnamefont {{Ananyeva}}},
  \bibinfo {author} {\bibfnamefont {S.~B.}\ \bibnamefont {{Anderson}}},
  \bibinfo {author} {\bibfnamefont {W.~G.}\ \bibnamefont {{Anderson}}},
  \bibinfo {author} {\bibfnamefont {S.~V.}\ \bibnamefont {{Angelova}}},
  \bibinfo {author} {\bibfnamefont {S.}~\bibnamefont {{Antier}}}, \bibinfo
  {author} {\bibfnamefont {S.}~\bibnamefont {{Appert}}}, \bibinfo {author}
  {\bibfnamefont {K.}~\bibnamefont {{Arai}}}, \bibinfo {author} {\bibfnamefont
  {M.~C.}\ \bibnamefont {{Araya}}}, \bibinfo {author} {\bibfnamefont {J.~S.}\
  \bibnamefont {{Areeda}}}, \bibinfo {author} {\bibfnamefont {N.}~\bibnamefont
  {{Arnaud}}}, \bibinfo {author} {\bibfnamefont {K.~G.}\ \bibnamefont
  {{Arun}}}, \bibinfo {author} {\bibfnamefont {S.}~\bibnamefont {{Ascenzi}}},
  \bibinfo {author} {\bibfnamefont {G.}~\bibnamefont {{Ashton}}}, \bibinfo
  {author} {\bibfnamefont {M.}~\bibnamefont {{Ast}}}, \bibinfo {author}
  {\bibfnamefont {S.~M.}\ \bibnamefont {{Aston}}}, \ and\ \bibinfo {author}
  {\bibfnamefont {P.}~\bibnamefont {{Astone}}},\ }\href {\doibase
  10.3847/2041-8213/aa91c9} {\bibfield  {journal} {\bibinfo  {journal} {\apjl}\
  }\textbf {\bibinfo {volume} {848}},\ \bibinfo {eid} {L12} (\bibinfo {year}
  {2017}{\natexlab{b}})},\ \Eprint {http://arxiv.org/abs/1710.05833}
  {arXiv:1710.05833 [astro-ph.HE]} \BibitemShut {NoStop}%
\bibitem [{\citenamefont {Metzger}(2020)}]{Metzger:2019zeh}%
  \BibitemOpen
  \bibfield  {author} {\bibinfo {author} {\bibfnamefont {B.~D.}\ \bibnamefont
  {Metzger}},\ }\href {\doibase 10.1007/s41114-019-0024-0} {\bibfield
  {journal} {\bibinfo  {journal} {Living Rev. Rel.}\ }\textbf {\bibinfo
  {volume} {23}},\ \bibinfo {pages} {1} (\bibinfo {year} {2020})},\ \Eprint
  {http://arxiv.org/abs/1910.01617} {arXiv:1910.01617 [astro-ph.HE]}
  \BibitemShut {NoStop}%
\bibitem [{\citenamefont {Nakar}(2020)}]{Nakar:2019fza}%
  \BibitemOpen
  \bibfield  {author} {\bibinfo {author} {\bibfnamefont {E.}~\bibnamefont
  {Nakar}},\ }\href {\doibase 10.1016/j.physrep.2020.08.008} {\bibfield
  {journal} {\bibinfo  {journal} {Phys. Rept.}\ }\textbf {\bibinfo {volume}
  {886}},\ \bibinfo {pages} {1} (\bibinfo {year} {2020})},\ \Eprint
  {http://arxiv.org/abs/1912.05659} {arXiv:1912.05659 [astro-ph.HE]}
  \BibitemShut {NoStop}%
\bibitem [{\citenamefont {Shibata}\ and\ \citenamefont
  {Hotokezaka}(2019)}]{Shibata:2019wef}%
  \BibitemOpen
  \bibfield  {author} {\bibinfo {author} {\bibfnamefont {M.}~\bibnamefont
  {Shibata}}\ and\ \bibinfo {author} {\bibfnamefont {K.}~\bibnamefont
  {Hotokezaka}},\ }\href {\doibase 10.1146/annurev-nucl-101918-023625}
  {\bibfield  {journal} {\bibinfo  {journal} {Ann. Rev. Nucl. Part. Sci.}\
  }\textbf {\bibinfo {volume} {69}},\ \bibinfo {pages} {41} (\bibinfo {year}
  {2019})},\ \Eprint {http://arxiv.org/abs/1908.02350} {arXiv:1908.02350
  [astro-ph.HE]} \BibitemShut {NoStop}%
\bibitem [{\citenamefont {Margutti}\ and\ \citenamefont
  {Chornock}(2021)}]{Margutti:2020xbo}%
  \BibitemOpen
  \bibfield  {author} {\bibinfo {author} {\bibfnamefont {R.}~\bibnamefont
  {Margutti}}\ and\ \bibinfo {author} {\bibfnamefont {R.}~\bibnamefont
  {Chornock}},\ }\href {\doibase 10.1146/annurev-astro-112420-030742}
  {\bibfield  {journal} {\bibinfo  {journal} {Ann. Rev. Astron. Astrophys.}\
  }\textbf {\bibinfo {volume} {59}},\ \bibinfo {pages} {155} (\bibinfo {year}
  {2021})},\ \Eprint {http://arxiv.org/abs/2012.04810} {arXiv:2012.04810
  [astro-ph.HE]} \BibitemShut {NoStop}%
\bibitem [{\citenamefont {Perego}\ \emph {et~al.}(2021)\citenamefont {Perego},
  \citenamefont {Thielemann},\ and\ \citenamefont {Cescutti}}]{Perego:2021dpw}%
  \BibitemOpen
  \bibfield  {author} {\bibinfo {author} {\bibfnamefont {A.}~\bibnamefont
  {Perego}}, \bibinfo {author} {\bibfnamefont {F.-K.}\ \bibnamefont
  {Thielemann}}, \ and\ \bibinfo {author} {\bibfnamefont {G.}~\bibnamefont
  {Cescutti}},\ }\enquote {\bibinfo {title} {{r-Process Nucleosynthesis from
  Compact Binary Mergers}},}\ \ (\bibinfo {year} {2021})\ \Eprint
  {http://arxiv.org/abs/2109.09162} {arXiv:2109.09162 [astro-ph.HE]}
  \BibitemShut {NoStop}%
\bibitem [{\citenamefont {{Kasen}}\ \emph {et~al.}(2017)\citenamefont
  {{Kasen}}, \citenamefont {{Metzger}}, \citenamefont {{Barnes}}, \citenamefont
  {{Quataert}},\ and\ \citenamefont {{Ramirez-Ruiz}}}]{Kasen+17}%
  \BibitemOpen
  \bibfield  {author} {\bibinfo {author} {\bibfnamefont {D.}~\bibnamefont
  {{Kasen}}}, \bibinfo {author} {\bibfnamefont {B.}~\bibnamefont {{Metzger}}},
  \bibinfo {author} {\bibfnamefont {J.}~\bibnamefont {{Barnes}}}, \bibinfo
  {author} {\bibfnamefont {E.}~\bibnamefont {{Quataert}}}, \ and\ \bibinfo
  {author} {\bibfnamefont {E.}~\bibnamefont {{Ramirez-Ruiz}}},\ }\href
  {\doibase 10.1038/nature24453} {\bibfield  {journal} {\bibinfo  {journal}
  {\nat}\ }\textbf {\bibinfo {volume} {551}},\ \bibinfo {pages} {80} (\bibinfo
  {year} {2017})},\ \Eprint {http://arxiv.org/abs/1710.05463} {arXiv:1710.05463
  [astro-ph.HE]} \BibitemShut {NoStop}%
\bibitem [{\citenamefont {{Drout}}\ \emph {et~al.}(2017)\citenamefont
  {{Drout}}, \citenamefont {{Piro}}, \citenamefont {{Shappee}}, \citenamefont
  {{Kilpatrick}}, \citenamefont {{Simon}}, \citenamefont {{Contreras}},
  \citenamefont {{Coulter}}, \citenamefont {{Foley}}, \citenamefont
  {{Siebert}}, \citenamefont {{Morrell}}, \citenamefont {{Boutsia}},
  \citenamefont {{Di Mille}}, \citenamefont {{Holoien}}, \citenamefont
  {{Kasen}}, \citenamefont {{Kollmeier}}, \citenamefont {{Madore}},
  \citenamefont {{Monson}}, \citenamefont {{Murguia-Berthier}}, \citenamefont
  {{Pan}}, \citenamefont {{Prochaska}}, \citenamefont {{Ramirez-Ruiz}},
  \citenamefont {{Rest}}, \citenamefont {{Adams}}, \citenamefont {{Alatalo}},
  \citenamefont {{Ba{\~n}ados}}, \citenamefont {{Baughman}}, \citenamefont
  {{Beers}}, \citenamefont {{Bernstein}}, \citenamefont {{Bitsakis}},
  \citenamefont {{Campillay}}, \citenamefont {{Hansen}}, \citenamefont
  {{Higgs}}, \citenamefont {{Ji}}, \citenamefont {{Maravelias}}, \citenamefont
  {{Marshall}}, \citenamefont {{Moni Bidin}}, \citenamefont {{Prieto}},
  \citenamefont {{Rasmussen}}, \citenamefont {{Rojas-Bravo}}, \citenamefont
  {{Strom}}, \citenamefont {{Ulloa}}, \citenamefont {{Vargas-Gonz{\'a}lez}},
  \citenamefont {{Wan}},\ and\ \citenamefont {{Whitten}}}]{Drout:2017ijr}%
  \BibitemOpen
  \bibfield  {author} {\bibinfo {author} {\bibfnamefont {M.~R.}\ \bibnamefont
  {{Drout}}}, \bibinfo {author} {\bibfnamefont {A.~L.}\ \bibnamefont {{Piro}}},
  \bibinfo {author} {\bibfnamefont {B.~J.}\ \bibnamefont {{Shappee}}}, \bibinfo
  {author} {\bibfnamefont {C.~D.}\ \bibnamefont {{Kilpatrick}}}, \bibinfo
  {author} {\bibfnamefont {J.~D.}\ \bibnamefont {{Simon}}}, \bibinfo {author}
  {\bibfnamefont {C.}~\bibnamefont {{Contreras}}}, \bibinfo {author}
  {\bibfnamefont {D.~A.}\ \bibnamefont {{Coulter}}}, \bibinfo {author}
  {\bibfnamefont {R.~J.}\ \bibnamefont {{Foley}}}, \bibinfo {author}
  {\bibfnamefont {M.~R.}\ \bibnamefont {{Siebert}}}, \bibinfo {author}
  {\bibfnamefont {N.}~\bibnamefont {{Morrell}}}, \bibinfo {author}
  {\bibfnamefont {K.}~\bibnamefont {{Boutsia}}}, \bibinfo {author}
  {\bibfnamefont {F.}~\bibnamefont {{Di Mille}}}, \bibinfo {author}
  {\bibfnamefont {T.~W.~S.}\ \bibnamefont {{Holoien}}}, \bibinfo {author}
  {\bibfnamefont {D.}~\bibnamefont {{Kasen}}}, \bibinfo {author} {\bibfnamefont
  {J.~A.}\ \bibnamefont {{Kollmeier}}}, \bibinfo {author} {\bibfnamefont
  {B.~F.}\ \bibnamefont {{Madore}}}, \bibinfo {author} {\bibfnamefont {A.~J.}\
  \bibnamefont {{Monson}}}, \bibinfo {author} {\bibfnamefont {A.}~\bibnamefont
  {{Murguia-Berthier}}}, \bibinfo {author} {\bibfnamefont {Y.~C.}\ \bibnamefont
  {{Pan}}}, \bibinfo {author} {\bibfnamefont {J.~X.}\ \bibnamefont
  {{Prochaska}}}, \bibinfo {author} {\bibfnamefont {E.}~\bibnamefont
  {{Ramirez-Ruiz}}}, \bibinfo {author} {\bibfnamefont {A.}~\bibnamefont
  {{Rest}}}, \bibinfo {author} {\bibfnamefont {C.}~\bibnamefont {{Adams}}},
  \bibinfo {author} {\bibfnamefont {K.}~\bibnamefont {{Alatalo}}}, \bibinfo
  {author} {\bibfnamefont {E.}~\bibnamefont {{Ba{\~n}ados}}}, \bibinfo {author}
  {\bibfnamefont {J.}~\bibnamefont {{Baughman}}}, \bibinfo {author}
  {\bibfnamefont {T.~C.}\ \bibnamefont {{Beers}}}, \bibinfo {author}
  {\bibfnamefont {R.~A.}\ \bibnamefont {{Bernstein}}}, \bibinfo {author}
  {\bibfnamefont {T.}~\bibnamefont {{Bitsakis}}}, \bibinfo {author}
  {\bibfnamefont {A.}~\bibnamefont {{Campillay}}}, \bibinfo {author}
  {\bibfnamefont {T.~T.}\ \bibnamefont {{Hansen}}}, \bibinfo {author}
  {\bibfnamefont {C.~R.}\ \bibnamefont {{Higgs}}}, \bibinfo {author}
  {\bibfnamefont {A.~P.}\ \bibnamefont {{Ji}}}, \bibinfo {author}
  {\bibfnamefont {G.}~\bibnamefont {{Maravelias}}}, \bibinfo {author}
  {\bibfnamefont {J.~L.}\ \bibnamefont {{Marshall}}}, \bibinfo {author}
  {\bibfnamefont {C.}~\bibnamefont {{Moni Bidin}}}, \bibinfo {author}
  {\bibfnamefont {J.~L.}\ \bibnamefont {{Prieto}}}, \bibinfo {author}
  {\bibfnamefont {K.~C.}\ \bibnamefont {{Rasmussen}}}, \bibinfo {author}
  {\bibfnamefont {C.}~\bibnamefont {{Rojas-Bravo}}}, \bibinfo {author}
  {\bibfnamefont {A.~L.}\ \bibnamefont {{Strom}}}, \bibinfo {author}
  {\bibfnamefont {N.}~\bibnamefont {{Ulloa}}}, \bibinfo {author} {\bibfnamefont
  {J.}~\bibnamefont {{Vargas-Gonz{\'a}lez}}}, \bibinfo {author} {\bibfnamefont
  {Z.}~\bibnamefont {{Wan}}}, \ and\ \bibinfo {author} {\bibfnamefont {D.~D.}\
  \bibnamefont {{Whitten}}},\ }\href {\doibase 10.1126/science.aaq0049}
  {\bibfield  {journal} {\bibinfo  {journal} {Science}\ }\textbf {\bibinfo
  {volume} {358}},\ \bibinfo {pages} {1570} (\bibinfo {year} {2017})},\ \Eprint
  {http://arxiv.org/abs/1710.05443} {arXiv:1710.05443 [astro-ph.HE]}
  \BibitemShut {NoStop}%
\bibitem [{\citenamefont {{Cowperthwaite}}\ \emph {et~al.}(2017)\citenamefont
  {{Cowperthwaite}}, \citenamefont {{Berger}}, \citenamefont {{Villar}},
  \citenamefont {{Metzger}}, \citenamefont {{Nicholl}}, \citenamefont
  {{Chornock}}, \citenamefont {{Blanchard}}, \citenamefont {{Fong}},
  \citenamefont {{Margutti}}, \citenamefont {{Soares-Santos}}, \citenamefont
  {{Alexander}}, \citenamefont {{Allam}}, \citenamefont {{Annis}},
  \citenamefont {{Brout}}, \citenamefont {{Brown}}, \citenamefont {{Butler}},
  \citenamefont {{Chen}}, \citenamefont {{Diehl}}, \citenamefont {{Doctor}},
  \citenamefont {{Drout}}, \citenamefont {{Eftekhari}}, \citenamefont {{Farr}},
  \citenamefont {{Finley}}, \citenamefont {{Foley}}, \citenamefont {{Frieman}},
  \citenamefont {{Fryer}}, \citenamefont {{Garc{\'\i}a-Bellido}}, \citenamefont
  {{Gill}}, \citenamefont {{Guillochon}}, \citenamefont {{Herner}},
  \citenamefont {{Holz}}, \citenamefont {{Kasen}}, \citenamefont {{Kessler}},
  \citenamefont {{Marriner}}, \citenamefont {{Matheson}}, \citenamefont
  {{Neilsen}}, \citenamefont {{Quataert}}, \citenamefont {{Palmese}},
  \citenamefont {{Rest}}, \citenamefont {{Sako}}, \citenamefont {{Scolnic}},
  \citenamefont {{Smith}}, \citenamefont {{Tucker}}, \citenamefont
  {{Williams}}, \citenamefont {{Balbinot}}, \citenamefont {{Carlin}},
  \citenamefont {{Cook}}, \citenamefont {{Durret}}, \citenamefont {{Li}},
  \citenamefont {{Lopes}}, \citenamefont {{Louren{\c{c}}o}}, \citenamefont
  {{Marshall}}, \citenamefont {{Medina}}, \citenamefont {{Muir}}, \citenamefont
  {{Mu{\~n}oz}}, \citenamefont {{Sauseda}}, \citenamefont {{Schlegel}},
  \citenamefont {{Secco}}, \citenamefont {{Vivas}}, \citenamefont {{Wester}},
  \citenamefont {{Zenteno}}, \citenamefont {{Zhang}}, \citenamefont {{Abbott}},
  \citenamefont {{Banerji}}, \citenamefont {{Bechtol}}, \citenamefont
  {{Benoit-L{\'e}vy}}, \citenamefont {{Bertin}}, \citenamefont
  {{Buckley-Geer}}, \citenamefont {{Burke}}, \citenamefont {{Capozzi}},
  \citenamefont {{Carnero Rosell}}, \citenamefont {{Carrasco Kind}},
  \citenamefont {{Castander}}, \citenamefont {{Crocce}}, \citenamefont
  {{Cunha}}, \citenamefont {{D'Andrea}}, \citenamefont {{da Costa}},
  \citenamefont {{Davis}}, \citenamefont {{DePoy}}, \citenamefont {{Desai}},
  \citenamefont {{Dietrich}}, \citenamefont {{Drlica-Wagner}}, \citenamefont
  {{Eifler}}, \citenamefont {{Evrard}}, \citenamefont {{Fernand ez}},
  \citenamefont {{Flaugher}}, \citenamefont {{Fosalba}}, \citenamefont
  {{Gaztanaga}}, \citenamefont {{Gerdes}}, \citenamefont {{Giannantonio}},
  \citenamefont {{Goldstein}}, \citenamefont {{Gruen}}, \citenamefont
  {{Gruendl}}, \citenamefont {{Gutierrez}}, \citenamefont {{Honscheid}},
  \citenamefont {{Jain}}, \citenamefont {{James}}, \citenamefont {{Jeltema}},
  \citenamefont {{Johnson}}, \citenamefont {{Johnson}}, \citenamefont {{Kent}},
  \citenamefont {{Krause}}, \citenamefont {{Kron}}, \citenamefont {{Kuehn}},
  \citenamefont {{Nuropatkin}}, \citenamefont {{Lahav}}, \citenamefont
  {{Lima}}, \citenamefont {{Lin}}, \citenamefont {{Maia}}, \citenamefont
  {{March}}, \citenamefont {{Martini}}, \citenamefont {{McMahon}},
  \citenamefont {{Menanteau}}, \citenamefont {{Miller}}, \citenamefont
  {{Miquel}}, \citenamefont {{Mohr}}, \citenamefont {{Neilsen}}, \citenamefont
  {{Nichol}}, \citenamefont {{Ogando}}, \citenamefont {{Plazas}}, \citenamefont
  {{Roe}}, \citenamefont {{Romer}}, \citenamefont {{Roodman}}, \citenamefont
  {{Rykoff}}, \citenamefont {{Sanchez}}, \citenamefont {{Scarpine}},
  \citenamefont {{Schindler}}, \citenamefont {{Schubnell}}, \citenamefont
  {{Sevilla-Noarbe}}, \citenamefont {{Smith}}, \citenamefont {{Smith}},
  \citenamefont {{Sobreira}}, \citenamefont {{Suchyta}}, \citenamefont
  {{Swanson}}, \citenamefont {{Tarle}}, \citenamefont {{Thomas}}, \citenamefont
  {{Thomas}}, \citenamefont {{Troxel}}, \citenamefont {{Vikram}}, \citenamefont
  {{Walker}}, \citenamefont {{Wechsler}}, \citenamefont {{Weller}},
  \citenamefont {{Yanny}},\ and\ \citenamefont {{Zuntz}}}]{Cowperthwaite+17}%
  \BibitemOpen
  \bibfield  {author} {\bibinfo {author} {\bibfnamefont {P.~S.}\ \bibnamefont
  {{Cowperthwaite}}}, \bibinfo {author} {\bibfnamefont {E.}~\bibnamefont
  {{Berger}}}, \bibinfo {author} {\bibfnamefont {V.~A.}\ \bibnamefont
  {{Villar}}}, \bibinfo {author} {\bibfnamefont {B.~D.}\ \bibnamefont
  {{Metzger}}}, \bibinfo {author} {\bibfnamefont {M.}~\bibnamefont
  {{Nicholl}}}, \bibinfo {author} {\bibfnamefont {R.}~\bibnamefont
  {{Chornock}}}, \bibinfo {author} {\bibfnamefont {P.~K.}\ \bibnamefont
  {{Blanchard}}}, \bibinfo {author} {\bibfnamefont {W.}~\bibnamefont {{Fong}}},
  \bibinfo {author} {\bibfnamefont {R.}~\bibnamefont {{Margutti}}}, \bibinfo
  {author} {\bibfnamefont {M.}~\bibnamefont {{Soares-Santos}}}, \bibinfo
  {author} {\bibfnamefont {K.~D.}\ \bibnamefont {{Alexander}}}, \bibinfo
  {author} {\bibfnamefont {S.}~\bibnamefont {{Allam}}}, \bibinfo {author}
  {\bibfnamefont {J.}~\bibnamefont {{Annis}}}, \bibinfo {author} {\bibfnamefont
  {D.}~\bibnamefont {{Brout}}}, \bibinfo {author} {\bibfnamefont {D.~A.}\
  \bibnamefont {{Brown}}}, \bibinfo {author} {\bibfnamefont {R.~E.}\
  \bibnamefont {{Butler}}}, \bibinfo {author} {\bibfnamefont {H.~Y.}\
  \bibnamefont {{Chen}}}, \bibinfo {author} {\bibfnamefont {H.~T.}\
  \bibnamefont {{Diehl}}}, \bibinfo {author} {\bibfnamefont {Z.}~\bibnamefont
  {{Doctor}}}, \bibinfo {author} {\bibfnamefont {M.~R.}\ \bibnamefont
  {{Drout}}}, \bibinfo {author} {\bibfnamefont {T.}~\bibnamefont
  {{Eftekhari}}}, \bibinfo {author} {\bibfnamefont {B.}~\bibnamefont {{Farr}}},
  \bibinfo {author} {\bibfnamefont {D.~A.}\ \bibnamefont {{Finley}}}, \bibinfo
  {author} {\bibfnamefont {R.~J.}\ \bibnamefont {{Foley}}}, \bibinfo {author}
  {\bibfnamefont {J.~A.}\ \bibnamefont {{Frieman}}}, \bibinfo {author}
  {\bibfnamefont {C.~L.}\ \bibnamefont {{Fryer}}}, \bibinfo {author}
  {\bibfnamefont {J.}~\bibnamefont {{Garc{\'\i}a-Bellido}}}, \bibinfo {author}
  {\bibfnamefont {M.~S.~S.}\ \bibnamefont {{Gill}}}, \bibinfo {author}
  {\bibfnamefont {J.}~\bibnamefont {{Guillochon}}}, \bibinfo {author}
  {\bibfnamefont {K.}~\bibnamefont {{Herner}}}, \bibinfo {author}
  {\bibfnamefont {D.~E.}\ \bibnamefont {{Holz}}}, \bibinfo {author}
  {\bibfnamefont {D.}~\bibnamefont {{Kasen}}}, \bibinfo {author} {\bibfnamefont
  {R.}~\bibnamefont {{Kessler}}}, \bibinfo {author} {\bibfnamefont
  {J.}~\bibnamefont {{Marriner}}}, \bibinfo {author} {\bibfnamefont
  {T.}~\bibnamefont {{Matheson}}}, \bibinfo {author} {\bibfnamefont
  {J.}~\bibnamefont {{Neilsen}}, \bibfnamefont {E.~H.}}, \bibinfo {author}
  {\bibfnamefont {E.}~\bibnamefont {{Quataert}}}, \bibinfo {author}
  {\bibfnamefont {A.}~\bibnamefont {{Palmese}}}, \bibinfo {author}
  {\bibfnamefont {A.}~\bibnamefont {{Rest}}}, \bibinfo {author} {\bibfnamefont
  {M.}~\bibnamefont {{Sako}}}, \bibinfo {author} {\bibfnamefont {D.~M.}\
  \bibnamefont {{Scolnic}}}, \bibinfo {author} {\bibfnamefont {N.}~\bibnamefont
  {{Smith}}}, \bibinfo {author} {\bibfnamefont {D.~L.}\ \bibnamefont
  {{Tucker}}}, \bibinfo {author} {\bibfnamefont {P.~K.~G.}\ \bibnamefont
  {{Williams}}}, \bibinfo {author} {\bibfnamefont {E.}~\bibnamefont
  {{Balbinot}}}, \bibinfo {author} {\bibfnamefont {J.~L.}\ \bibnamefont
  {{Carlin}}}, \bibinfo {author} {\bibfnamefont {E.~R.}\ \bibnamefont
  {{Cook}}}, \bibinfo {author} {\bibfnamefont {F.}~\bibnamefont {{Durret}}},
  \bibinfo {author} {\bibfnamefont {T.~S.}\ \bibnamefont {{Li}}}, \bibinfo
  {author} {\bibfnamefont {P.~A.~A.}\ \bibnamefont {{Lopes}}}, \bibinfo
  {author} {\bibfnamefont {A.~C.~C.}\ \bibnamefont {{Louren{\c{c}}o}}},
  \bibinfo {author} {\bibfnamefont {J.~L.}\ \bibnamefont {{Marshall}}},
  \bibinfo {author} {\bibfnamefont {G.~E.}\ \bibnamefont {{Medina}}}, \bibinfo
  {author} {\bibfnamefont {J.}~\bibnamefont {{Muir}}}, \bibinfo {author}
  {\bibfnamefont {R.~R.}\ \bibnamefont {{Mu{\~n}oz}}}, \bibinfo {author}
  {\bibfnamefont {M.}~\bibnamefont {{Sauseda}}}, \bibinfo {author}
  {\bibfnamefont {D.~J.}\ \bibnamefont {{Schlegel}}}, \bibinfo {author}
  {\bibfnamefont {L.~F.}\ \bibnamefont {{Secco}}}, \bibinfo {author}
  {\bibfnamefont {A.~K.}\ \bibnamefont {{Vivas}}}, \bibinfo {author}
  {\bibfnamefont {W.}~\bibnamefont {{Wester}}}, \bibinfo {author}
  {\bibfnamefont {A.}~\bibnamefont {{Zenteno}}}, \bibinfo {author}
  {\bibfnamefont {Y.}~\bibnamefont {{Zhang}}}, \bibinfo {author} {\bibfnamefont
  {T.~M.~C.}\ \bibnamefont {{Abbott}}}, \bibinfo {author} {\bibfnamefont
  {M.}~\bibnamefont {{Banerji}}}, \bibinfo {author} {\bibfnamefont
  {K.}~\bibnamefont {{Bechtol}}}, \bibinfo {author} {\bibfnamefont
  {A.}~\bibnamefont {{Benoit-L{\'e}vy}}}, \bibinfo {author} {\bibfnamefont
  {E.}~\bibnamefont {{Bertin}}}, \bibinfo {author} {\bibfnamefont
  {E.}~\bibnamefont {{Buckley-Geer}}}, \bibinfo {author} {\bibfnamefont
  {D.~L.}\ \bibnamefont {{Burke}}}, \bibinfo {author} {\bibfnamefont
  {D.}~\bibnamefont {{Capozzi}}}, \bibinfo {author} {\bibfnamefont
  {A.}~\bibnamefont {{Carnero Rosell}}}, \bibinfo {author} {\bibfnamefont
  {M.}~\bibnamefont {{Carrasco Kind}}}, \bibinfo {author} {\bibfnamefont
  {F.~J.}\ \bibnamefont {{Castander}}}, \bibinfo {author} {\bibfnamefont
  {M.}~\bibnamefont {{Crocce}}}, \bibinfo {author} {\bibfnamefont {C.~E.}\
  \bibnamefont {{Cunha}}}, \bibinfo {author} {\bibfnamefont {C.~B.}\
  \bibnamefont {{D'Andrea}}}, \bibinfo {author} {\bibfnamefont {L.~N.}\
  \bibnamefont {{da Costa}}}, \bibinfo {author} {\bibfnamefont
  {C.}~\bibnamefont {{Davis}}}, \bibinfo {author} {\bibfnamefont {D.~L.}\
  \bibnamefont {{DePoy}}}, \bibinfo {author} {\bibfnamefont {S.}~\bibnamefont
  {{Desai}}}, \bibinfo {author} {\bibfnamefont {J.~P.}\ \bibnamefont
  {{Dietrich}}}, \bibinfo {author} {\bibfnamefont {A.}~\bibnamefont
  {{Drlica-Wagner}}}, \bibinfo {author} {\bibfnamefont {T.~F.}\ \bibnamefont
  {{Eifler}}}, \bibinfo {author} {\bibfnamefont {A.~E.}\ \bibnamefont
  {{Evrard}}}, \bibinfo {author} {\bibfnamefont {E.}~\bibnamefont {{Fernand
  ez}}}, \bibinfo {author} {\bibfnamefont {B.}~\bibnamefont {{Flaugher}}},
  \bibinfo {author} {\bibfnamefont {P.}~\bibnamefont {{Fosalba}}}, \bibinfo
  {author} {\bibfnamefont {E.}~\bibnamefont {{Gaztanaga}}}, \bibinfo {author}
  {\bibfnamefont {D.~W.}\ \bibnamefont {{Gerdes}}}, \bibinfo {author}
  {\bibfnamefont {T.}~\bibnamefont {{Giannantonio}}}, \bibinfo {author}
  {\bibfnamefont {D.~A.}\ \bibnamefont {{Goldstein}}}, \bibinfo {author}
  {\bibfnamefont {D.}~\bibnamefont {{Gruen}}}, \bibinfo {author} {\bibfnamefont
  {R.~A.}\ \bibnamefont {{Gruendl}}}, \bibinfo {author} {\bibfnamefont
  {G.}~\bibnamefont {{Gutierrez}}}, \bibinfo {author} {\bibfnamefont
  {K.}~\bibnamefont {{Honscheid}}}, \bibinfo {author} {\bibfnamefont
  {B.}~\bibnamefont {{Jain}}}, \bibinfo {author} {\bibfnamefont {D.~J.}\
  \bibnamefont {{James}}}, \bibinfo {author} {\bibfnamefont {T.}~\bibnamefont
  {{Jeltema}}}, \bibinfo {author} {\bibfnamefont {M.~W.~G.}\ \bibnamefont
  {{Johnson}}}, \bibinfo {author} {\bibfnamefont {M.~D.}\ \bibnamefont
  {{Johnson}}}, \bibinfo {author} {\bibfnamefont {S.}~\bibnamefont {{Kent}}},
  \bibinfo {author} {\bibfnamefont {E.}~\bibnamefont {{Krause}}}, \bibinfo
  {author} {\bibfnamefont {R.}~\bibnamefont {{Kron}}}, \bibinfo {author}
  {\bibfnamefont {K.}~\bibnamefont {{Kuehn}}}, \bibinfo {author} {\bibfnamefont
  {N.}~\bibnamefont {{Nuropatkin}}}, \bibinfo {author} {\bibfnamefont
  {O.}~\bibnamefont {{Lahav}}}, \bibinfo {author} {\bibfnamefont
  {M.}~\bibnamefont {{Lima}}}, \bibinfo {author} {\bibfnamefont
  {H.}~\bibnamefont {{Lin}}}, \bibinfo {author} {\bibfnamefont {M.~A.~G.}\
  \bibnamefont {{Maia}}}, \bibinfo {author} {\bibfnamefont {M.}~\bibnamefont
  {{March}}}, \bibinfo {author} {\bibfnamefont {P.}~\bibnamefont {{Martini}}},
  \bibinfo {author} {\bibfnamefont {R.~G.}\ \bibnamefont {{McMahon}}}, \bibinfo
  {author} {\bibfnamefont {F.}~\bibnamefont {{Menanteau}}}, \bibinfo {author}
  {\bibfnamefont {C.~J.}\ \bibnamefont {{Miller}}}, \bibinfo {author}
  {\bibfnamefont {R.}~\bibnamefont {{Miquel}}}, \bibinfo {author}
  {\bibfnamefont {J.~J.}\ \bibnamefont {{Mohr}}}, \bibinfo {author}
  {\bibfnamefont {E.}~\bibnamefont {{Neilsen}}}, \bibinfo {author}
  {\bibfnamefont {R.~C.}\ \bibnamefont {{Nichol}}}, \bibinfo {author}
  {\bibfnamefont {R.~L.~C.}\ \bibnamefont {{Ogando}}}, \bibinfo {author}
  {\bibfnamefont {A.~A.}\ \bibnamefont {{Plazas}}}, \bibinfo {author}
  {\bibfnamefont {N.}~\bibnamefont {{Roe}}}, \bibinfo {author} {\bibfnamefont
  {A.~K.}\ \bibnamefont {{Romer}}}, \bibinfo {author} {\bibfnamefont
  {A.}~\bibnamefont {{Roodman}}}, \bibinfo {author} {\bibfnamefont {E.~S.}\
  \bibnamefont {{Rykoff}}}, \bibinfo {author} {\bibfnamefont {E.}~\bibnamefont
  {{Sanchez}}}, \bibinfo {author} {\bibfnamefont {V.}~\bibnamefont
  {{Scarpine}}}, \bibinfo {author} {\bibfnamefont {R.}~\bibnamefont
  {{Schindler}}}, \bibinfo {author} {\bibfnamefont {M.}~\bibnamefont
  {{Schubnell}}}, \bibinfo {author} {\bibfnamefont {I.}~\bibnamefont
  {{Sevilla-Noarbe}}}, \bibinfo {author} {\bibfnamefont {M.}~\bibnamefont
  {{Smith}}}, \bibinfo {author} {\bibfnamefont {R.~C.}\ \bibnamefont
  {{Smith}}}, \bibinfo {author} {\bibfnamefont {F.}~\bibnamefont {{Sobreira}}},
  \bibinfo {author} {\bibfnamefont {E.}~\bibnamefont {{Suchyta}}}, \bibinfo
  {author} {\bibfnamefont {M.~E.~C.}\ \bibnamefont {{Swanson}}}, \bibinfo
  {author} {\bibfnamefont {G.}~\bibnamefont {{Tarle}}}, \bibinfo {author}
  {\bibfnamefont {D.}~\bibnamefont {{Thomas}}}, \bibinfo {author}
  {\bibfnamefont {R.~C.}\ \bibnamefont {{Thomas}}}, \bibinfo {author}
  {\bibfnamefont {M.~A.}\ \bibnamefont {{Troxel}}}, \bibinfo {author}
  {\bibfnamefont {V.}~\bibnamefont {{Vikram}}}, \bibinfo {author}
  {\bibfnamefont {A.~R.}\ \bibnamefont {{Walker}}}, \bibinfo {author}
  {\bibfnamefont {R.~H.}\ \bibnamefont {{Wechsler}}}, \bibinfo {author}
  {\bibfnamefont {J.}~\bibnamefont {{Weller}}}, \bibinfo {author}
  {\bibfnamefont {B.}~\bibnamefont {{Yanny}}}, \ and\ \bibinfo {author}
  {\bibfnamefont {J.}~\bibnamefont {{Zuntz}}},\ }\href {\doibase
  10.3847/2041-8213/aa8fc7} {\bibfield  {journal} {\bibinfo  {journal} {\apjl}\
  }\textbf {\bibinfo {volume} {848}},\ \bibinfo {eid} {L17} (\bibinfo {year}
  {2017})},\ \Eprint {http://arxiv.org/abs/1710.05840} {arXiv:1710.05840
  [astro-ph.HE]} \BibitemShut {NoStop}%
\bibitem [{\citenamefont {{Villar}}\ \emph {et~al.}(2017)\citenamefont
  {{Villar}}, \citenamefont {{Guillochon}}, \citenamefont {{Berger}},
  \citenamefont {{Metzger}}, \citenamefont {{Cowperthwaite}}, \citenamefont
  {{Nicholl}}, \citenamefont {{Alexander}}, \citenamefont {{Blanchard}},
  \citenamefont {{Chornock}}, \citenamefont {{Eftekhari}}, \citenamefont
  {{Fong}}, \citenamefont {{Margutti}},\ and\ \citenamefont
  {{Williams}}}]{Villar+17}%
  \BibitemOpen
  \bibfield  {author} {\bibinfo {author} {\bibfnamefont {V.~A.}\ \bibnamefont
  {{Villar}}}, \bibinfo {author} {\bibfnamefont {J.}~\bibnamefont
  {{Guillochon}}}, \bibinfo {author} {\bibfnamefont {E.}~\bibnamefont
  {{Berger}}}, \bibinfo {author} {\bibfnamefont {B.~D.}\ \bibnamefont
  {{Metzger}}}, \bibinfo {author} {\bibfnamefont {P.~S.}\ \bibnamefont
  {{Cowperthwaite}}}, \bibinfo {author} {\bibfnamefont {M.}~\bibnamefont
  {{Nicholl}}}, \bibinfo {author} {\bibfnamefont {K.~D.}\ \bibnamefont
  {{Alexander}}}, \bibinfo {author} {\bibfnamefont {P.~K.}\ \bibnamefont
  {{Blanchard}}}, \bibinfo {author} {\bibfnamefont {R.}~\bibnamefont
  {{Chornock}}}, \bibinfo {author} {\bibfnamefont {T.}~\bibnamefont
  {{Eftekhari}}}, \bibinfo {author} {\bibfnamefont {W.}~\bibnamefont {{Fong}}},
  \bibinfo {author} {\bibfnamefont {R.}~\bibnamefont {{Margutti}}}, \ and\
  \bibinfo {author} {\bibfnamefont {P.~K.~G.}\ \bibnamefont {{Williams}}},\
  }\href {\doibase 10.3847/2041-8213/aa9c84} {\bibfield  {journal} {\bibinfo
  {journal} {\apjl}\ }\textbf {\bibinfo {volume} {851}},\ \bibinfo {eid} {L21}
  (\bibinfo {year} {2017})},\ \Eprint {http://arxiv.org/abs/1710.11576}
  {arXiv:1710.11576 [astro-ph.HE]} \BibitemShut {NoStop}%
\bibitem [{\citenamefont {{Kawaguchi}}\ \emph {et~al.}(2018)\citenamefont
  {{Kawaguchi}}, \citenamefont {{Shibata}},\ and\ \citenamefont
  {{Tanaka}}}]{Kawaguchi+18}%
  \BibitemOpen
  \bibfield  {author} {\bibinfo {author} {\bibfnamefont {K.}~\bibnamefont
  {{Kawaguchi}}}, \bibinfo {author} {\bibfnamefont {M.}~\bibnamefont
  {{Shibata}}}, \ and\ \bibinfo {author} {\bibfnamefont {M.}~\bibnamefont
  {{Tanaka}}},\ }\href {\doibase 10.3847/2041-8213/aade02} {\bibfield
  {journal} {\bibinfo  {journal} {\apjl}\ }\textbf {\bibinfo {volume} {865}},\
  \bibinfo {eid} {L21} (\bibinfo {year} {2018})},\ \Eprint
  {http://arxiv.org/abs/1806.04088} {arXiv:1806.04088 [astro-ph.HE]}
  \BibitemShut {NoStop}%
\bibitem [{\citenamefont {Watson}\ \emph {et~al.}(2019)\citenamefont {Watson}
  \emph {et~al.}}]{Watson:2019xjv}%
  \BibitemOpen
  \bibfield  {author} {\bibinfo {author} {\bibfnamefont {D.}~\bibnamefont
  {Watson}} \emph {et~al.},\ }\href {\doibase 10.1038/s41586-019-1676-3}
  {\bibfield  {journal} {\bibinfo  {journal} {Nature}\ }\textbf {\bibinfo
  {volume} {574}},\ \bibinfo {pages} {497} (\bibinfo {year} {2019})},\ \Eprint
  {http://arxiv.org/abs/1910.10510} {arXiv:1910.10510 [astro-ph.HE]}
  \BibitemShut {NoStop}%
\bibitem [{\citenamefont {Domoto}\ \emph {et~al.}(2021)\citenamefont {Domoto},
  \citenamefont {Tanaka}, \citenamefont {Wanajo},\ and\ \citenamefont
  {Kawaguchi}}]{Domoto:2021xfq}%
  \BibitemOpen
  \bibfield  {author} {\bibinfo {author} {\bibfnamefont {N.}~\bibnamefont
  {Domoto}}, \bibinfo {author} {\bibfnamefont {M.}~\bibnamefont {Tanaka}},
  \bibinfo {author} {\bibfnamefont {S.}~\bibnamefont {Wanajo}}, \ and\ \bibinfo
  {author} {\bibfnamefont {K.}~\bibnamefont {Kawaguchi}},\ }\href {\doibase
  10.3847/1538-4357/abf358} {\bibfield  {journal} {\bibinfo  {journal}
  {Astrophys. J.}\ }\textbf {\bibinfo {volume} {913}},\ \bibinfo {pages} {26}
  (\bibinfo {year} {2021})},\ \Eprint {http://arxiv.org/abs/2103.15284}
  {arXiv:2103.15284 [astro-ph.HE]} \BibitemShut {NoStop}%
\bibitem [{\citenamefont {Gillanders}\ \emph {et~al.}(2021)\citenamefont
  {Gillanders}, \citenamefont {McCann}, \citenamefont {Sim}, \citenamefont
  {Smartt},\ and\ \citenamefont {Ballance}}]{Gillanders:2021qvx}%
  \BibitemOpen
  \bibfield  {author} {\bibinfo {author} {\bibfnamefont {J.~H.}\ \bibnamefont
  {Gillanders}}, \bibinfo {author} {\bibfnamefont {M.}~\bibnamefont {McCann}},
  \bibinfo {author} {\bibfnamefont {S.~A.}\ \bibnamefont {Sim}}, \bibinfo
  {author} {\bibfnamefont {S.~A. S. S.~J.}\ \bibnamefont {Smartt}}, \ and\
  \bibinfo {author} {\bibfnamefont {C.~P.}\ \bibnamefont {Ballance}},\ }\href
  {\doibase 10.1093/mnras/stab1861} {\bibfield  {journal} {\bibinfo  {journal}
  {Mon. Not. Roy. Astron. Soc.}\ }\textbf {\bibinfo {volume} {506}},\ \bibinfo
  {pages} {3560} (\bibinfo {year} {2021})},\ \Eprint
  {http://arxiv.org/abs/2101.08271} {arXiv:2101.08271 [astro-ph.HE]}
  \BibitemShut {NoStop}%
\bibitem [{\citenamefont {Perego}\ \emph {et~al.}(2022)\citenamefont {Perego}
  \emph {et~al.}}]{Perego:2020evn}%
  \BibitemOpen
  \bibfield  {author} {\bibinfo {author} {\bibfnamefont {A.}~\bibnamefont
  {Perego}} \emph {et~al.},\ }\href {\doibase 10.3847/1538-4357/ac3751}
  {\bibfield  {journal} {\bibinfo  {journal} {Astrophys. J.}\ }\textbf
  {\bibinfo {volume} {925}},\ \bibinfo {pages} {22} (\bibinfo {year} {2022})},\
  \Eprint {http://arxiv.org/abs/2009.08988} {arXiv:2009.08988 [astro-ph.HE]}
  \BibitemShut {NoStop}%
\bibitem [{\citenamefont {Kasliwal}\ \emph {et~al.}(2022)\citenamefont
  {Kasliwal} \emph {et~al.}}]{Kasliwal:2018fwk}%
  \BibitemOpen
  \bibfield  {author} {\bibinfo {author} {\bibfnamefont {M.~M.}\ \bibnamefont
  {Kasliwal}} \emph {et~al.},\ }\href {\doibase 10.1093/mnrasl/slz007}
  {\bibfield  {journal} {\bibinfo  {journal} {Mon. Not. Roy. Astron. Soc.}\
  }\textbf {\bibinfo {volume} {510}},\ \bibinfo {pages} {L7} (\bibinfo {year}
  {2022})},\ \Eprint {http://arxiv.org/abs/1812.08708} {arXiv:1812.08708
  [astro-ph.HE]} \BibitemShut {NoStop}%
\bibitem [{\citenamefont {{Sneden}}\ \emph {et~al.}(2008)\citenamefont
  {{Sneden}}, \citenamefont {{Cowan}},\ and\ \citenamefont
  {{Gallino}}}]{Sneden2008}%
  \BibitemOpen
  \bibfield  {author} {\bibinfo {author} {\bibfnamefont {C.}~\bibnamefont
  {{Sneden}}}, \bibinfo {author} {\bibfnamefont {J.~J.}\ \bibnamefont
  {{Cowan}}}, \ and\ \bibinfo {author} {\bibfnamefont {R.}~\bibnamefont
  {{Gallino}}},\ }\href {\doibase 10.1146/annurev.astro.46.060407.145207}
  {\bibfield  {journal} {\bibinfo  {journal} {\araa}\ }\textbf {\bibinfo
  {volume} {46}},\ \bibinfo {pages} {241} (\bibinfo {year} {2008})}\BibitemShut
  {NoStop}%
\bibitem [{\citenamefont {{Cowan}}\ \emph {et~al.}(2021)\citenamefont
  {{Cowan}}, \citenamefont {{Sneden}}, \citenamefont {{Lawler}}, \citenamefont
  {{Aprahamian}}, \citenamefont {{Wiescher}}, \citenamefont {{Langanke}},
  \citenamefont {{Mart{\'\i}nez-Pinedo}},\ and\ \citenamefont
  {{Thielemann}}}]{Cowan2019}%
  \BibitemOpen
  \bibfield  {author} {\bibinfo {author} {\bibfnamefont {J.~J.}\ \bibnamefont
  {{Cowan}}}, \bibinfo {author} {\bibfnamefont {C.}~\bibnamefont {{Sneden}}},
  \bibinfo {author} {\bibfnamefont {J.~E.}\ \bibnamefont {{Lawler}}}, \bibinfo
  {author} {\bibfnamefont {A.}~\bibnamefont {{Aprahamian}}}, \bibinfo {author}
  {\bibfnamefont {M.}~\bibnamefont {{Wiescher}}}, \bibinfo {author}
  {\bibfnamefont {K.}~\bibnamefont {{Langanke}}}, \bibinfo {author}
  {\bibfnamefont {G.}~\bibnamefont {{Mart{\'\i}nez-Pinedo}}}, \ and\ \bibinfo
  {author} {\bibfnamefont {F.-K.}\ \bibnamefont {{Thielemann}}},\ }\href
  {\doibase 10.1103/RevModPhys.93.015002} {\bibfield  {journal} {\bibinfo
  {journal} {Reviews of Modern Physics}\ }\textbf {\bibinfo {volume} {93}},\
  \bibinfo {eid} {015002} (\bibinfo {year} {2021})},\ \Eprint
  {http://arxiv.org/abs/1901.01410} {arXiv:1901.01410 [astro-ph.HE]}
  \BibitemShut {NoStop}%
\bibitem [{\citenamefont {Fowler}\ and\ \citenamefont
  {Hoyle}(1960)}]{Fowler:1960}%
  \BibitemOpen
  \bibfield  {author} {\bibinfo {author} {\bibfnamefont {W.~A.}\ \bibnamefont
  {Fowler}}\ and\ \bibinfo {author} {\bibfnamefont {F.}~\bibnamefont {Hoyle}},\
  }\href {\doibase https://doi.org/10.1016/0003-4916(60)90025-7} {\bibfield
  {journal} {\bibinfo  {journal} {Annals of Physics}\ }\textbf {\bibinfo
  {volume} {10}},\ \bibinfo {pages} {280} (\bibinfo {year} {1960})}\BibitemShut
  {NoStop}%
\bibitem [{\citenamefont {{Schramm}}\ and\ \citenamefont
  {{Wasserburg}}(1970)}]{Schramm:1970}%
  \BibitemOpen
  \bibfield  {author} {\bibinfo {author} {\bibfnamefont {D.~N.}\ \bibnamefont
  {{Schramm}}}\ and\ \bibinfo {author} {\bibfnamefont {G.~J.}\ \bibnamefont
  {{Wasserburg}}},\ }\href {\doibase 10.1086/150634} {\bibfield  {journal}
  {\bibinfo  {journal} {\apj}\ }\textbf {\bibinfo {volume} {162}},\ \bibinfo
  {pages} {57} (\bibinfo {year} {1970})}\BibitemShut {NoStop}%
\bibitem [{\citenamefont {{Cowan}}\ \emph {et~al.}(1991)\citenamefont
  {{Cowan}}, \citenamefont {{Thielemann}},\ and\ \citenamefont
  {{Truran}}}]{Cowan:1991}%
  \BibitemOpen
  \bibfield  {author} {\bibinfo {author} {\bibfnamefont {J.~J.}\ \bibnamefont
  {{Cowan}}}, \bibinfo {author} {\bibfnamefont {F.-K.}\ \bibnamefont
  {{Thielemann}}}, \ and\ \bibinfo {author} {\bibfnamefont {J.~W.}\
  \bibnamefont {{Truran}}},\ }\href {\doibase 10.1016/0370-1573(91)90070-3}
  {\bibfield  {journal} {\bibinfo  {journal} {\physrep}\ }\textbf {\bibinfo
  {volume} {208}},\ \bibinfo {pages} {267} (\bibinfo {year}
  {1991})}\BibitemShut {NoStop}%
\bibitem [{\citenamefont {Meyer}\ and\ \citenamefont
  {Truran}(2000)}]{Meyer:2000}%
  \BibitemOpen
  \bibfield  {author} {\bibinfo {author} {\bibfnamefont {B.~S.}\ \bibnamefont
  {Meyer}}\ and\ \bibinfo {author} {\bibfnamefont {J.~W.}\ \bibnamefont
  {Truran}},\ }\href {\doibase https://doi.org/10.1016/S0370-1573(00)00012-0}
  {\bibfield  {journal} {\bibinfo  {journal} {Physics Reports}\ }\textbf
  {\bibinfo {volume} {333-334}},\ \bibinfo {pages} {1} (\bibinfo {year}
  {2000})}\BibitemShut {NoStop}%
\bibitem [{\citenamefont {Panov}\ \emph {et~al.}(2017)\citenamefont {Panov},
  \citenamefont {Lutostansky}, \citenamefont {Eichler},\ and\ \citenamefont
  {Thielemann}}]{Panov:2017lbx}%
  \BibitemOpen
  \bibfield  {author} {\bibinfo {author} {\bibfnamefont {I.~V.}\ \bibnamefont
  {Panov}}, \bibinfo {author} {\bibfnamefont {Y.~S.}\ \bibnamefont
  {Lutostansky}}, \bibinfo {author} {\bibfnamefont {M.}~\bibnamefont
  {Eichler}}, \ and\ \bibinfo {author} {\bibfnamefont {F.~K.}\ \bibnamefont
  {Thielemann}},\ }\href {\doibase 10.1134/S1063778817040202} {\bibfield
  {journal} {\bibinfo  {journal} {Phys. Atom. Nucl.}\ }\textbf {\bibinfo
  {volume} {80}},\ \bibinfo {pages} {657} (\bibinfo {year} {2017})}\BibitemShut
  {NoStop}%
\bibitem [{\citenamefont {Wu}\ \emph {et~al.}(2021)\citenamefont {Wu},
  \citenamefont {Zhao}, \citenamefont {Zhang},\ and\ \citenamefont
  {Meng}}]{Wu:2021jzv}%
  \BibitemOpen
  \bibfield  {author} {\bibinfo {author} {\bibfnamefont {X.~H.}\ \bibnamefont
  {Wu}}, \bibinfo {author} {\bibfnamefont {P.~W.}\ \bibnamefont {Zhao}},
  \bibinfo {author} {\bibfnamefont {S.~Q.}\ \bibnamefont {Zhang}}, \ and\
  \bibinfo {author} {\bibfnamefont {J.}~\bibnamefont {Meng}},\ }\href@noop {}
  {\  (\bibinfo {year} {2021})},\ \Eprint {http://arxiv.org/abs/2108.06104}
  {arXiv:2108.06104 [nucl-th]} \BibitemShut {NoStop}%
\bibitem [{\citenamefont {Wallner}\ \emph {et~al.}(2015)\citenamefont
  {Wallner}, \citenamefont {Faestermann}, \citenamefont {Feige}, \citenamefont
  {Feldstein}, \citenamefont {Knie}, \citenamefont {Korschinek}, \citenamefont
  {Kutschera}, \citenamefont {Ofan}, \citenamefont {Paul}, \citenamefont
  {Quinto},\ and\ \citenamefont {et~al.}}]{Wallner_2015}%
  \BibitemOpen
  \bibfield  {author} {\bibinfo {author} {\bibfnamefont {A.}~\bibnamefont
  {Wallner}}, \bibinfo {author} {\bibfnamefont {T.}~\bibnamefont
  {Faestermann}}, \bibinfo {author} {\bibfnamefont {J.}~\bibnamefont {Feige}},
  \bibinfo {author} {\bibfnamefont {C.}~\bibnamefont {Feldstein}}, \bibinfo
  {author} {\bibfnamefont {K.}~\bibnamefont {Knie}}, \bibinfo {author}
  {\bibfnamefont {G.}~\bibnamefont {Korschinek}}, \bibinfo {author}
  {\bibfnamefont {W.}~\bibnamefont {Kutschera}}, \bibinfo {author}
  {\bibfnamefont {A.}~\bibnamefont {Ofan}}, \bibinfo {author} {\bibfnamefont
  {M.}~\bibnamefont {Paul}}, \bibinfo {author} {\bibfnamefont {F.}~\bibnamefont
  {Quinto}}, \ and\ \bibinfo {author} {\bibnamefont {et~al.}},\ }\href
  {\doibase 10.1038/ncomms6956} {\bibfield  {journal} {\bibinfo  {journal}
  {Nature Communications}\ }\textbf {\bibinfo {volume} {6}} (\bibinfo {year}
  {2015}),\ 10.1038/ncomms6956}\BibitemShut {NoStop}%
\bibitem [{\citenamefont {{Hotokezaka}}\ \emph {et~al.}(2015)\citenamefont
  {{Hotokezaka}}, \citenamefont {{Piran}},\ and\ \citenamefont
  {{Paul}}}]{hotokezaka2015}%
  \BibitemOpen
  \bibfield  {author} {\bibinfo {author} {\bibfnamefont {K.}~\bibnamefont
  {{Hotokezaka}}}, \bibinfo {author} {\bibfnamefont {T.}~\bibnamefont
  {{Piran}}}, \ and\ \bibinfo {author} {\bibfnamefont {M.}~\bibnamefont
  {{Paul}}},\ }\href {\doibase 10.1038/nphys3574} {\bibfield  {journal}
  {\bibinfo  {journal} {Nature Physics}\ }\textbf {\bibinfo {volume} {11}},\
  \bibinfo {pages} {1042} (\bibinfo {year} {2015})},\ \Eprint
  {http://arxiv.org/abs/1510.00711} {arXiv:1510.00711 [astro-ph.HE]}
  \BibitemShut {NoStop}%
\bibitem [{\citenamefont {Bartos}\ and\ \citenamefont
  {Marka}(2019)}]{Bartos:2019cec}%
  \BibitemOpen
  \bibfield  {author} {\bibinfo {author} {\bibfnamefont {I.}~\bibnamefont
  {Bartos}}\ and\ \bibinfo {author} {\bibfnamefont {S.}~\bibnamefont {Marka}},\
  }\href {\doibase 10.1038/s41586-019-1113-7} {\bibfield  {journal} {\bibinfo
  {journal} {Nature}\ }\textbf {\bibinfo {volume} {569}},\ \bibinfo {pages}
  {85} (\bibinfo {year} {2019})}\BibitemShut {NoStop}%
\bibitem [{\citenamefont {{C{\^o}t{\'e}}}\ \emph {et~al.}(2021)\citenamefont
  {{C{\^o}t{\'e}}}, \citenamefont {{Eichler}}, \citenamefont {{Yag{\"u}e
  L{\'o}pez}}, \citenamefont {{Vassh}}, \citenamefont {{Mumpower}},
  \citenamefont {{Vil{\'a}gos}}, \citenamefont {{So{\'o}s}}, \citenamefont
  {{Arcones}}, \citenamefont {{Sprouse}}, \citenamefont {{Surman}},
  \citenamefont {{Pignatari}}, \citenamefont {{Pet{\H{o}}}}, \citenamefont
  {{Wehmeyer}}, \citenamefont {{Rauscher}},\ and\ \citenamefont
  {{Lugaro}}}]{Cote2021}%
  \BibitemOpen
  \bibfield  {author} {\bibinfo {author} {\bibfnamefont {B.}~\bibnamefont
  {{C{\^o}t{\'e}}}}, \bibinfo {author} {\bibfnamefont {M.}~\bibnamefont
  {{Eichler}}}, \bibinfo {author} {\bibfnamefont {A.}~\bibnamefont {{Yag{\"u}e
  L{\'o}pez}}}, \bibinfo {author} {\bibfnamefont {N.}~\bibnamefont {{Vassh}}},
  \bibinfo {author} {\bibfnamefont {M.~R.}\ \bibnamefont {{Mumpower}}},
  \bibinfo {author} {\bibfnamefont {B.}~\bibnamefont {{Vil{\'a}gos}}}, \bibinfo
  {author} {\bibfnamefont {B.}~\bibnamefont {{So{\'o}s}}}, \bibinfo {author}
  {\bibfnamefont {A.}~\bibnamefont {{Arcones}}}, \bibinfo {author}
  {\bibfnamefont {T.~M.}\ \bibnamefont {{Sprouse}}}, \bibinfo {author}
  {\bibfnamefont {R.}~\bibnamefont {{Surman}}}, \bibinfo {author}
  {\bibfnamefont {M.}~\bibnamefont {{Pignatari}}}, \bibinfo {author}
  {\bibfnamefont {M.~K.}\ \bibnamefont {{Pet{\H{o}}}}}, \bibinfo {author}
  {\bibfnamefont {B.}~\bibnamefont {{Wehmeyer}}}, \bibinfo {author}
  {\bibfnamefont {T.}~\bibnamefont {{Rauscher}}}, \ and\ \bibinfo {author}
  {\bibfnamefont {M.}~\bibnamefont {{Lugaro}}},\ }\href {\doibase
  10.1126/science.aba1111} {\bibfield  {journal} {\bibinfo  {journal}
  {Science}\ }\textbf {\bibinfo {volume} {371}},\ \bibinfo {pages} {945}
  (\bibinfo {year} {2021})},\ \Eprint {http://arxiv.org/abs/2006.04833}
  {arXiv:2006.04833 [astro-ph.SR]} \BibitemShut {NoStop}%
\bibitem [{\citenamefont {{Wallner}}\ \emph {et~al.}(2021)\citenamefont
  {{Wallner}}, \citenamefont {{Froehlich}}, \citenamefont {{Hotchkis}},
  \citenamefont {{Kinoshita}}, \citenamefont {{Paul}}, \citenamefont
  {{Martschini}}, \citenamefont {{Pavetich}}, \citenamefont {{Tims}},
  \citenamefont {{Kivel}}, \citenamefont {{Schumann}}, \citenamefont {{Honda}},
  \citenamefont {{Matsuzaki}},\ and\ \citenamefont
  {{Yamagata}}}]{Wallner:2021}%
  \BibitemOpen
  \bibfield  {author} {\bibinfo {author} {\bibfnamefont {A.}~\bibnamefont
  {{Wallner}}}, \bibinfo {author} {\bibfnamefont {M.~B.}\ \bibnamefont
  {{Froehlich}}}, \bibinfo {author} {\bibfnamefont {M.~A.~C.}\ \bibnamefont
  {{Hotchkis}}}, \bibinfo {author} {\bibfnamefont {N.}~\bibnamefont
  {{Kinoshita}}}, \bibinfo {author} {\bibfnamefont {M.}~\bibnamefont {{Paul}}},
  \bibinfo {author} {\bibfnamefont {M.}~\bibnamefont {{Martschini}}}, \bibinfo
  {author} {\bibfnamefont {S.}~\bibnamefont {{Pavetich}}}, \bibinfo {author}
  {\bibfnamefont {S.~G.}\ \bibnamefont {{Tims}}}, \bibinfo {author}
  {\bibfnamefont {N.}~\bibnamefont {{Kivel}}}, \bibinfo {author} {\bibfnamefont
  {D.}~\bibnamefont {{Schumann}}}, \bibinfo {author} {\bibfnamefont
  {M.}~\bibnamefont {{Honda}}}, \bibinfo {author} {\bibfnamefont
  {H.}~\bibnamefont {{Matsuzaki}}}, \ and\ \bibinfo {author} {\bibfnamefont
  {T.}~\bibnamefont {{Yamagata}}},\ }\href {\doibase 10.1126/science.aax3972}
  {\bibfield  {journal} {\bibinfo  {journal} {Science}\ }\textbf {\bibinfo
  {volume} {372}},\ \bibinfo {pages} {742} (\bibinfo {year}
  {2021})}\BibitemShut {NoStop}%
\bibitem [{\citenamefont {{Banerjee}}\ \emph {et~al.}(2022)\citenamefont
  {{Banerjee}}, \citenamefont {{Wu}},\ and\ \citenamefont {{S K}}}]{bwj2022}%
  \BibitemOpen
  \bibfield  {author} {\bibinfo {author} {\bibfnamefont {P.}~\bibnamefont
  {{Banerjee}}}, \bibinfo {author} {\bibfnamefont {M.-R.}\ \bibnamefont
  {{Wu}}}, \ and\ \bibinfo {author} {\bibfnamefont {J.}~\bibnamefont {{S K}}},\
  }\href {\doibase 10.1093/mnras/stac318} {\bibfield  {journal} {\bibinfo
  {journal} {\mnras}\ }\textbf {\bibinfo {volume} {512}},\ \bibinfo {pages}
  {4948} (\bibinfo {year} {2022})},\ \Eprint {http://arxiv.org/abs/2110.05449}
  {arXiv:2110.05449 [astro-ph.GA]} \BibitemShut {NoStop}%
\bibitem [{\citenamefont {{Goriely}}\ \emph {et~al.}(2011)\citenamefont
  {{Goriely}}, \citenamefont {{Bauswein}},\ and\ \citenamefont
  {{Janka}}}]{Goriely2011}%
  \BibitemOpen
  \bibfield  {author} {\bibinfo {author} {\bibfnamefont {S.}~\bibnamefont
  {{Goriely}}}, \bibinfo {author} {\bibfnamefont {A.}~\bibnamefont
  {{Bauswein}}}, \ and\ \bibinfo {author} {\bibfnamefont {H.-T.}\ \bibnamefont
  {{Janka}}},\ }\href {\doibase 10.1088/2041-8205/738/2/L32} {\bibfield
  {journal} {\bibinfo  {journal} {\apjl}\ }\textbf {\bibinfo {volume} {738}},\
  \bibinfo {eid} {L32} (\bibinfo {year} {2011})},\ \Eprint
  {http://arxiv.org/abs/1107.0899} {arXiv:1107.0899 [astro-ph.SR]} \BibitemShut
  {NoStop}%
\bibitem [{\citenamefont {{Korobkin}}\ \emph {et~al.}(2012)\citenamefont
  {{Korobkin}}, \citenamefont {{Rosswog}}, \citenamefont {{Arcones}},\ and\
  \citenamefont {{Winteler}}}]{Korobkin2012}%
  \BibitemOpen
  \bibfield  {author} {\bibinfo {author} {\bibfnamefont {O.}~\bibnamefont
  {{Korobkin}}}, \bibinfo {author} {\bibfnamefont {S.}~\bibnamefont
  {{Rosswog}}}, \bibinfo {author} {\bibfnamefont {A.}~\bibnamefont
  {{Arcones}}}, \ and\ \bibinfo {author} {\bibfnamefont {C.}~\bibnamefont
  {{Winteler}}},\ }\href {\doibase 10.1111/j.1365-2966.2012.21859.x} {\bibfield
   {journal} {\bibinfo  {journal} {\mnras}\ }\textbf {\bibinfo {volume}
  {426}},\ \bibinfo {pages} {1940} (\bibinfo {year} {2012})},\ \Eprint
  {http://arxiv.org/abs/1206.2379} {arXiv:1206.2379 [astro-ph.SR]} \BibitemShut
  {NoStop}%
\bibitem [{\citenamefont {Wanajo}\ \emph {et~al.}(2014)\citenamefont {Wanajo},
  \citenamefont {Sekiguchi}, \citenamefont {Nishimura}, \citenamefont {Kiuchi},
  \citenamefont {Kyutoku},\ and\ \citenamefont {Shibata}}]{Wanajo:2014wha}%
  \BibitemOpen
  \bibfield  {author} {\bibinfo {author} {\bibfnamefont {S.}~\bibnamefont
  {Wanajo}}, \bibinfo {author} {\bibfnamefont {Y.}~\bibnamefont {Sekiguchi}},
  \bibinfo {author} {\bibfnamefont {N.}~\bibnamefont {Nishimura}}, \bibinfo
  {author} {\bibfnamefont {K.}~\bibnamefont {Kiuchi}}, \bibinfo {author}
  {\bibfnamefont {K.}~\bibnamefont {Kyutoku}}, \ and\ \bibinfo {author}
  {\bibfnamefont {M.}~\bibnamefont {Shibata}},\ }\href {\doibase
  10.1088/2041-8205/789/2/L39} {\bibfield  {journal} {\bibinfo  {journal}
  {\apjl}\ }\textbf {\bibinfo {volume} {789}},\ \bibinfo {pages} {L39}
  (\bibinfo {year} {2014})},\ \Eprint {http://arxiv.org/abs/1402.7317}
  {arXiv:1402.7317 [astro-ph.SR]} \BibitemShut {NoStop}%
\bibitem [{\citenamefont {Just}\ \emph {et~al.}(2015)\citenamefont {Just},
  \citenamefont {Bauswein}, \citenamefont {Pulpillo}, \citenamefont {Goriely},\
  and\ \citenamefont {Janka}}]{Just:2014fka}%
  \BibitemOpen
  \bibfield  {author} {\bibinfo {author} {\bibfnamefont {O.}~\bibnamefont
  {Just}}, \bibinfo {author} {\bibfnamefont {A.}~\bibnamefont {Bauswein}},
  \bibinfo {author} {\bibfnamefont {R.~A.}\ \bibnamefont {Pulpillo}}, \bibinfo
  {author} {\bibfnamefont {S.}~\bibnamefont {Goriely}}, \ and\ \bibinfo
  {author} {\bibfnamefont {H.~T.}\ \bibnamefont {Janka}},\ }\href {\doibase
  10.1093/mnras/stv009} {\bibfield  {journal} {\bibinfo  {journal} {\mnras}\
  }\textbf {\bibinfo {volume} {448}},\ \bibinfo {pages} {541} (\bibinfo {year}
  {2015})},\ \Eprint {http://arxiv.org/abs/1406.2687} {arXiv:1406.2687
  [astro-ph.SR]} \BibitemShut {NoStop}%
\bibitem [{\citenamefont {Lippuner}\ and\ \citenamefont
  {Roberts}(2015)}]{Lippuner:2015gwa}%
  \BibitemOpen
  \bibfield  {author} {\bibinfo {author} {\bibfnamefont {J.}~\bibnamefont
  {Lippuner}}\ and\ \bibinfo {author} {\bibfnamefont {L.~F.}\ \bibnamefont
  {Roberts}},\ }\href {\doibase 10.1088/0004-637X/815/2/82} {\bibfield
  {journal} {\bibinfo  {journal} {Astrophys. J.}\ }\textbf {\bibinfo {volume}
  {815}},\ \bibinfo {pages} {82} (\bibinfo {year} {2015})},\ \Eprint
  {http://arxiv.org/abs/1508.03133} {arXiv:1508.03133 [astro-ph.HE]}
  \BibitemShut {NoStop}%
\bibitem [{\citenamefont {Wu}\ \emph {et~al.}(2016)\citenamefont {Wu},
  \citenamefont {Fernández}, \citenamefont {Martínez-Pinedo},\ and\
  \citenamefont {Metzger}}]{Wu:2016pnw}%
  \BibitemOpen
  \bibfield  {author} {\bibinfo {author} {\bibfnamefont {M.-R.}\ \bibnamefont
  {Wu}}, \bibinfo {author} {\bibfnamefont {R.}~\bibnamefont {Fernández}},
  \bibinfo {author} {\bibfnamefont {G.}~\bibnamefont {Martínez-Pinedo}}, \
  and\ \bibinfo {author} {\bibfnamefont {B.~D.}\ \bibnamefont {Metzger}},\
  }\href {\doibase 10.1093/mnras/stw2156} {\bibfield  {journal} {\bibinfo
  {journal} {\mnras}\ }\textbf {\bibinfo {volume} {463}},\ \bibinfo {pages}
  {2323} (\bibinfo {year} {2016})},\ \Eprint {http://arxiv.org/abs/1607.05290}
  {arXiv:1607.05290 [astro-ph.HE]} \BibitemShut {NoStop}%
\bibitem [{\citenamefont {Siegel}\ and\ \citenamefont
  {Metzger}(2017)}]{Siegel:2017nub}%
  \BibitemOpen
  \bibfield  {author} {\bibinfo {author} {\bibfnamefont {D.~M.}\ \bibnamefont
  {Siegel}}\ and\ \bibinfo {author} {\bibfnamefont {B.~D.}\ \bibnamefont
  {Metzger}},\ }\href {\doibase 10.1103/PhysRevLett.119.231102} {\bibfield
  {journal} {\bibinfo  {journal} {Phys. Rev. Lett.}\ }\textbf {\bibinfo
  {volume} {119}},\ \bibinfo {pages} {231102} (\bibinfo {year} {2017})},\
  \Eprint {http://arxiv.org/abs/1705.05473} {arXiv:1705.05473 [astro-ph.HE]}
  \BibitemShut {NoStop}%
\bibitem [{\citenamefont {Holmbeck}\ \emph
  {et~al.}(2019{\natexlab{a}})\citenamefont {Holmbeck}, \citenamefont {Surman},
  \citenamefont {Sprouse}, \citenamefont {Mumpower}, \citenamefont {Vassh},
  \citenamefont {Beers},\ and\ \citenamefont {Kawano}}]{Holmbeck:2018xet}%
  \BibitemOpen
  \bibfield  {author} {\bibinfo {author} {\bibfnamefont {E.~M.}\ \bibnamefont
  {Holmbeck}}, \bibinfo {author} {\bibfnamefont {R.}~\bibnamefont {Surman}},
  \bibinfo {author} {\bibfnamefont {T.~M.}\ \bibnamefont {Sprouse}}, \bibinfo
  {author} {\bibfnamefont {M.~R.}\ \bibnamefont {Mumpower}}, \bibinfo {author}
  {\bibfnamefont {N.}~\bibnamefont {Vassh}}, \bibinfo {author} {\bibfnamefont
  {T.~C.}\ \bibnamefont {Beers}}, \ and\ \bibinfo {author} {\bibfnamefont
  {T.}~\bibnamefont {Kawano}},\ }\href {\doibase 10.3847/1538-4357/aaefef}
  {\bibfield  {journal} {\bibinfo  {journal} {Astrophys. J.}\ }\textbf
  {\bibinfo {volume} {870}},\ \bibinfo {pages} {23} (\bibinfo {year}
  {2019}{\natexlab{a}})},\ \Eprint {http://arxiv.org/abs/1807.06662}
  {arXiv:1807.06662 [astro-ph.SR]} \BibitemShut {NoStop}%
\bibitem [{\citenamefont {Holmbeck}\ \emph
  {et~al.}(2019{\natexlab{b}})\citenamefont {Holmbeck}, \citenamefont {Frebel},
  \citenamefont {McLaughlin}, \citenamefont {Mumpower}, \citenamefont
  {Sprouse},\ and\ \citenamefont {Surman}}]{Holmbeck:2019xnd}%
  \BibitemOpen
  \bibfield  {author} {\bibinfo {author} {\bibfnamefont {E.~M.}\ \bibnamefont
  {Holmbeck}}, \bibinfo {author} {\bibfnamefont {A.}~\bibnamefont {Frebel}},
  \bibinfo {author} {\bibfnamefont {G.~C.}\ \bibnamefont {McLaughlin}},
  \bibinfo {author} {\bibfnamefont {M.~R.}\ \bibnamefont {Mumpower}}, \bibinfo
  {author} {\bibfnamefont {T.~M.}\ \bibnamefont {Sprouse}}, \ and\ \bibinfo
  {author} {\bibfnamefont {R.}~\bibnamefont {Surman}},\ }\href {\doibase
  10.3847/1538-4357/ab2a01} {\  (\bibinfo {year} {2019}{\natexlab{b}}),\
  10.3847/1538-4357/ab2a01},\ \Eprint {http://arxiv.org/abs/1904.02139}
  {arXiv:1904.02139 [astro-ph.HE]} \BibitemShut {NoStop}%
\bibitem [{\citenamefont {Eichler}\ \emph {et~al.}(2019)\citenamefont
  {Eichler}, \citenamefont {Sayar}, \citenamefont {Arcones},\ and\
  \citenamefont {Rauscher}}]{Eichler:2019rzj}%
  \BibitemOpen
  \bibfield  {author} {\bibinfo {author} {\bibfnamefont {M.}~\bibnamefont
  {Eichler}}, \bibinfo {author} {\bibfnamefont {W.}~\bibnamefont {Sayar}},
  \bibinfo {author} {\bibfnamefont {A.}~\bibnamefont {Arcones}}, \ and\
  \bibinfo {author} {\bibfnamefont {T.}~\bibnamefont {Rauscher}},\ }\href
  {\doibase 10.3847/1538-4357/ab24cf} {\bibfield  {journal} {\bibinfo
  {journal} {Astrophys. J.}\ }\textbf {\bibinfo {volume} {879}},\ \bibinfo
  {pages} {47} (\bibinfo {year} {2019})},\ \Eprint
  {http://arxiv.org/abs/1904.07013} {arXiv:1904.07013 [astro-ph.HE]}
  \BibitemShut {NoStop}%
\bibitem [{\citenamefont {Fern\'andez}\ \emph {et~al.}(2020)\citenamefont
  {Fern\'andez}, \citenamefont {Foucart},\ and\ \citenamefont
  {Lippuner}}]{Fernandez:2020oow}%
  \BibitemOpen
  \bibfield  {author} {\bibinfo {author} {\bibfnamefont {R.}~\bibnamefont
  {Fern\'andez}}, \bibinfo {author} {\bibfnamefont {F.}~\bibnamefont
  {Foucart}}, \ and\ \bibinfo {author} {\bibfnamefont {J.}~\bibnamefont
  {Lippuner}},\ }\href {\doibase 10.1093/mnras/staa2209} {\bibfield  {journal}
  {\bibinfo  {journal} {Mon. Not. Roy. Astron. Soc.}\ }\textbf {\bibinfo
  {volume} {497}},\ \bibinfo {pages} {3221} (\bibinfo {year} {2020})},\ \Eprint
  {http://arxiv.org/abs/2005.14208} {arXiv:2005.14208 [astro-ph.HE]}
  \BibitemShut {NoStop}%
\bibitem [{\citenamefont {Nedora}\ \emph {et~al.}(2021)\citenamefont {Nedora},
  \citenamefont {Bernuzzi}, \citenamefont {Radice}, \citenamefont {Daszuta},
  \citenamefont {Endrizzi}, \citenamefont {Perego}, \citenamefont {Prakash},
  \citenamefont {Safarzadeh}, \citenamefont {Schianchi},\ and\ \citenamefont
  {Logoteta}}]{Nedora:2020hxc}%
  \BibitemOpen
  \bibfield  {author} {\bibinfo {author} {\bibfnamefont {V.}~\bibnamefont
  {Nedora}}, \bibinfo {author} {\bibfnamefont {S.}~\bibnamefont {Bernuzzi}},
  \bibinfo {author} {\bibfnamefont {D.}~\bibnamefont {Radice}}, \bibinfo
  {author} {\bibfnamefont {B.}~\bibnamefont {Daszuta}}, \bibinfo {author}
  {\bibfnamefont {A.}~\bibnamefont {Endrizzi}}, \bibinfo {author}
  {\bibfnamefont {A.}~\bibnamefont {Perego}}, \bibinfo {author} {\bibfnamefont
  {A.}~\bibnamefont {Prakash}}, \bibinfo {author} {\bibfnamefont
  {M.}~\bibnamefont {Safarzadeh}}, \bibinfo {author} {\bibfnamefont
  {F.}~\bibnamefont {Schianchi}}, \ and\ \bibinfo {author} {\bibfnamefont
  {D.}~\bibnamefont {Logoteta}},\ }\href {\doibase 10.3847/1538-4357/abc9be}
  {\bibfield  {journal} {\bibinfo  {journal} {Astrophys. J.}\ }\textbf
  {\bibinfo {volume} {906}},\ \bibinfo {pages} {98} (\bibinfo {year} {2021})},\
  \Eprint {http://arxiv.org/abs/2008.04333} {arXiv:2008.04333 [astro-ph.HE]}
  \BibitemShut {NoStop}%
\bibitem [{\citenamefont {Kullmann}\ \emph {et~al.}(2022)\citenamefont
  {Kullmann}, \citenamefont {Goriely}, \citenamefont {Just}, \citenamefont
  {Ardevol-Pulpillo}, \citenamefont {Bauswein},\ and\ \citenamefont
  {Janka}}]{Kullmann:2021gvo}%
  \BibitemOpen
  \bibfield  {author} {\bibinfo {author} {\bibfnamefont {I.}~\bibnamefont
  {Kullmann}}, \bibinfo {author} {\bibfnamefont {S.}~\bibnamefont {Goriely}},
  \bibinfo {author} {\bibfnamefont {O.}~\bibnamefont {Just}}, \bibinfo {author}
  {\bibfnamefont {R.}~\bibnamefont {Ardevol-Pulpillo}}, \bibinfo {author}
  {\bibfnamefont {A.}~\bibnamefont {Bauswein}}, \ and\ \bibinfo {author}
  {\bibfnamefont {H.~T.}\ \bibnamefont {Janka}},\ }\href {\doibase
  10.1093/mnras/stab3393} {\bibfield  {journal} {\bibinfo  {journal} {Mon. Not.
  Roy. Astron. Soc.}\ }\textbf {\bibinfo {volume} {510}},\ \bibinfo {pages}
  {2804} (\bibinfo {year} {2022})},\ \Eprint {http://arxiv.org/abs/2109.02509}
  {arXiv:2109.02509 [astro-ph.HE]} \BibitemShut {NoStop}%
\bibitem [{\citenamefont {Fujibayashi}\ \emph {et~al.}(2022)\citenamefont
  {Fujibayashi}, \citenamefont {Kiuchi}, \citenamefont {Wanajo}, \citenamefont
  {Kyutoku}, \citenamefont {Sekiguchi},\ and\ \citenamefont
  {Shibata}}]{Fujibayashi:2022ftg}%
  \BibitemOpen
  \bibfield  {author} {\bibinfo {author} {\bibfnamefont {S.}~\bibnamefont
  {Fujibayashi}}, \bibinfo {author} {\bibfnamefont {K.}~\bibnamefont {Kiuchi}},
  \bibinfo {author} {\bibfnamefont {S.}~\bibnamefont {Wanajo}}, \bibinfo
  {author} {\bibfnamefont {K.}~\bibnamefont {Kyutoku}}, \bibinfo {author}
  {\bibfnamefont {Y.}~\bibnamefont {Sekiguchi}}, \ and\ \bibinfo {author}
  {\bibfnamefont {M.}~\bibnamefont {Shibata}},\ }\href@noop {} {\  (\bibinfo
  {year} {2022})},\ \Eprint {http://arxiv.org/abs/2205.05557} {arXiv:2205.05557
  [astro-ph.HE]} \BibitemShut {NoStop}%
\bibitem [{\citenamefont {{Woosley}}\ and\ \citenamefont
  {{Hoffman}}(1992)}]{Woosley:1992ApJ}%
  \BibitemOpen
  \bibfield  {author} {\bibinfo {author} {\bibfnamefont {S.~E.}\ \bibnamefont
  {{Woosley}}}\ and\ \bibinfo {author} {\bibfnamefont {R.~D.}\ \bibnamefont
  {{Hoffman}}},\ }\href {\doibase 10.1086/171644} {\bibfield  {journal}
  {\bibinfo  {journal} {\apj}\ }\textbf {\bibinfo {volume} {395}},\ \bibinfo
  {pages} {202} (\bibinfo {year} {1992})}\BibitemShut {NoStop}%
\bibitem [{\citenamefont {{Wu}}\ \emph {et~al.}(2019)\citenamefont {{Wu}},
  \citenamefont {{Barnes}}, \citenamefont {{Mart{\'{\i}}nez-Pinedo}},\ and\
  \citenamefont {{Metzger}}}]{Wu+19}%
  \BibitemOpen
  \bibfield  {author} {\bibinfo {author} {\bibfnamefont {M.-R.}\ \bibnamefont
  {{Wu}}}, \bibinfo {author} {\bibfnamefont {J.}~\bibnamefont {{Barnes}}},
  \bibinfo {author} {\bibfnamefont {G.}~\bibnamefont
  {{Mart{\'{\i}}nez-Pinedo}}}, \ and\ \bibinfo {author} {\bibfnamefont {B.~D.}\
  \bibnamefont {{Metzger}}},\ }\href {\doibase 10.1103/PhysRevLett.122.062701}
  {\bibfield  {journal} {\bibinfo  {journal} {\prl}\ }\textbf {\bibinfo
  {volume} {122}},\ \bibinfo {eid} {062701} (\bibinfo {year} {2019})},\ \Eprint
  {http://arxiv.org/abs/1808.10459} {arXiv:1808.10459 [astro-ph.HE]}
  \BibitemShut {NoStop}%
\bibitem [{\citenamefont {Goriely}\ and\ \citenamefont
  {Clerbaux}(1999)}]{Goriely:1999jq}%
  \BibitemOpen
  \bibfield  {author} {\bibinfo {author} {\bibfnamefont {S.}~\bibnamefont
  {Goriely}}\ and\ \bibinfo {author} {\bibfnamefont {B.}~\bibnamefont
  {Clerbaux}},\ }\href@noop {} {\bibfield  {journal} {\bibinfo  {journal}
  {Astron. Astrophys.}\ }\textbf {\bibinfo {volume} {346}},\ \bibinfo {pages}
  {798} (\bibinfo {year} {1999})},\ \Eprint
  {http://arxiv.org/abs/astro-ph/9904409} {arXiv:astro-ph/9904409} \BibitemShut
  {NoStop}%
\bibitem [{\citenamefont {Panov}\ \emph {et~al.}(2013)\citenamefont {Panov},
  \citenamefont {Korneev}, \citenamefont {Martinez-Pinedo},\ and\ \citenamefont
  {Thielemann}}]{Panov:2013tfa}%
  \BibitemOpen
  \bibfield  {author} {\bibinfo {author} {\bibfnamefont {I.~V.}\ \bibnamefont
  {Panov}}, \bibinfo {author} {\bibfnamefont {I.~Y.}\ \bibnamefont {Korneev}},
  \bibinfo {author} {\bibfnamefont {G.}~\bibnamefont {Martinez-Pinedo}}, \ and\
  \bibinfo {author} {\bibfnamefont {F.~K.}\ \bibnamefont {Thielemann}},\ }\href
  {\doibase 10.1134/S1063773713030043} {\bibfield  {journal} {\bibinfo
  {journal} {Astron. Lett.}\ }\textbf {\bibinfo {volume} {39}},\ \bibinfo
  {pages} {150} (\bibinfo {year} {2013})}\BibitemShut {NoStop}%
\bibitem [{\citenamefont {Eichler}\ \emph {et~al.}(2015)\citenamefont {Eichler}
  \emph {et~al.}}]{Eichler:2014kma}%
  \BibitemOpen
  \bibfield  {author} {\bibinfo {author} {\bibfnamefont {M.}~\bibnamefont
  {Eichler}} \emph {et~al.},\ }\href {\doibase 10.1088/0004-637X/808/1/30}
  {\bibfield  {journal} {\bibinfo  {journal} {Astrophys. J.}\ }\textbf
  {\bibinfo {volume} {808}},\ \bibinfo {pages} {30} (\bibinfo {year} {2015})},\
  \Eprint {http://arxiv.org/abs/1411.0974} {arXiv:1411.0974 [astro-ph.HE]}
  \BibitemShut {NoStop}%
\bibitem [{\citenamefont {{Mendoza-Temis}}\ \emph {et~al.}(2015)\citenamefont
  {{Mendoza-Temis}}, \citenamefont {{Wu}}, \citenamefont {{Langanke}},
  \citenamefont {{Mart{\'\i}nez-Pinedo}}, \citenamefont {{Bauswein}},\ and\
  \citenamefont {{Janka}}}]{Mendoza2015}%
  \BibitemOpen
  \bibfield  {author} {\bibinfo {author} {\bibfnamefont {J.~d.~J.}\
  \bibnamefont {{Mendoza-Temis}}}, \bibinfo {author} {\bibfnamefont {M.-R.}\
  \bibnamefont {{Wu}}}, \bibinfo {author} {\bibfnamefont {K.}~\bibnamefont
  {{Langanke}}}, \bibinfo {author} {\bibfnamefont {G.}~\bibnamefont
  {{Mart{\'\i}nez-Pinedo}}}, \bibinfo {author} {\bibfnamefont {A.}~\bibnamefont
  {{Bauswein}}}, \ and\ \bibinfo {author} {\bibfnamefont {H.-T.}\ \bibnamefont
  {{Janka}}},\ }\href {\doibase 10.1103/PhysRevC.92.055805} {\bibfield
  {journal} {\bibinfo  {journal} {\prc}\ }\textbf {\bibinfo {volume} {92}},\
  \bibinfo {eid} {055805} (\bibinfo {year} {2015})},\ \Eprint
  {http://arxiv.org/abs/1409.6135} {arXiv:1409.6135 [astro-ph.HE]} \BibitemShut
  {NoStop}%
\bibitem [{\citenamefont {Goriely}\ and\ \citenamefont
  {Janka}(2016)}]{Goriely:2016gfe}%
  \BibitemOpen
  \bibfield  {author} {\bibinfo {author} {\bibfnamefont {S.}~\bibnamefont
  {Goriely}}\ and\ \bibinfo {author} {\bibfnamefont {H.~T.}\ \bibnamefont
  {Janka}},\ }\href {\doibase 10.1093/mnras/stw946} {\bibfield  {journal}
  {\bibinfo  {journal} {Mon. Not. Roy. Astron. Soc.}\ }\textbf {\bibinfo
  {volume} {459}},\ \bibinfo {pages} {4174} (\bibinfo {year} {2016})},\ \Eprint
  {http://arxiv.org/abs/1603.04282} {arXiv:1603.04282 [astro-ph.SR]}
  \BibitemShut {NoStop}%
\bibitem [{\citenamefont {Vassh}\ \emph {et~al.}(2019)\citenamefont {Vassh}
  \emph {et~al.}}]{Vassh:2018wcf}%
  \BibitemOpen
  \bibfield  {author} {\bibinfo {author} {\bibfnamefont {N.}~\bibnamefont
  {Vassh}} \emph {et~al.},\ }\href {\doibase 10.1088/1361-6471/ab0bea}
  {\bibfield  {journal} {\bibinfo  {journal} {J. Phys. G}\ }\textbf {\bibinfo
  {volume} {46}},\ \bibinfo {pages} {065202} (\bibinfo {year} {2019})},\
  \Eprint {http://arxiv.org/abs/1810.08133} {arXiv:1810.08133 [nucl-th]}
  \BibitemShut {NoStop}%
\bibitem [{\citenamefont {Zhu}\ \emph {et~al.}(2018)\citenamefont {Zhu} \emph
  {et~al.}}]{Zhu:2018oay}%
  \BibitemOpen
  \bibfield  {author} {\bibinfo {author} {\bibfnamefont {Y.}~\bibnamefont
  {Zhu}} \emph {et~al.},\ }\href {\doibase 10.3847/2041-8213/aad5de} {\bibfield
   {journal} {\bibinfo  {journal} {Astrophys. J. Lett.}\ }\textbf {\bibinfo
  {volume} {863}},\ \bibinfo {pages} {L23} (\bibinfo {year} {2018})},\ \Eprint
  {http://arxiv.org/abs/1806.09724} {arXiv:1806.09724 [astro-ph.HE]}
  \BibitemShut {NoStop}%
\bibitem [{\citenamefont {Giuliani}\ \emph {et~al.}(2020)\citenamefont
  {Giuliani}, \citenamefont {Mart\'\i{}nez-Pinedo}, \citenamefont {Wu},\ and\
  \citenamefont {Robledo}}]{Giuliani:2019oot}%
  \BibitemOpen
  \bibfield  {author} {\bibinfo {author} {\bibfnamefont {S.~A.}\ \bibnamefont
  {Giuliani}}, \bibinfo {author} {\bibfnamefont {G.}~\bibnamefont
  {Mart\'\i{}nez-Pinedo}}, \bibinfo {author} {\bibfnamefont {M.-R.}\
  \bibnamefont {Wu}}, \ and\ \bibinfo {author} {\bibfnamefont {L.~M.}\
  \bibnamefont {Robledo}},\ }\href {\doibase 10.1103/PhysRevC.102.045804}
  {\bibfield  {journal} {\bibinfo  {journal} {Phys. Rev. C}\ }\textbf {\bibinfo
  {volume} {102}},\ \bibinfo {pages} {045804} (\bibinfo {year} {2020})},\
  \Eprint {http://arxiv.org/abs/1904.03733} {arXiv:1904.03733 [nucl-th]}
  \BibitemShut {NoStop}%
\bibitem [{\citenamefont {Zhu}\ \emph {et~al.}(2021)\citenamefont {Zhu},
  \citenamefont {Lund}, \citenamefont {Barnes}, \citenamefont {Sprouse},
  \citenamefont {Vassh}, \citenamefont {McLaughlin}, \citenamefont {Mumpower},\
  and\ \citenamefont {Surman}}]{Zhu:2020eyk}%
  \BibitemOpen
  \bibfield  {author} {\bibinfo {author} {\bibfnamefont {Y.~L.}\ \bibnamefont
  {Zhu}}, \bibinfo {author} {\bibfnamefont {K.}~\bibnamefont {Lund}}, \bibinfo
  {author} {\bibfnamefont {J.}~\bibnamefont {Barnes}}, \bibinfo {author}
  {\bibfnamefont {T.~M.}\ \bibnamefont {Sprouse}}, \bibinfo {author}
  {\bibfnamefont {N.}~\bibnamefont {Vassh}}, \bibinfo {author} {\bibfnamefont
  {G.~C.}\ \bibnamefont {McLaughlin}}, \bibinfo {author} {\bibfnamefont
  {M.~R.}\ \bibnamefont {Mumpower}}, \ and\ \bibinfo {author} {\bibfnamefont
  {R.}~\bibnamefont {Surman}},\ }\href {\doibase 10.3847/1538-4357/abc69e}
  {\bibfield  {journal} {\bibinfo  {journal} {Astrophys. J.}\ }\textbf
  {\bibinfo {volume} {906}},\ \bibinfo {pages} {94} (\bibinfo {year} {2021})},\
  \Eprint {http://arxiv.org/abs/2010.03668} {arXiv:2010.03668 [astro-ph.HE]}
  \BibitemShut {NoStop}%
\bibitem [{\citenamefont {Lema\^\i{}tre}\ \emph {et~al.}(2021)\citenamefont
  {Lema\^\i{}tre}, \citenamefont {Goriely}, \citenamefont {Bauswein},\ and\
  \citenamefont {Janka}}]{Lemaitre:2021yje}%
  \BibitemOpen
  \bibfield  {author} {\bibinfo {author} {\bibfnamefont {J.~F.}\ \bibnamefont
  {Lema\^\i{}tre}}, \bibinfo {author} {\bibfnamefont {S.}~\bibnamefont
  {Goriely}}, \bibinfo {author} {\bibfnamefont {A.}~\bibnamefont {Bauswein}}, \
  and\ \bibinfo {author} {\bibfnamefont {H.~T.}\ \bibnamefont {Janka}},\ }\href
  {\doibase 10.1103/PhysRevC.103.025806} {\bibfield  {journal} {\bibinfo
  {journal} {Phys. Rev. C}\ }\textbf {\bibinfo {volume} {103}},\ \bibinfo
  {pages} {025806} (\bibinfo {year} {2021})},\ \Eprint
  {http://arxiv.org/abs/2102.03686} {arXiv:2102.03686 [nucl-th]} \BibitemShut
  {NoStop}%
\bibitem [{\citenamefont {Moller}\ \emph {et~al.}(1995)\citenamefont {Moller},
  \citenamefont {Nix}, \citenamefont {Myers},\ and\ \citenamefont
  {Swiatecki}}]{Moller:1993ed}%
  \BibitemOpen
  \bibfield  {author} {\bibinfo {author} {\bibfnamefont {P.}~\bibnamefont
  {Moller}}, \bibinfo {author} {\bibfnamefont {J.~R.}\ \bibnamefont {Nix}},
  \bibinfo {author} {\bibfnamefont {W.~D.}\ \bibnamefont {Myers}}, \ and\
  \bibinfo {author} {\bibfnamefont {W.~J.}\ \bibnamefont {Swiatecki}},\ }\href
  {\doibase 10.1006/adnd.1995.1002} {\bibfield  {journal} {\bibinfo  {journal}
  {Atom. Data Nucl. Data Tabl.}\ }\textbf {\bibinfo {volume} {59}},\ \bibinfo
  {pages} {185} (\bibinfo {year} {1995})},\ \Eprint
  {http://arxiv.org/abs/nucl-th/9308022} {arXiv:nucl-th/9308022} \BibitemShut
  {NoStop}%
\bibitem [{\citenamefont {Duflo}\ and\ \citenamefont
  {Zuker}(1995)}]{Duflo:1995ep}%
  \BibitemOpen
  \bibfield  {author} {\bibinfo {author} {\bibfnamefont {J.}~\bibnamefont
  {Duflo}}\ and\ \bibinfo {author} {\bibfnamefont {A.~P.}\ \bibnamefont
  {Zuker}},\ }\href {\doibase 10.1103/PhysRevC.52.R23} {\bibfield  {journal}
  {\bibinfo  {journal} {Phys. Rev. C}\ }\textbf {\bibinfo {volume} {52}},\
  \bibinfo {pages} {R23} (\bibinfo {year} {1995})},\ \Eprint
  {http://arxiv.org/abs/nucl-th/9505011} {arXiv:nucl-th/9505011} \BibitemShut
  {NoStop}%
\bibitem [{\citenamefont {Moller}\ \emph {et~al.}(2003)\citenamefont {Moller},
  \citenamefont {Pfeiffer},\ and\ \citenamefont {Kratz}}]{Moller:2003fn}%
  \BibitemOpen
  \bibfield  {author} {\bibinfo {author} {\bibfnamefont {P.}~\bibnamefont
  {Moller}}, \bibinfo {author} {\bibfnamefont {B.}~\bibnamefont {Pfeiffer}}, \
  and\ \bibinfo {author} {\bibfnamefont {K.-L.}\ \bibnamefont {Kratz}},\ }\href
  {\doibase 10.1103/PhysRevC.67.055802} {\bibfield  {journal} {\bibinfo
  {journal} {Phys. Rev. C}\ }\textbf {\bibinfo {volume} {67}},\ \bibinfo
  {pages} {055802} (\bibinfo {year} {2003})}\BibitemShut {NoStop}%
\bibitem [{\citenamefont {Marketin}\ \emph {et~al.}(2016)\citenamefont
  {Marketin}, \citenamefont {Huther},\ and\ \citenamefont
  {Mart\'\i{}nez-Pinedo}}]{Marketin:2015gya}%
  \BibitemOpen
  \bibfield  {author} {\bibinfo {author} {\bibfnamefont {T.}~\bibnamefont
  {Marketin}}, \bibinfo {author} {\bibfnamefont {L.}~\bibnamefont {Huther}}, \
  and\ \bibinfo {author} {\bibfnamefont {G.}~\bibnamefont
  {Mart\'\i{}nez-Pinedo}},\ }\href {\doibase 10.1103/PhysRevC.93.025805}
  {\bibfield  {journal} {\bibinfo  {journal} {Phys. Rev. C}\ }\textbf {\bibinfo
  {volume} {93}},\ \bibinfo {pages} {025805} (\bibinfo {year} {2016})},\
  \Eprint {http://arxiv.org/abs/1507.07442} {arXiv:1507.07442 [nucl-th]}
  \BibitemShut {NoStop}%
\bibitem [{\citenamefont {{Arnould}}\ \emph {et~al.}(2007)\citenamefont
  {{Arnould}}, \citenamefont {{Goriely}},\ and\ \citenamefont
  {{Takahashi}}}]{arnould2007}%
  \BibitemOpen
  \bibfield  {author} {\bibinfo {author} {\bibfnamefont {M.}~\bibnamefont
  {{Arnould}}}, \bibinfo {author} {\bibfnamefont {S.}~\bibnamefont
  {{Goriely}}}, \ and\ \bibinfo {author} {\bibfnamefont {K.}~\bibnamefont
  {{Takahashi}}},\ }\href {\doibase 10.1016/j.physrep.2007.06.002} {\bibfield
  {journal} {\bibinfo  {journal} {\physrep}\ }\textbf {\bibinfo {volume}
  {450}},\ \bibinfo {pages} {97} (\bibinfo {year} {2007})},\ \Eprint
  {http://arxiv.org/abs/0705.4512} {arXiv:0705.4512 [astro-ph]} \BibitemShut
  {NoStop}%
\bibitem [{\citenamefont {{Kratz}}\ \emph {et~al.}(1993)\citenamefont
  {{Kratz}}, \citenamefont {{Bitouzet}}, \citenamefont {{Thielemann}},
  \citenamefont {{Moeller}},\ and\ \citenamefont {{Pfeiffer}}}]{Kratz:1993ApJ}%
  \BibitemOpen
  \bibfield  {author} {\bibinfo {author} {\bibfnamefont {K.-L.}\ \bibnamefont
  {{Kratz}}}, \bibinfo {author} {\bibfnamefont {J.-P.}\ \bibnamefont
  {{Bitouzet}}}, \bibinfo {author} {\bibfnamefont {F.-K.}\ \bibnamefont
  {{Thielemann}}}, \bibinfo {author} {\bibfnamefont {P.}~\bibnamefont
  {{Moeller}}}, \ and\ \bibinfo {author} {\bibfnamefont {B.}~\bibnamefont
  {{Pfeiffer}}},\ }\href {\doibase 10.1086/172196} {\bibfield  {journal}
  {\bibinfo  {journal} {\apj}\ }\textbf {\bibinfo {volume} {403}},\ \bibinfo
  {pages} {216} (\bibinfo {year} {1993})}\BibitemShut {NoStop}%
\bibitem [{\citenamefont {Li}\ and\ \citenamefont
  {Paczynski}(1998)}]{Li:1998bw}%
  \BibitemOpen
  \bibfield  {author} {\bibinfo {author} {\bibfnamefont {L.-X.}\ \bibnamefont
  {Li}}\ and\ \bibinfo {author} {\bibfnamefont {B.}~\bibnamefont {Paczynski}},\
  }\href {\doibase 10.1086/311680} {\bibfield  {journal} {\bibinfo  {journal}
  {Astrophys. J. Lett.}\ }\textbf {\bibinfo {volume} {507}},\ \bibinfo {pages}
  {L59} (\bibinfo {year} {1998})},\ \Eprint
  {http://arxiv.org/abs/astro-ph/9807272} {arXiv:astro-ph/9807272} \BibitemShut
  {NoStop}%
\bibitem [{\citenamefont {Metzger}\ \emph {et~al.}(2010)\citenamefont
  {Metzger}, \citenamefont {Martinez-Pinedo}, \citenamefont {Darbha},
  \citenamefont {Quataert}, \citenamefont {Arcones}, \citenamefont {Kasen},
  \citenamefont {Thomas}, \citenamefont {Nugent}, \citenamefont {Panov},\ and\
  \citenamefont {Zinner}}]{Metzger:2010sy}%
  \BibitemOpen
  \bibfield  {author} {\bibinfo {author} {\bibfnamefont {B.~D.}\ \bibnamefont
  {Metzger}}, \bibinfo {author} {\bibfnamefont {G.}~\bibnamefont
  {Martinez-Pinedo}}, \bibinfo {author} {\bibfnamefont {S.}~\bibnamefont
  {Darbha}}, \bibinfo {author} {\bibfnamefont {E.}~\bibnamefont {Quataert}},
  \bibinfo {author} {\bibfnamefont {A.}~\bibnamefont {Arcones}}, \bibinfo
  {author} {\bibfnamefont {D.}~\bibnamefont {Kasen}}, \bibinfo {author}
  {\bibfnamefont {R.}~\bibnamefont {Thomas}}, \bibinfo {author} {\bibfnamefont
  {P.}~\bibnamefont {Nugent}}, \bibinfo {author} {\bibfnamefont {I.~V.}\
  \bibnamefont {Panov}}, \ and\ \bibinfo {author} {\bibfnamefont {N.~T.}\
  \bibnamefont {Zinner}},\ }\href {\doibase 10.1111/j.1365-2966.2010.16864.x}
  {\bibfield  {journal} {\bibinfo  {journal} {Mon. Not. Roy. Astron. Soc.}\
  }\textbf {\bibinfo {volume} {406}},\ \bibinfo {pages} {2650} (\bibinfo {year}
  {2010})},\ \Eprint {http://arxiv.org/abs/1001.5029} {arXiv:1001.5029
  [astro-ph.HE]} \BibitemShut {NoStop}%
\bibitem [{\citenamefont {Barnes}\ and\ \citenamefont
  {Kasen}(2013)}]{Barnes:2013wka}%
  \BibitemOpen
  \bibfield  {author} {\bibinfo {author} {\bibfnamefont {J.}~\bibnamefont
  {Barnes}}\ and\ \bibinfo {author} {\bibfnamefont {D.}~\bibnamefont {Kasen}},\
  }\href {\doibase 10.1088/0004-637X/775/1/18} {\bibfield  {journal} {\bibinfo
  {journal} {Astrophys. J.}\ }\textbf {\bibinfo {volume} {775}},\ \bibinfo
  {pages} {18} (\bibinfo {year} {2013})},\ \Eprint
  {http://arxiv.org/abs/1303.5787} {arXiv:1303.5787 [astro-ph.HE]} \BibitemShut
  {NoStop}%
\bibitem [{\citenamefont {Tanaka}\ and\ \citenamefont
  {Hotokezaka}(2013)}]{Tanaka:2013ana}%
  \BibitemOpen
  \bibfield  {author} {\bibinfo {author} {\bibfnamefont {M.}~\bibnamefont
  {Tanaka}}\ and\ \bibinfo {author} {\bibfnamefont {K.}~\bibnamefont
  {Hotokezaka}},\ }\href {\doibase 10.1088/0004-637X/775/2/113} {\bibfield
  {journal} {\bibinfo  {journal} {Astrophys. J.}\ }\textbf {\bibinfo {volume}
  {775}},\ \bibinfo {pages} {113} (\bibinfo {year} {2013})},\ \Eprint
  {http://arxiv.org/abs/1306.3742} {arXiv:1306.3742 [astro-ph.HE]} \BibitemShut
  {NoStop}%
\bibitem [{\citenamefont {Barnes}\ \emph {et~al.}(2016)\citenamefont {Barnes},
  \citenamefont {Kasen}, \citenamefont {Wu},\ and\ \citenamefont
  {Mart\'\i{}nez-Pinedo}}]{Barnes:2016umi}%
  \BibitemOpen
  \bibfield  {author} {\bibinfo {author} {\bibfnamefont {J.}~\bibnamefont
  {Barnes}}, \bibinfo {author} {\bibfnamefont {D.}~\bibnamefont {Kasen}},
  \bibinfo {author} {\bibfnamefont {M.-R.}\ \bibnamefont {Wu}}, \ and\ \bibinfo
  {author} {\bibfnamefont {G.}~\bibnamefont {Mart\'\i{}nez-Pinedo}},\ }\href
  {\doibase 10.3847/0004-637X/829/2/110} {\bibfield  {journal} {\bibinfo
  {journal} {Astrophys. J.}\ }\textbf {\bibinfo {volume} {829}},\ \bibinfo
  {pages} {110} (\bibinfo {year} {2016})},\ \Eprint
  {http://arxiv.org/abs/1605.07218} {arXiv:1605.07218 [astro-ph.HE]}
  \BibitemShut {NoStop}%
\bibitem [{\citenamefont {Wollaeger}\ \emph {et~al.}(2018)\citenamefont
  {Wollaeger}, \citenamefont {Korobkin}, \citenamefont {Fontes}, \citenamefont
  {Rosswog}, \citenamefont {Even}, \citenamefont {Fryer}, \citenamefont
  {Sollerman}, \citenamefont {Hungerford}, \citenamefont {van Rossum},\ and\
  \citenamefont {Wollaber}}]{Wollaeger:2017ahm}%
  \BibitemOpen
  \bibfield  {author} {\bibinfo {author} {\bibfnamefont {R.~T.}\ \bibnamefont
  {Wollaeger}}, \bibinfo {author} {\bibfnamefont {O.}~\bibnamefont {Korobkin}},
  \bibinfo {author} {\bibfnamefont {C.~J.}\ \bibnamefont {Fontes}}, \bibinfo
  {author} {\bibfnamefont {S.~K.}\ \bibnamefont {Rosswog}}, \bibinfo {author}
  {\bibfnamefont {W.~P.}\ \bibnamefont {Even}}, \bibinfo {author}
  {\bibfnamefont {C.~L.}\ \bibnamefont {Fryer}}, \bibinfo {author}
  {\bibfnamefont {J.}~\bibnamefont {Sollerman}}, \bibinfo {author}
  {\bibfnamefont {A.~L.}\ \bibnamefont {Hungerford}}, \bibinfo {author}
  {\bibfnamefont {D.~R.}\ \bibnamefont {van Rossum}}, \ and\ \bibinfo {author}
  {\bibfnamefont {A.~B.}\ \bibnamefont {Wollaber}},\ }\href {\doibase
  10.1093/mnras/sty1018} {\bibfield  {journal} {\bibinfo  {journal} {Mon. Not.
  Roy. Astron. Soc.}\ }\textbf {\bibinfo {volume} {478}},\ \bibinfo {pages}
  {3298} (\bibinfo {year} {2018})},\ \Eprint {http://arxiv.org/abs/1705.07084}
  {arXiv:1705.07084 [astro-ph.HE]} \BibitemShut {NoStop}%
\bibitem [{\citenamefont {Tanaka}\ \emph {et~al.}(2017)\citenamefont {Tanaka}
  \emph {et~al.}}]{Tanaka:2017qxj}%
  \BibitemOpen
  \bibfield  {author} {\bibinfo {author} {\bibfnamefont {M.}~\bibnamefont
  {Tanaka}} \emph {et~al.},\ }\href {\doibase 10.1093/pasj/psx121} {\bibfield
  {journal} {\bibinfo  {journal} {Publ. Astron. Soc. Jap.}\ }\textbf {\bibinfo
  {volume} {69}},\ \bibinfo {pages} {psx12} (\bibinfo {year} {2017})},\ \Eprint
  {http://arxiv.org/abs/1710.05850} {arXiv:1710.05850 [astro-ph.HE]}
  \BibitemShut {NoStop}%
\bibitem [{\citenamefont {Waxman}\ \emph {et~al.}(2018)\citenamefont {Waxman},
  \citenamefont {Ofek}, \citenamefont {Kushnir},\ and\ \citenamefont
  {Gal-Yam}}]{Waxman:2017sqv}%
  \BibitemOpen
  \bibfield  {author} {\bibinfo {author} {\bibfnamefont {E.}~\bibnamefont
  {Waxman}}, \bibinfo {author} {\bibfnamefont {E.~O.}\ \bibnamefont {Ofek}},
  \bibinfo {author} {\bibfnamefont {D.}~\bibnamefont {Kushnir}}, \ and\
  \bibinfo {author} {\bibfnamefont {A.}~\bibnamefont {Gal-Yam}},\ }\href
  {\doibase 10.1093/mnras/sty2441} {\bibfield  {journal} {\bibinfo  {journal}
  {Mon. Not. Roy. Astron. Soc.}\ }\textbf {\bibinfo {volume} {481}},\ \bibinfo
  {pages} {3423} (\bibinfo {year} {2018})},\ \Eprint
  {http://arxiv.org/abs/1711.09638} {arXiv:1711.09638 [astro-ph.HE]}
  \BibitemShut {NoStop}%
\bibitem [{\citenamefont {Coughlin}\ \emph {et~al.}(2018)\citenamefont
  {Coughlin} \emph {et~al.}}]{Coughlin:2018miv}%
  \BibitemOpen
  \bibfield  {author} {\bibinfo {author} {\bibfnamefont {M.~W.}\ \bibnamefont
  {Coughlin}} \emph {et~al.},\ }\href {\doibase 10.1093/mnras/sty2174}
  {\bibfield  {journal} {\bibinfo  {journal} {Mon. Not. Roy. Astron. Soc.}\
  }\textbf {\bibinfo {volume} {480}},\ \bibinfo {pages} {3871} (\bibinfo {year}
  {2018})},\ \Eprint {http://arxiv.org/abs/1805.09371} {arXiv:1805.09371
  [astro-ph.HE]} \BibitemShut {NoStop}%
\bibitem [{\citenamefont {Hotokezaka}\ and\ \citenamefont
  {Nakar}(2019)}]{Hotokezaka:2019uwo}%
  \BibitemOpen
  \bibfield  {author} {\bibinfo {author} {\bibfnamefont {K.}~\bibnamefont
  {Hotokezaka}}\ and\ \bibinfo {author} {\bibfnamefont {E.}~\bibnamefont
  {Nakar}},\ }\href {\doibase 10.3847/1538-4357/ab6a98} {\  (\bibinfo {year}
  {2019}),\ 10.3847/1538-4357/ab6a98},\ \Eprint
  {http://arxiv.org/abs/1909.02581} {arXiv:1909.02581 [astro-ph.HE]}
  \BibitemShut {NoStop}%
\bibitem [{\citenamefont {{Kawaguchi}}\ \emph {et~al.}(2020)\citenamefont
  {{Kawaguchi}}, \citenamefont {{Shibata}},\ and\ \citenamefont
  {{Tanaka}}}]{Kawaguchi:2020}%
  \BibitemOpen
  \bibfield  {author} {\bibinfo {author} {\bibfnamefont {K.}~\bibnamefont
  {{Kawaguchi}}}, \bibinfo {author} {\bibfnamefont {M.}~\bibnamefont
  {{Shibata}}}, \ and\ \bibinfo {author} {\bibfnamefont {M.}~\bibnamefont
  {{Tanaka}}},\ }\href {\doibase 10.3847/1538-4357/ab61f6} {\bibfield
  {journal} {\bibinfo  {journal} {\apj}\ }\textbf {\bibinfo {volume} {889}},\
  \bibinfo {eid} {171} (\bibinfo {year} {2020})},\ \Eprint
  {http://arxiv.org/abs/1908.05815} {arXiv:1908.05815 [astro-ph.HE]}
  \BibitemShut {NoStop}%
\bibitem [{\citenamefont {Ristic}\ \emph {et~al.}(2022)\citenamefont {Ristic},
  \citenamefont {Champion}, \citenamefont {O'Shaughnessy}, \citenamefont
  {Wollaeger}, \citenamefont {Korobkin}, \citenamefont {Chase}, \citenamefont
  {Fryer}, \citenamefont {Hungerford},\ and\ \citenamefont
  {Fontes}}]{Ristic:2021ksz}%
  \BibitemOpen
  \bibfield  {author} {\bibinfo {author} {\bibfnamefont {M.}~\bibnamefont
  {Ristic}}, \bibinfo {author} {\bibfnamefont {E.}~\bibnamefont {Champion}},
  \bibinfo {author} {\bibfnamefont {R.}~\bibnamefont {O'Shaughnessy}}, \bibinfo
  {author} {\bibfnamefont {R.}~\bibnamefont {Wollaeger}}, \bibinfo {author}
  {\bibfnamefont {O.}~\bibnamefont {Korobkin}}, \bibinfo {author}
  {\bibfnamefont {E.~A.}\ \bibnamefont {Chase}}, \bibinfo {author}
  {\bibfnamefont {C.~L.}\ \bibnamefont {Fryer}}, \bibinfo {author}
  {\bibfnamefont {A.~L.}\ \bibnamefont {Hungerford}}, \ and\ \bibinfo {author}
  {\bibfnamefont {C.~J.}\ \bibnamefont {Fontes}},\ }\href {\doibase
  10.1103/PhysRevResearch.4.013046} {\bibfield  {journal} {\bibinfo  {journal}
  {Phys. Rev. Res.}\ }\textbf {\bibinfo {volume} {4}},\ \bibinfo {pages}
  {013046} (\bibinfo {year} {2022})},\ \Eprint
  {http://arxiv.org/abs/2105.07013} {arXiv:2105.07013 [astro-ph.HE]}
  \BibitemShut {NoStop}%
\bibitem [{\citenamefont {Breschi}\ \emph {et~al.}(2021)\citenamefont
  {Breschi}, \citenamefont {Perego}, \citenamefont {Bernuzzi}, \citenamefont
  {Del~Pozzo}, \citenamefont {Nedora}, \citenamefont {Radice},\ and\
  \citenamefont {Vescovi}}]{Breschi:2021tbm}%
  \BibitemOpen
  \bibfield  {author} {\bibinfo {author} {\bibfnamefont {M.}~\bibnamefont
  {Breschi}}, \bibinfo {author} {\bibfnamefont {A.}~\bibnamefont {Perego}},
  \bibinfo {author} {\bibfnamefont {S.}~\bibnamefont {Bernuzzi}}, \bibinfo
  {author} {\bibfnamefont {W.}~\bibnamefont {Del~Pozzo}}, \bibinfo {author}
  {\bibfnamefont {V.}~\bibnamefont {Nedora}}, \bibinfo {author} {\bibfnamefont
  {D.}~\bibnamefont {Radice}}, \ and\ \bibinfo {author} {\bibfnamefont
  {D.}~\bibnamefont {Vescovi}},\ }\href {\doibase 10.1093/mnras/stab1287}
  {\bibfield  {journal} {\bibinfo  {journal} {Mon. Not. Roy. Astron. Soc.}\
  }\textbf {\bibinfo {volume} {505}},\ \bibinfo {pages} {1661} (\bibinfo {year}
  {2021})},\ \Eprint {http://arxiv.org/abs/2101.01201} {arXiv:2101.01201
  [astro-ph.HE]} \BibitemShut {NoStop}%
\bibitem [{\citenamefont {Rosswog}\ \emph {et~al.}(2017)\citenamefont
  {Rosswog}, \citenamefont {Feindt}, \citenamefont {Korobkin}, \citenamefont
  {Wu}, \citenamefont {Sollerman}, \citenamefont {Goobar},\ and\ \citenamefont
  {Martinez-Pinedo}}]{Rosswog:2016dhy}%
  \BibitemOpen
  \bibfield  {author} {\bibinfo {author} {\bibfnamefont {S.}~\bibnamefont
  {Rosswog}}, \bibinfo {author} {\bibfnamefont {U.}~\bibnamefont {Feindt}},
  \bibinfo {author} {\bibfnamefont {O.}~\bibnamefont {Korobkin}}, \bibinfo
  {author} {\bibfnamefont {M.~R.}\ \bibnamefont {Wu}}, \bibinfo {author}
  {\bibfnamefont {J.}~\bibnamefont {Sollerman}}, \bibinfo {author}
  {\bibfnamefont {A.}~\bibnamefont {Goobar}}, \ and\ \bibinfo {author}
  {\bibfnamefont {G.}~\bibnamefont {Martinez-Pinedo}},\ }\href {\doibase
  10.1088/1361-6382/aa68a9} {\bibfield  {journal} {\bibinfo  {journal} {Class.
  Quant. Grav.}\ }\textbf {\bibinfo {volume} {34}},\ \bibinfo {pages} {104001}
  (\bibinfo {year} {2017})},\ \Eprint {http://arxiv.org/abs/1611.09822}
  {arXiv:1611.09822 [astro-ph.HE]} \BibitemShut {NoStop}%
\bibitem [{\citenamefont {{Wanajo}}(2018)}]{Wanajo2018}%
  \BibitemOpen
  \bibfield  {author} {\bibinfo {author} {\bibfnamefont {S.}~\bibnamefont
  {{Wanajo}}},\ }\href {\doibase 10.3847/1538-4357/aae0f2} {\bibfield
  {journal} {\bibinfo  {journal} {\apj}\ }\textbf {\bibinfo {volume} {868}},\
  \bibinfo {eid} {65} (\bibinfo {year} {2018})},\ \Eprint
  {http://arxiv.org/abs/1808.03763} {arXiv:1808.03763 [astro-ph.HE]}
  \BibitemShut {NoStop}%
\bibitem [{\citenamefont {Barnes}\ \emph {et~al.}(2021)\citenamefont {Barnes},
  \citenamefont {Zhu}, \citenamefont {Lund}, \citenamefont {Sprouse},
  \citenamefont {Vassh}, \citenamefont {McLaughlin}, \citenamefont {Mumpower},\
  and\ \citenamefont {Surman}}]{Barnes:2020nfi}%
  \BibitemOpen
  \bibfield  {author} {\bibinfo {author} {\bibfnamefont {J.}~\bibnamefont
  {Barnes}}, \bibinfo {author} {\bibfnamefont {Y.~L.}\ \bibnamefont {Zhu}},
  \bibinfo {author} {\bibfnamefont {K.~A.}\ \bibnamefont {Lund}}, \bibinfo
  {author} {\bibfnamefont {T.~M.}\ \bibnamefont {Sprouse}}, \bibinfo {author}
  {\bibfnamefont {N.}~\bibnamefont {Vassh}}, \bibinfo {author} {\bibfnamefont
  {G.~C.}\ \bibnamefont {McLaughlin}}, \bibinfo {author} {\bibfnamefont
  {M.~R.}\ \bibnamefont {Mumpower}}, \ and\ \bibinfo {author} {\bibfnamefont
  {R.}~\bibnamefont {Surman}},\ }\href {\doibase 10.3847/1538-4357/ac0aec}
  {\bibfield  {journal} {\bibinfo  {journal} {Astrophys. J.}\ }\textbf
  {\bibinfo {volume} {918}},\ \bibinfo {pages} {44} (\bibinfo {year} {2021})},\
  \Eprint {http://arxiv.org/abs/2010.11182} {arXiv:2010.11182 [astro-ph.HE]}
  \BibitemShut {NoStop}%
\bibitem [{\citenamefont {Hotokezaka}\ \emph {et~al.}(2016)\citenamefont
  {Hotokezaka}, \citenamefont {Wanajo}, \citenamefont {Tanaka}, \citenamefont
  {Bamba}, \citenamefont {Terada},\ and\ \citenamefont
  {Piran}}]{Hotokezaka:2015cma}%
  \BibitemOpen
  \bibfield  {author} {\bibinfo {author} {\bibfnamefont {K.}~\bibnamefont
  {Hotokezaka}}, \bibinfo {author} {\bibfnamefont {S.}~\bibnamefont {Wanajo}},
  \bibinfo {author} {\bibfnamefont {M.}~\bibnamefont {Tanaka}}, \bibinfo
  {author} {\bibfnamefont {A.}~\bibnamefont {Bamba}}, \bibinfo {author}
  {\bibfnamefont {Y.}~\bibnamefont {Terada}}, \ and\ \bibinfo {author}
  {\bibfnamefont {T.}~\bibnamefont {Piran}},\ }\href {\doibase
  10.1093/mnras/stw404} {\bibfield  {journal} {\bibinfo  {journal} {Mon. Not.
  Roy. Astron. Soc.}\ }\textbf {\bibinfo {volume} {459}},\ \bibinfo {pages}
  {35} (\bibinfo {year} {2016})},\ \Eprint {http://arxiv.org/abs/1511.05580}
  {arXiv:1511.05580 [astro-ph.HE]} \BibitemShut {NoStop}%
\bibitem [{\citenamefont {Hotokezaka}\ \emph {et~al.}(2021)\citenamefont
  {Hotokezaka}, \citenamefont {Tanaka}, \citenamefont {Kato},\ and\
  \citenamefont {Gaigalas}}]{Hotokezaka:2021ofe}%
  \BibitemOpen
  \bibfield  {author} {\bibinfo {author} {\bibfnamefont {K.}~\bibnamefont
  {Hotokezaka}}, \bibinfo {author} {\bibfnamefont {M.}~\bibnamefont {Tanaka}},
  \bibinfo {author} {\bibfnamefont {D.}~\bibnamefont {Kato}}, \ and\ \bibinfo
  {author} {\bibfnamefont {G.}~\bibnamefont {Gaigalas}},\ }\href {\doibase
  10.1093/mnras/stab1975} {\bibfield  {journal} {\bibinfo  {journal} {Mon. Not.
  Roy. Astron. Soc.}\ }\textbf {\bibinfo {volume} {506}},\ \bibinfo {pages}
  {5863} (\bibinfo {year} {2021})},\ \Eprint {http://arxiv.org/abs/2102.07879}
  {arXiv:2102.07879 [astro-ph.HE]} \BibitemShut {NoStop}%
\bibitem [{\citenamefont {Pognan}\ \emph {et~al.}(2022)\citenamefont {Pognan},
  \citenamefont {Jerkstrand},\ and\ \citenamefont {Grumer}}]{Pognan:2021wpy}%
  \BibitemOpen
  \bibfield  {author} {\bibinfo {author} {\bibfnamefont {Q.}~\bibnamefont
  {Pognan}}, \bibinfo {author} {\bibfnamefont {A.}~\bibnamefont {Jerkstrand}},
  \ and\ \bibinfo {author} {\bibfnamefont {J.}~\bibnamefont {Grumer}},\ }\href
  {\doibase 10.1093/mnras/stab3674} {\bibfield  {journal} {\bibinfo  {journal}
  {Mon. Not. Roy. Astron. Soc.}\ }\textbf {\bibinfo {volume} {510}},\ \bibinfo
  {pages} {3806} (\bibinfo {year} {2022})},\ \Eprint
  {http://arxiv.org/abs/2112.07484} {arXiv:2112.07484 [astro-ph.HE]}
  \BibitemShut {NoStop}%
\bibitem [{\citenamefont {Li}(2019)}]{Li:2018wee}%
  \BibitemOpen
  \bibfield  {author} {\bibinfo {author} {\bibfnamefont {L.-X.}\ \bibnamefont
  {Li}},\ }\href {\doibase 10.3847/1538-4357/aaf961} {\bibfield  {journal}
  {\bibinfo  {journal} {Astrophys. J.}\ }\textbf {\bibinfo {volume} {872}},\
  \bibinfo {pages} {19} (\bibinfo {year} {2019})},\ \Eprint
  {http://arxiv.org/abs/1808.09833} {arXiv:1808.09833 [astro-ph.HE]}
  \BibitemShut {NoStop}%
\bibitem [{\citenamefont {Korobkin}\ \emph {et~al.}(2019)\citenamefont
  {Korobkin} \emph {et~al.}}]{Korobkin:2019uxw}%
  \BibitemOpen
  \bibfield  {author} {\bibinfo {author} {\bibfnamefont {O.}~\bibnamefont
  {Korobkin}} \emph {et~al.},\ }\href {\doibase 10.3847/1538-4357/ab64d8} {\
  (\bibinfo {year} {2019}),\ 10.3847/1538-4357/ab64d8},\ \Eprint
  {http://arxiv.org/abs/1905.05089} {arXiv:1905.05089 [astro-ph.HE]}
  \BibitemShut {NoStop}%
\bibitem [{\citenamefont {Wang}\ \emph {et~al.}(2020)\citenamefont {Wang},
  \citenamefont {Vassh}, \citenamefont {Sprouse}, \citenamefont {Mumpower},
  \citenamefont {Vogt}, \citenamefont {Randrup},\ and\ \citenamefont
  {Surman}}]{Wang:2020qkn}%
  \BibitemOpen
  \bibfield  {author} {\bibinfo {author} {\bibfnamefont {X.}~\bibnamefont
  {Wang}}, \bibinfo {author} {\bibfnamefont {N.}~\bibnamefont {Vassh}},
  \bibinfo {author} {\bibfnamefont {T.}~\bibnamefont {Sprouse}}, \bibinfo
  {author} {\bibfnamefont {M.}~\bibnamefont {Mumpower}}, \bibinfo {author}
  {\bibfnamefont {R.}~\bibnamefont {Vogt}}, \bibinfo {author} {\bibfnamefont
  {J.}~\bibnamefont {Randrup}}, \ and\ \bibinfo {author} {\bibfnamefont
  {R.}~\bibnamefont {Surman}} (\bibinfo {collaboration} {FIRE}),\ }\href
  {\doibase 10.3847/2041-8213/abbe18} {\bibfield  {journal} {\bibinfo
  {journal} {Astrophys. J. Lett.}\ }\textbf {\bibinfo {volume} {903}},\
  \bibinfo {pages} {L3} (\bibinfo {year} {2020})},\ \Eprint
  {http://arxiv.org/abs/2008.03335} {arXiv:2008.03335 [astro-ph.HE]}
  \BibitemShut {NoStop}%
\bibitem [{\citenamefont {Chen}\ \emph {et~al.}(2021)\citenamefont {Chen},
  \citenamefont {Li}, \citenamefont {Lin},\ and\ \citenamefont
  {Liang}}]{Chen:2021tob}%
  \BibitemOpen
  \bibfield  {author} {\bibinfo {author} {\bibfnamefont {M.-H.}\ \bibnamefont
  {Chen}}, \bibinfo {author} {\bibfnamefont {L.-X.}\ \bibnamefont {Li}},
  \bibinfo {author} {\bibfnamefont {D.-B.}\ \bibnamefont {Lin}}, \ and\
  \bibinfo {author} {\bibfnamefont {E.-W.}\ \bibnamefont {Liang}},\ }\href
  {\doibase 10.3847/1538-4357/ac1267} {\bibfield  {journal} {\bibinfo
  {journal} {Astrophys. J.}\ }\textbf {\bibinfo {volume} {919}},\ \bibinfo
  {pages} {59} (\bibinfo {year} {2021})},\ \Eprint
  {http://arxiv.org/abs/2107.02982} {arXiv:2107.02982 [astro-ph.HE]}
  \BibitemShut {NoStop}%
\bibitem [{\citenamefont {Chen}\ \emph {et~al.}(2022)\citenamefont {Chen},
  \citenamefont {Hu},\ and\ \citenamefont {Liang}}]{Chen:2022nsj}%
  \BibitemOpen
  \bibfield  {author} {\bibinfo {author} {\bibfnamefont {M.-H.}\ \bibnamefont
  {Chen}}, \bibinfo {author} {\bibfnamefont {R.-C.}\ \bibnamefont {Hu}}, \ and\
  \bibinfo {author} {\bibfnamefont {E.-W.}\ \bibnamefont {Liang}},\ }\href@noop
  {} {\  (\bibinfo {year} {2022})},\ \Eprint {http://arxiv.org/abs/2204.13269}
  {arXiv:2204.13269 [astro-ph.HE]} \BibitemShut {NoStop}%
\bibitem [{\citenamefont {Wu}\ \emph {et~al.}(2019)\citenamefont {Wu},
  \citenamefont {Banerjee}, \citenamefont {Metzger}, \citenamefont
  {Mart\'\i{}nez-Pinedo}, \citenamefont {Aramaki}, \citenamefont {Burns},
  \citenamefont {Hailey}, \citenamefont {Barnes},\ and\ \citenamefont
  {Karagiorgi}}]{Wu:2019xrq}%
  \BibitemOpen
  \bibfield  {author} {\bibinfo {author} {\bibfnamefont {M.-R.}\ \bibnamefont
  {Wu}}, \bibinfo {author} {\bibfnamefont {P.}~\bibnamefont {Banerjee}},
  \bibinfo {author} {\bibfnamefont {B.~D.}\ \bibnamefont {Metzger}}, \bibinfo
  {author} {\bibfnamefont {G.}~\bibnamefont {Mart\'\i{}nez-Pinedo}}, \bibinfo
  {author} {\bibfnamefont {T.}~\bibnamefont {Aramaki}}, \bibinfo {author}
  {\bibfnamefont {E.}~\bibnamefont {Burns}}, \bibinfo {author} {\bibfnamefont
  {C.~J.}\ \bibnamefont {Hailey}}, \bibinfo {author} {\bibfnamefont
  {J.}~\bibnamefont {Barnes}}, \ and\ \bibinfo {author} {\bibfnamefont
  {G.}~\bibnamefont {Karagiorgi}},\ }\href {\doibase 10.3847/1538-4357/ab2593}
  {\bibfield  {journal} {\bibinfo  {journal} {Astrophys. J.}\ }\textbf
  {\bibinfo {volume} {880}},\ \bibinfo {pages} {23} (\bibinfo {year} {2019})},\
  \Eprint {http://arxiv.org/abs/1905.03793} {arXiv:1905.03793 [astro-ph.HE]}
  \BibitemShut {NoStop}%
\bibitem [{\citenamefont {Terada}\ \emph {et~al.}(2022)\citenamefont {Terada},
  \citenamefont {Miwa}, \citenamefont {Ohsumi}, \citenamefont {Fujimoto},
  \citenamefont {Katsuda}, \citenamefont {Bamba},\ and\ \citenamefont
  {Yamazaki}}]{Terada:2022hut}%
  \BibitemOpen
  \bibfield  {author} {\bibinfo {author} {\bibfnamefont {Y.}~\bibnamefont
  {Terada}}, \bibinfo {author} {\bibfnamefont {Y.}~\bibnamefont {Miwa}},
  \bibinfo {author} {\bibfnamefont {H.}~\bibnamefont {Ohsumi}}, \bibinfo
  {author} {\bibfnamefont {S.-i.}\ \bibnamefont {Fujimoto}}, \bibinfo {author}
  {\bibfnamefont {S.}~\bibnamefont {Katsuda}}, \bibinfo {author} {\bibfnamefont
  {A.}~\bibnamefont {Bamba}}, \ and\ \bibinfo {author} {\bibfnamefont
  {R.}~\bibnamefont {Yamazaki}},\ }\href@noop {} {\  (\bibinfo {year}
  {2022})},\ \Eprint {http://arxiv.org/abs/2205.05407} {arXiv:2205.05407
  [astro-ph.HE]} \BibitemShut {NoStop}%
\bibitem [{\citenamefont {{Suda}}\ \emph {et~al.}(2008)\citenamefont {{Suda}},
  \citenamefont {{Katsuta}}, \citenamefont {{Yamada}}, \citenamefont {{Suwa}},
  \citenamefont {{Ishizuka}}, \citenamefont {{Komiya}}, \citenamefont
  {{Sorai}}, \citenamefont {{Aikawa}},\ and\ \citenamefont
  {{Fujimoto}}}]{SAGA}%
  \BibitemOpen
  \bibfield  {author} {\bibinfo {author} {\bibfnamefont {T.}~\bibnamefont
  {{Suda}}}, \bibinfo {author} {\bibfnamefont {Y.}~\bibnamefont {{Katsuta}}},
  \bibinfo {author} {\bibfnamefont {S.}~\bibnamefont {{Yamada}}}, \bibinfo
  {author} {\bibfnamefont {T.}~\bibnamefont {{Suwa}}}, \bibinfo {author}
  {\bibfnamefont {C.}~\bibnamefont {{Ishizuka}}}, \bibinfo {author}
  {\bibfnamefont {Y.}~\bibnamefont {{Komiya}}}, \bibinfo {author}
  {\bibfnamefont {K.}~\bibnamefont {{Sorai}}}, \bibinfo {author} {\bibfnamefont
  {M.}~\bibnamefont {{Aikawa}}}, \ and\ \bibinfo {author} {\bibfnamefont
  {M.~Y.}\ \bibnamefont {{Fujimoto}}},\ }\href {\doibase
  10.1093/pasj/60.5.1159} {\bibfield  {journal} {\bibinfo  {journal} {\pasj}\
  }\textbf {\bibinfo {volume} {60}},\ \bibinfo {pages} {1159} (\bibinfo {year}
  {2008})},\ \Eprint {http://arxiv.org/abs/0806.3697} {arXiv:0806.3697
  [astro-ph]} \BibitemShut {NoStop}%
\bibitem [{\citenamefont {{Johnson}}\ and\ \citenamefont
  {{Bolte}}(2001)}]{johnson2001}%
  \BibitemOpen
  \bibfield  {author} {\bibinfo {author} {\bibfnamefont {J.~A.}\ \bibnamefont
  {{Johnson}}}\ and\ \bibinfo {author} {\bibfnamefont {M.}~\bibnamefont
  {{Bolte}}},\ }\href {\doibase 10.1086/321386} {\bibfield  {journal} {\bibinfo
   {journal} {\apj}\ }\textbf {\bibinfo {volume} {554}},\ \bibinfo {pages}
  {888} (\bibinfo {year} {2001})},\ \Eprint
  {http://arxiv.org/abs/astro-ph/0103299} {arXiv:astro-ph/0103299 [astro-ph]}
  \BibitemShut {NoStop}%
\bibitem [{\citenamefont {{Cowan}}\ \emph {et~al.}(2002)\citenamefont
  {{Cowan}}, \citenamefont {{Sneden}}, \citenamefont {{Burles}}, \citenamefont
  {{Ivans}}, \citenamefont {{Beers}}, \citenamefont {{Truran}}, \citenamefont
  {{Lawler}}, \citenamefont {{Primas}}, \citenamefont {{Fuller}}, \citenamefont
  {{Pfeiffer}},\ and\ \citenamefont {{Kratz}}}]{cowan2002}%
  \BibitemOpen
  \bibfield  {author} {\bibinfo {author} {\bibfnamefont {J.~J.}\ \bibnamefont
  {{Cowan}}}, \bibinfo {author} {\bibfnamefont {C.}~\bibnamefont {{Sneden}}},
  \bibinfo {author} {\bibfnamefont {S.}~\bibnamefont {{Burles}}}, \bibinfo
  {author} {\bibfnamefont {I.~I.}\ \bibnamefont {{Ivans}}}, \bibinfo {author}
  {\bibfnamefont {T.~C.}\ \bibnamefont {{Beers}}}, \bibinfo {author}
  {\bibfnamefont {J.~W.}\ \bibnamefont {{Truran}}}, \bibinfo {author}
  {\bibfnamefont {J.~E.}\ \bibnamefont {{Lawler}}}, \bibinfo {author}
  {\bibfnamefont {F.}~\bibnamefont {{Primas}}}, \bibinfo {author}
  {\bibfnamefont {G.~M.}\ \bibnamefont {{Fuller}}}, \bibinfo {author}
  {\bibfnamefont {B.}~\bibnamefont {{Pfeiffer}}}, \ and\ \bibinfo {author}
  {\bibfnamefont {K.-L.}\ \bibnamefont {{Kratz}}},\ }\href {\doibase
  10.1086/340347} {\bibfield  {journal} {\bibinfo  {journal} {\apj}\ }\textbf
  {\bibinfo {volume} {572}},\ \bibinfo {pages} {861} (\bibinfo {year}
  {2002})},\ \Eprint {http://arxiv.org/abs/astro-ph/0202429}
  {arXiv:astro-ph/0202429 [astro-ph]} \BibitemShut {NoStop}%
\bibitem [{\citenamefont {{Hill}}\ \emph {et~al.}(2002)\citenamefont {{Hill}},
  \citenamefont {{Plez}}, \citenamefont {{Cayrel}}, \citenamefont {{Beers}},
  \citenamefont {{Nordstr{\"o}m}}, \citenamefont {{Andersen}}, \citenamefont
  {{Spite}}, \citenamefont {{Spite}}, \citenamefont {{Barbuy}}, \citenamefont
  {{Bonifacio}}, \citenamefont {{Depagne}}, \citenamefont {{Fran{\c{c}}ois}},\
  and\ \citenamefont {{Primas}}}]{hill2002}%
  \BibitemOpen
  \bibfield  {author} {\bibinfo {author} {\bibfnamefont {V.}~\bibnamefont
  {{Hill}}}, \bibinfo {author} {\bibfnamefont {B.}~\bibnamefont {{Plez}}},
  \bibinfo {author} {\bibfnamefont {R.}~\bibnamefont {{Cayrel}}}, \bibinfo
  {author} {\bibfnamefont {T.~C.}\ \bibnamefont {{Beers}}}, \bibinfo {author}
  {\bibfnamefont {B.}~\bibnamefont {{Nordstr{\"o}m}}}, \bibinfo {author}
  {\bibfnamefont {J.}~\bibnamefont {{Andersen}}}, \bibinfo {author}
  {\bibfnamefont {M.}~\bibnamefont {{Spite}}}, \bibinfo {author} {\bibfnamefont
  {F.}~\bibnamefont {{Spite}}}, \bibinfo {author} {\bibfnamefont
  {B.}~\bibnamefont {{Barbuy}}}, \bibinfo {author} {\bibfnamefont
  {P.}~\bibnamefont {{Bonifacio}}}, \bibinfo {author} {\bibfnamefont
  {E.}~\bibnamefont {{Depagne}}}, \bibinfo {author} {\bibfnamefont
  {P.}~\bibnamefont {{Fran{\c{c}}ois}}}, \ and\ \bibinfo {author}
  {\bibfnamefont {F.}~\bibnamefont {{Primas}}},\ }\href {\doibase
  10.1051/0004-6361:20020434} {\bibfield  {journal} {\bibinfo  {journal}
  {\aap}\ }\textbf {\bibinfo {volume} {387}},\ \bibinfo {pages} {560} (\bibinfo
  {year} {2002})},\ \Eprint {http://arxiv.org/abs/astro-ph/0203462}
  {arXiv:astro-ph/0203462 [astro-ph]} \BibitemShut {NoStop}%
\bibitem [{\citenamefont {{Honda}}\ \emph {et~al.}(2004)\citenamefont
  {{Honda}}, \citenamefont {{Aoki}}, \citenamefont {{Kajino}}, \citenamefont
  {{Ando}}, \citenamefont {{Beers}}, \citenamefont {{Izumiura}}, \citenamefont
  {{Sadakane}},\ and\ \citenamefont {{Takada-Hidai}}}]{honda2004}%
  \BibitemOpen
  \bibfield  {author} {\bibinfo {author} {\bibfnamefont {S.}~\bibnamefont
  {{Honda}}}, \bibinfo {author} {\bibfnamefont {W.}~\bibnamefont {{Aoki}}},
  \bibinfo {author} {\bibfnamefont {T.}~\bibnamefont {{Kajino}}}, \bibinfo
  {author} {\bibfnamefont {H.}~\bibnamefont {{Ando}}}, \bibinfo {author}
  {\bibfnamefont {T.~C.}\ \bibnamefont {{Beers}}}, \bibinfo {author}
  {\bibfnamefont {H.}~\bibnamefont {{Izumiura}}}, \bibinfo {author}
  {\bibfnamefont {K.}~\bibnamefont {{Sadakane}}}, \ and\ \bibinfo {author}
  {\bibfnamefont {M.}~\bibnamefont {{Takada-Hidai}}},\ }\href {\doibase
  10.1086/383406} {\bibfield  {journal} {\bibinfo  {journal} {\apj}\ }\textbf
  {\bibinfo {volume} {607}},\ \bibinfo {pages} {474} (\bibinfo {year}
  {2004})},\ \Eprint {http://arxiv.org/abs/astro-ph/0402298}
  {arXiv:astro-ph/0402298 [astro-ph]} \BibitemShut {NoStop}%
\bibitem [{\citenamefont {{Ivans}}\ \emph {et~al.}(2006)\citenamefont
  {{Ivans}}, \citenamefont {{Simmerer}}, \citenamefont {{Sneden}},
  \citenamefont {{Lawler}}, \citenamefont {{Cowan}}, \citenamefont
  {{Gallino}},\ and\ \citenamefont {{Bisterzo}}}]{ivans2006}%
  \BibitemOpen
  \bibfield  {author} {\bibinfo {author} {\bibfnamefont {I.~I.}\ \bibnamefont
  {{Ivans}}}, \bibinfo {author} {\bibfnamefont {J.}~\bibnamefont {{Simmerer}}},
  \bibinfo {author} {\bibfnamefont {C.}~\bibnamefont {{Sneden}}}, \bibinfo
  {author} {\bibfnamefont {J.~E.}\ \bibnamefont {{Lawler}}}, \bibinfo {author}
  {\bibfnamefont {J.~J.}\ \bibnamefont {{Cowan}}}, \bibinfo {author}
  {\bibfnamefont {R.}~\bibnamefont {{Gallino}}}, \ and\ \bibinfo {author}
  {\bibfnamefont {S.}~\bibnamefont {{Bisterzo}}},\ }\href {\doibase
  10.1086/504069} {\bibfield  {journal} {\bibinfo  {journal} {\apj}\ }\textbf
  {\bibinfo {volume} {645}},\ \bibinfo {pages} {613} (\bibinfo {year}
  {2006})},\ \Eprint {http://arxiv.org/abs/astro-ph/0604180}
  {arXiv:astro-ph/0604180 [astro-ph]} \BibitemShut {NoStop}%
\bibitem [{\citenamefont {{Frebel}}\ \emph {et~al.}(2007)\citenamefont
  {{Frebel}}, \citenamefont {{Christlieb}}, \citenamefont {{Norris}},
  \citenamefont {{Thom}}, \citenamefont {{Beers}},\ and\ \citenamefont
  {{Rhee}}}]{frebel2007}%
  \BibitemOpen
  \bibfield  {author} {\bibinfo {author} {\bibfnamefont {A.}~\bibnamefont
  {{Frebel}}}, \bibinfo {author} {\bibfnamefont {N.}~\bibnamefont
  {{Christlieb}}}, \bibinfo {author} {\bibfnamefont {J.~E.}\ \bibnamefont
  {{Norris}}}, \bibinfo {author} {\bibfnamefont {C.}~\bibnamefont {{Thom}}},
  \bibinfo {author} {\bibfnamefont {T.~C.}\ \bibnamefont {{Beers}}}, \ and\
  \bibinfo {author} {\bibfnamefont {J.}~\bibnamefont {{Rhee}}},\ }\href
  {\doibase 10.1086/518122} {\bibfield  {journal} {\bibinfo  {journal} {\apjl}\
  }\textbf {\bibinfo {volume} {660}},\ \bibinfo {pages} {L117} (\bibinfo {year}
  {2007})},\ \Eprint {http://arxiv.org/abs/astro-ph/0703414}
  {arXiv:astro-ph/0703414 [astro-ph]} \BibitemShut {NoStop}%
\bibitem [{\citenamefont {{Aoki}}\ \emph {et~al.}(2007)\citenamefont {{Aoki}},
  \citenamefont {{Honda}}, \citenamefont {{Sadakane}},\ and\ \citenamefont
  {{Arimoto}}}]{aoki2007}%
  \BibitemOpen
  \bibfield  {author} {\bibinfo {author} {\bibfnamefont {W.}~\bibnamefont
  {{Aoki}}}, \bibinfo {author} {\bibfnamefont {S.}~\bibnamefont {{Honda}}},
  \bibinfo {author} {\bibfnamefont {K.}~\bibnamefont {{Sadakane}}}, \ and\
  \bibinfo {author} {\bibfnamefont {N.}~\bibnamefont {{Arimoto}}},\ }\href
  {\doibase 10.1093/pasj/59.3.L15} {\bibfield  {journal} {\bibinfo  {journal}
  {\pasj}\ }\textbf {\bibinfo {volume} {59}},\ \bibinfo {pages} {L15} (\bibinfo
  {year} {2007})},\ \Eprint {http://arxiv.org/abs/0704.3104} {arXiv:0704.3104
  [astro-ph]} \BibitemShut {NoStop}%
\bibitem [{\citenamefont {{Lai}}\ \emph {et~al.}(2008)\citenamefont {{Lai}},
  \citenamefont {{Bolte}}, \citenamefont {{Johnson}}, \citenamefont
  {{Lucatello}}, \citenamefont {{Heger}},\ and\ \citenamefont
  {{Woosley}}}]{lai2008}%
  \BibitemOpen
  \bibfield  {author} {\bibinfo {author} {\bibfnamefont {D.~K.}\ \bibnamefont
  {{Lai}}}, \bibinfo {author} {\bibfnamefont {M.}~\bibnamefont {{Bolte}}},
  \bibinfo {author} {\bibfnamefont {J.~A.}\ \bibnamefont {{Johnson}}}, \bibinfo
  {author} {\bibfnamefont {S.}~\bibnamefont {{Lucatello}}}, \bibinfo {author}
  {\bibfnamefont {A.}~\bibnamefont {{Heger}}}, \ and\ \bibinfo {author}
  {\bibfnamefont {S.~E.}\ \bibnamefont {{Woosley}}},\ }\href {\doibase
  10.1086/588811} {\bibfield  {journal} {\bibinfo  {journal} {\apj}\ }\textbf
  {\bibinfo {volume} {681}},\ \bibinfo {pages} {1524} (\bibinfo {year}
  {2008})},\ \Eprint {http://arxiv.org/abs/0804.1370} {arXiv:0804.1370
  [astro-ph]} \BibitemShut {NoStop}%
\bibitem [{\citenamefont {{Roederer}}\ \emph {et~al.}(2009)\citenamefont
  {{Roederer}}, \citenamefont {{Kratz}}, \citenamefont {{Frebel}},
  \citenamefont {{Christlieb}}, \citenamefont {{Pfeiffer}}, \citenamefont
  {{Cowan}},\ and\ \citenamefont {{Sneden}}}]{roederer2009}%
  \BibitemOpen
  \bibfield  {author} {\bibinfo {author} {\bibfnamefont {I.~U.}\ \bibnamefont
  {{Roederer}}}, \bibinfo {author} {\bibfnamefont {K.-L.}\ \bibnamefont
  {{Kratz}}}, \bibinfo {author} {\bibfnamefont {A.}~\bibnamefont {{Frebel}}},
  \bibinfo {author} {\bibfnamefont {N.}~\bibnamefont {{Christlieb}}}, \bibinfo
  {author} {\bibfnamefont {B.}~\bibnamefont {{Pfeiffer}}}, \bibinfo {author}
  {\bibfnamefont {J.~J.}\ \bibnamefont {{Cowan}}}, \ and\ \bibinfo {author}
  {\bibfnamefont {C.}~\bibnamefont {{Sneden}}},\ }\href {\doibase
  10.1088/0004-637X/698/2/1963} {\bibfield  {journal} {\bibinfo  {journal}
  {\apj}\ }\textbf {\bibinfo {volume} {698}},\ \bibinfo {pages} {1963}
  (\bibinfo {year} {2009})},\ \Eprint {http://arxiv.org/abs/0904.3105}
  {arXiv:0904.3105 [astro-ph.SR]} \BibitemShut {NoStop}%
\bibitem [{\citenamefont {{Roederer}}\ \emph {et~al.}(2010)\citenamefont
  {{Roederer}}, \citenamefont {{Sneden}}, \citenamefont {{Thompson}},
  \citenamefont {{Preston}},\ and\ \citenamefont {{Shectman}}}]{roederer2010}%
  \BibitemOpen
  \bibfield  {author} {\bibinfo {author} {\bibfnamefont {I.~U.}\ \bibnamefont
  {{Roederer}}}, \bibinfo {author} {\bibfnamefont {C.}~\bibnamefont
  {{Sneden}}}, \bibinfo {author} {\bibfnamefont {I.~B.}\ \bibnamefont
  {{Thompson}}}, \bibinfo {author} {\bibfnamefont {G.~W.}\ \bibnamefont
  {{Preston}}}, \ and\ \bibinfo {author} {\bibfnamefont {S.~A.}\ \bibnamefont
  {{Shectman}}},\ }\href {\doibase 10.1088/0004-637X/711/2/573} {\bibfield
  {journal} {\bibinfo  {journal} {\apj}\ }\textbf {\bibinfo {volume} {711}},\
  \bibinfo {pages} {573} (\bibinfo {year} {2010})},\ \Eprint
  {http://arxiv.org/abs/1001.1745} {arXiv:1001.1745 [astro-ph.GA]} \BibitemShut
  {NoStop}%
\bibitem [{\citenamefont {{Mashonkina}}\ \emph {et~al.}(2010)\citenamefont
  {{Mashonkina}}, \citenamefont {{Christlieb}}, \citenamefont {{Barklem}},
  \citenamefont {{Hill}}, \citenamefont {{Beers}},\ and\ \citenamefont
  {{Velichko}}}]{mashonkina2010}%
  \BibitemOpen
  \bibfield  {author} {\bibinfo {author} {\bibfnamefont {L.}~\bibnamefont
  {{Mashonkina}}}, \bibinfo {author} {\bibfnamefont {N.}~\bibnamefont
  {{Christlieb}}}, \bibinfo {author} {\bibfnamefont {P.~S.}\ \bibnamefont
  {{Barklem}}}, \bibinfo {author} {\bibfnamefont {V.}~\bibnamefont {{Hill}}},
  \bibinfo {author} {\bibfnamefont {T.~C.}\ \bibnamefont {{Beers}}}, \ and\
  \bibinfo {author} {\bibfnamefont {A.}~\bibnamefont {{Velichko}}},\ }\href
  {\doibase 10.1051/0004-6361/200913825} {\bibfield  {journal} {\bibinfo
  {journal} {\aap}\ }\textbf {\bibinfo {volume} {516}},\ \bibinfo {eid} {A46}
  (\bibinfo {year} {2010})},\ \Eprint {http://arxiv.org/abs/1003.3571}
  {arXiv:1003.3571 [astro-ph.SR]} \BibitemShut {NoStop}%
\bibitem [{\citenamefont {{Roederer}}\ \emph {et~al.}(2012)\citenamefont
  {{Roederer}}, \citenamefont {{Lawler}}, \citenamefont {{Sobeck}},
  \citenamefont {{Beers}}, \citenamefont {{Cowan}}, \citenamefont {{Frebel}},
  \citenamefont {{Ivans}}, \citenamefont {{Schatz}}, \citenamefont {{Sneden}},\
  and\ \citenamefont {{Thompson}}}]{roederer2012}%
  \BibitemOpen
  \bibfield  {author} {\bibinfo {author} {\bibfnamefont {I.~U.}\ \bibnamefont
  {{Roederer}}}, \bibinfo {author} {\bibfnamefont {J.~E.}\ \bibnamefont
  {{Lawler}}}, \bibinfo {author} {\bibfnamefont {J.~S.}\ \bibnamefont
  {{Sobeck}}}, \bibinfo {author} {\bibfnamefont {T.~C.}\ \bibnamefont
  {{Beers}}}, \bibinfo {author} {\bibfnamefont {J.~J.}\ \bibnamefont
  {{Cowan}}}, \bibinfo {author} {\bibfnamefont {A.}~\bibnamefont {{Frebel}}},
  \bibinfo {author} {\bibfnamefont {I.~I.}\ \bibnamefont {{Ivans}}}, \bibinfo
  {author} {\bibfnamefont {H.}~\bibnamefont {{Schatz}}}, \bibinfo {author}
  {\bibfnamefont {C.}~\bibnamefont {{Sneden}}}, \ and\ \bibinfo {author}
  {\bibfnamefont {I.~B.}\ \bibnamefont {{Thompson}}},\ }\href {\doibase
  10.1088/0067-0049/203/2/27} {\bibfield  {journal} {\bibinfo  {journal}
  {\apjs}\ }\textbf {\bibinfo {volume} {203}},\ \bibinfo {eid} {27} (\bibinfo
  {year} {2012})},\ \Eprint {http://arxiv.org/abs/1210.6387} {arXiv:1210.6387
  [astro-ph.SR]} \BibitemShut {NoStop}%
\bibitem [{\citenamefont {{Roederer}}\ \emph {et~al.}(2014)\citenamefont
  {{Roederer}}, \citenamefont {{Preston}}, \citenamefont {{Thompson}},
  \citenamefont {{Shectman}}, \citenamefont {{Sneden}}, \citenamefont
  {{Burley}},\ and\ \citenamefont {{Kelson}}}]{roederer2014}%
  \BibitemOpen
  \bibfield  {author} {\bibinfo {author} {\bibfnamefont {I.~U.}\ \bibnamefont
  {{Roederer}}}, \bibinfo {author} {\bibfnamefont {G.~W.}\ \bibnamefont
  {{Preston}}}, \bibinfo {author} {\bibfnamefont {I.~B.}\ \bibnamefont
  {{Thompson}}}, \bibinfo {author} {\bibfnamefont {S.~A.}\ \bibnamefont
  {{Shectman}}}, \bibinfo {author} {\bibfnamefont {C.}~\bibnamefont
  {{Sneden}}}, \bibinfo {author} {\bibfnamefont {G.~S.}\ \bibnamefont
  {{Burley}}}, \ and\ \bibinfo {author} {\bibfnamefont {D.~D.}\ \bibnamefont
  {{Kelson}}},\ }\href {\doibase 10.1088/0004-6256/147/6/136} {\bibfield
  {journal} {\bibinfo  {journal} {\aj}\ }\textbf {\bibinfo {volume} {147}},\
  \bibinfo {eid} {136} (\bibinfo {year} {2014})},\ \Eprint
  {http://arxiv.org/abs/1403.6853} {arXiv:1403.6853 [astro-ph.SR]} \BibitemShut
  {NoStop}%
\bibitem [{\citenamefont {{Siqueira Mello}}\ \emph {et~al.}(2014)\citenamefont
  {{Siqueira Mello}}, \citenamefont {{Hill}}, \citenamefont {{Barbuy}},
  \citenamefont {{Spite}}, \citenamefont {{Spite}}, \citenamefont {{Beers}},
  \citenamefont {{Caffau}}, \citenamefont {{Bonifacio}}, \citenamefont
  {{Cayrel}}, \citenamefont {{Fran{\c{c}}ois}}, \citenamefont {{Schatz}},\ and\
  \citenamefont {{Wanajo}}}]{siqueira2014}%
  \BibitemOpen
  \bibfield  {author} {\bibinfo {author} {\bibfnamefont {C.}~\bibnamefont
  {{Siqueira Mello}}}, \bibinfo {author} {\bibfnamefont {V.}~\bibnamefont
  {{Hill}}}, \bibinfo {author} {\bibfnamefont {B.}~\bibnamefont {{Barbuy}}},
  \bibinfo {author} {\bibfnamefont {M.}~\bibnamefont {{Spite}}}, \bibinfo
  {author} {\bibfnamefont {F.}~\bibnamefont {{Spite}}}, \bibinfo {author}
  {\bibfnamefont {T.~C.}\ \bibnamefont {{Beers}}}, \bibinfo {author}
  {\bibfnamefont {E.}~\bibnamefont {{Caffau}}}, \bibinfo {author}
  {\bibfnamefont {P.}~\bibnamefont {{Bonifacio}}}, \bibinfo {author}
  {\bibfnamefont {R.}~\bibnamefont {{Cayrel}}}, \bibinfo {author}
  {\bibfnamefont {P.}~\bibnamefont {{Fran{\c{c}}ois}}}, \bibinfo {author}
  {\bibfnamefont {H.}~\bibnamefont {{Schatz}}}, \ and\ \bibinfo {author}
  {\bibfnamefont {S.}~\bibnamefont {{Wanajo}}},\ }\href {\doibase
  10.1051/0004-6361/201423826} {\bibfield  {journal} {\bibinfo  {journal}
  {\aap}\ }\textbf {\bibinfo {volume} {565}},\ \bibinfo {eid} {A93} (\bibinfo
  {year} {2014})},\ \Eprint {http://arxiv.org/abs/1404.0234} {arXiv:1404.0234
  [astro-ph.SR]} \BibitemShut {NoStop}%
\bibitem [{\citenamefont {{Mashonkina, L.}}\ \emph {et~al.}(2014)\citenamefont
  {{Mashonkina, L.}}, \citenamefont {{Christlieb, N.}},\ and\ \citenamefont
  {{Eriksson, K.}}}]{mashokina2014}%
  \BibitemOpen
  \bibfield  {author} {\bibinfo {author} {\bibnamefont {{Mashonkina, L.}}},
  \bibinfo {author} {\bibnamefont {{Christlieb, N.}}}, \ and\ \bibinfo {author}
  {\bibnamefont {{Eriksson, K.}}},\ }\href {\doibase
  10.1051/0004-6361/201424017} {\bibfield  {journal} {\bibinfo  {journal}
  {A\&A}\ }\textbf {\bibinfo {volume} {569}},\ \bibinfo {pages} {A43} (\bibinfo
  {year} {2014})}\BibitemShut {NoStop}%
\bibitem [{\citenamefont {{Placco}}\ \emph {et~al.}(2017)\citenamefont
  {{Placco}}, \citenamefont {{Holmbeck}}, \citenamefont {{Frebel}},
  \citenamefont {{Beers}}, \citenamefont {{Surman}}, \citenamefont {{Ji}},
  \citenamefont {{Ezzeddine}}, \citenamefont {{Points}}, \citenamefont
  {{Kaleida}}, \citenamefont {{Hansen}}, \citenamefont {{Sakari}},\ and\
  \citenamefont {{Casey}}}]{placco2017}%
  \BibitemOpen
  \bibfield  {author} {\bibinfo {author} {\bibfnamefont {V.~M.}\ \bibnamefont
  {{Placco}}}, \bibinfo {author} {\bibfnamefont {E.~M.}\ \bibnamefont
  {{Holmbeck}}}, \bibinfo {author} {\bibfnamefont {A.}~\bibnamefont
  {{Frebel}}}, \bibinfo {author} {\bibfnamefont {T.~C.}\ \bibnamefont
  {{Beers}}}, \bibinfo {author} {\bibfnamefont {R.~A.}\ \bibnamefont
  {{Surman}}}, \bibinfo {author} {\bibfnamefont {A.~P.}\ \bibnamefont {{Ji}}},
  \bibinfo {author} {\bibfnamefont {R.}~\bibnamefont {{Ezzeddine}}}, \bibinfo
  {author} {\bibfnamefont {S.~D.}\ \bibnamefont {{Points}}}, \bibinfo {author}
  {\bibfnamefont {C.~C.}\ \bibnamefont {{Kaleida}}}, \bibinfo {author}
  {\bibfnamefont {T.~T.}\ \bibnamefont {{Hansen}}}, \bibinfo {author}
  {\bibfnamefont {C.~M.}\ \bibnamefont {{Sakari}}}, \ and\ \bibinfo {author}
  {\bibfnamefont {A.~R.}\ \bibnamefont {{Casey}}},\ }\href {\doibase
  10.3847/1538-4357/aa78ef} {\bibfield  {journal} {\bibinfo  {journal} {\apj}\
  }\textbf {\bibinfo {volume} {844}},\ \bibinfo {eid} {18} (\bibinfo {year}
  {2017})},\ \Eprint {http://arxiv.org/abs/1706.02934} {arXiv:1706.02934
  [astro-ph.SR]} \BibitemShut {NoStop}%
\bibitem [{\citenamefont {{Holmbeck}}\ \emph {et~al.}(2018)\citenamefont
  {{Holmbeck}}, \citenamefont {{Beers}}, \citenamefont {{Roederer}},
  \citenamefont {{Placco}}, \citenamefont {{Hansen}}, \citenamefont {{Sakari}},
  \citenamefont {{Sneden}}, \citenamefont {{Liu}}, \citenamefont {{Lee}},
  \citenamefont {{Cowan}},\ and\ \citenamefont {{Frebel}}}]{Holmbeck2018obs}%
  \BibitemOpen
  \bibfield  {author} {\bibinfo {author} {\bibfnamefont {E.~M.}\ \bibnamefont
  {{Holmbeck}}}, \bibinfo {author} {\bibfnamefont {T.~C.}\ \bibnamefont
  {{Beers}}}, \bibinfo {author} {\bibfnamefont {I.~U.}\ \bibnamefont
  {{Roederer}}}, \bibinfo {author} {\bibfnamefont {V.~M.}\ \bibnamefont
  {{Placco}}}, \bibinfo {author} {\bibfnamefont {T.~T.}\ \bibnamefont
  {{Hansen}}}, \bibinfo {author} {\bibfnamefont {C.~M.}\ \bibnamefont
  {{Sakari}}}, \bibinfo {author} {\bibfnamefont {C.}~\bibnamefont {{Sneden}}},
  \bibinfo {author} {\bibfnamefont {C.}~\bibnamefont {{Liu}}}, \bibinfo
  {author} {\bibfnamefont {Y.~S.}\ \bibnamefont {{Lee}}}, \bibinfo {author}
  {\bibfnamefont {J.~J.}\ \bibnamefont {{Cowan}}}, \ and\ \bibinfo {author}
  {\bibfnamefont {A.}~\bibnamefont {{Frebel}}},\ }\href {\doibase
  10.3847/2041-8213/aac722} {\bibfield  {journal} {\bibinfo  {journal} {\apjl}\
  }\textbf {\bibinfo {volume} {859}},\ \bibinfo {eid} {L24} (\bibinfo {year}
  {2018})},\ \Eprint {http://arxiv.org/abs/1805.11925} {arXiv:1805.11925
  [astro-ph.SR]} \BibitemShut {NoStop}%
\bibitem [{\citenamefont {{Sakari}}\ \emph
  {et~al.}(2018{\natexlab{a}})\citenamefont {{Sakari}}, \citenamefont
  {{Placco}}, \citenamefont {{Farrell}}, \citenamefont {{Roederer}},
  \citenamefont {{Wallerstein}}, \citenamefont {{Beers}}, \citenamefont
  {{Ezzeddine}}, \citenamefont {{Frebel}}, \citenamefont {{Hansen}},
  \citenamefont {{Holmbeck}}, \citenamefont {{Sneden}}, \citenamefont
  {{Cowan}}, \citenamefont {{Venn}}, \citenamefont {{Davis}}, \citenamefont
  {{Matijevi{\v{c}}}}, \citenamefont {{Wyse}}, \citenamefont
  {{Bland-Hawthorn}}, \citenamefont {{Chiappini}}, \citenamefont {{Freeman}},
  \citenamefont {{Gibson}}, \citenamefont {{Grebel}}, \citenamefont {{Helmi}},
  \citenamefont {{Kordopatis}}, \citenamefont {{Kunder}}, \citenamefont
  {{Navarro}}, \citenamefont {{Reid}}, \citenamefont {{Seabroke}},
  \citenamefont {{Steinmetz}},\ and\ \citenamefont {{Watson}}}]{sakari2018}%
  \BibitemOpen
  \bibfield  {author} {\bibinfo {author} {\bibfnamefont {C.~M.}\ \bibnamefont
  {{Sakari}}}, \bibinfo {author} {\bibfnamefont {V.~M.}\ \bibnamefont
  {{Placco}}}, \bibinfo {author} {\bibfnamefont {E.~M.}\ \bibnamefont
  {{Farrell}}}, \bibinfo {author} {\bibfnamefont {I.~U.}\ \bibnamefont
  {{Roederer}}}, \bibinfo {author} {\bibfnamefont {G.}~\bibnamefont
  {{Wallerstein}}}, \bibinfo {author} {\bibfnamefont {T.~C.}\ \bibnamefont
  {{Beers}}}, \bibinfo {author} {\bibfnamefont {R.}~\bibnamefont
  {{Ezzeddine}}}, \bibinfo {author} {\bibfnamefont {A.}~\bibnamefont
  {{Frebel}}}, \bibinfo {author} {\bibfnamefont {T.}~\bibnamefont {{Hansen}}},
  \bibinfo {author} {\bibfnamefont {E.~M.}\ \bibnamefont {{Holmbeck}}},
  \bibinfo {author} {\bibfnamefont {C.}~\bibnamefont {{Sneden}}}, \bibinfo
  {author} {\bibfnamefont {J.~J.}\ \bibnamefont {{Cowan}}}, \bibinfo {author}
  {\bibfnamefont {K.~A.}\ \bibnamefont {{Venn}}}, \bibinfo {author}
  {\bibfnamefont {C.~E.}\ \bibnamefont {{Davis}}}, \bibinfo {author}
  {\bibfnamefont {G.}~\bibnamefont {{Matijevi{\v{c}}}}}, \bibinfo {author}
  {\bibfnamefont {R.~F.~G.}\ \bibnamefont {{Wyse}}}, \bibinfo {author}
  {\bibfnamefont {J.}~\bibnamefont {{Bland-Hawthorn}}}, \bibinfo {author}
  {\bibfnamefont {C.}~\bibnamefont {{Chiappini}}}, \bibinfo {author}
  {\bibfnamefont {K.~C.}\ \bibnamefont {{Freeman}}}, \bibinfo {author}
  {\bibfnamefont {B.~K.}\ \bibnamefont {{Gibson}}}, \bibinfo {author}
  {\bibfnamefont {E.~K.}\ \bibnamefont {{Grebel}}}, \bibinfo {author}
  {\bibfnamefont {A.}~\bibnamefont {{Helmi}}}, \bibinfo {author} {\bibfnamefont
  {G.}~\bibnamefont {{Kordopatis}}}, \bibinfo {author} {\bibfnamefont
  {A.}~\bibnamefont {{Kunder}}}, \bibinfo {author} {\bibfnamefont
  {J.}~\bibnamefont {{Navarro}}}, \bibinfo {author} {\bibfnamefont
  {W.}~\bibnamefont {{Reid}}}, \bibinfo {author} {\bibfnamefont
  {G.}~\bibnamefont {{Seabroke}}}, \bibinfo {author} {\bibfnamefont
  {M.}~\bibnamefont {{Steinmetz}}}, \ and\ \bibinfo {author} {\bibfnamefont
  {F.}~\bibnamefont {{Watson}}},\ }\href {\doibase 10.3847/1538-4357/aae9df}
  {\bibfield  {journal} {\bibinfo  {journal} {\apj}\ }\textbf {\bibinfo
  {volume} {868}},\ \bibinfo {eid} {110} (\bibinfo {year}
  {2018}{\natexlab{a}})},\ \Eprint {http://arxiv.org/abs/1809.09156}
  {arXiv:1809.09156 [astro-ph.SR]} \BibitemShut {NoStop}%
\bibitem [{\citenamefont {{Roederer}}\ \emph {et~al.}(2018)\citenamefont
  {{Roederer}}, \citenamefont {{Sakari}}, \citenamefont {{Placco}},
  \citenamefont {{Beers}}, \citenamefont {{Ezzeddine}}, \citenamefont
  {{Frebel}},\ and\ \citenamefont {{Hansen}}}]{roederer2018}%
  \BibitemOpen
  \bibfield  {author} {\bibinfo {author} {\bibfnamefont {I.~U.}\ \bibnamefont
  {{Roederer}}}, \bibinfo {author} {\bibfnamefont {C.~M.}\ \bibnamefont
  {{Sakari}}}, \bibinfo {author} {\bibfnamefont {V.~M.}\ \bibnamefont
  {{Placco}}}, \bibinfo {author} {\bibfnamefont {T.~C.}\ \bibnamefont
  {{Beers}}}, \bibinfo {author} {\bibfnamefont {R.}~\bibnamefont
  {{Ezzeddine}}}, \bibinfo {author} {\bibfnamefont {A.}~\bibnamefont
  {{Frebel}}}, \ and\ \bibinfo {author} {\bibfnamefont {T.~T.}\ \bibnamefont
  {{Hansen}}},\ }\href {\doibase 10.3847/1538-4357/aadd92} {\bibfield
  {journal} {\bibinfo  {journal} {\apj}\ }\textbf {\bibinfo {volume} {865}},\
  \bibinfo {eid} {129} (\bibinfo {year} {2018})},\ \Eprint
  {http://arxiv.org/abs/1808.09469} {arXiv:1808.09469 [astro-ph.SR]}
  \BibitemShut {NoStop}%
\bibitem [{\citenamefont {{Cain}}\ \emph {et~al.}(2018)\citenamefont {{Cain}},
  \citenamefont {{Frebel}}, \citenamefont {{Gull}}, \citenamefont {{Ji}},
  \citenamefont {{Placco}}, \citenamefont {{Beers}}, \citenamefont
  {{Mel{\'e}ndez}}, \citenamefont {{Ezzeddine}}, \citenamefont {{Casey}},
  \citenamefont {{Hansen}}, \citenamefont {{Roederer}},\ and\ \citenamefont
  {{Sakari}}}]{cain2018}%
  \BibitemOpen
  \bibfield  {author} {\bibinfo {author} {\bibfnamefont {M.}~\bibnamefont
  {{Cain}}}, \bibinfo {author} {\bibfnamefont {A.}~\bibnamefont {{Frebel}}},
  \bibinfo {author} {\bibfnamefont {M.}~\bibnamefont {{Gull}}}, \bibinfo
  {author} {\bibfnamefont {A.~P.}\ \bibnamefont {{Ji}}}, \bibinfo {author}
  {\bibfnamefont {V.~M.}\ \bibnamefont {{Placco}}}, \bibinfo {author}
  {\bibfnamefont {T.~C.}\ \bibnamefont {{Beers}}}, \bibinfo {author}
  {\bibfnamefont {J.}~\bibnamefont {{Mel{\'e}ndez}}}, \bibinfo {author}
  {\bibfnamefont {R.}~\bibnamefont {{Ezzeddine}}}, \bibinfo {author}
  {\bibfnamefont {A.~R.}\ \bibnamefont {{Casey}}}, \bibinfo {author}
  {\bibfnamefont {T.~T.}\ \bibnamefont {{Hansen}}}, \bibinfo {author}
  {\bibfnamefont {I.~U.}\ \bibnamefont {{Roederer}}}, \ and\ \bibinfo {author}
  {\bibfnamefont {C.}~\bibnamefont {{Sakari}}},\ }\href {\doibase
  10.3847/1538-4357/aad37d} {\bibfield  {journal} {\bibinfo  {journal} {\apj}\
  }\textbf {\bibinfo {volume} {864}},\ \bibinfo {eid} {43} (\bibinfo {year}
  {2018})},\ \Eprint {http://arxiv.org/abs/1807.03734} {arXiv:1807.03734
  [astro-ph.SR]} \BibitemShut {NoStop}%
\bibitem [{\citenamefont {{Gull}}\ \emph {et~al.}(2018)\citenamefont {{Gull}},
  \citenamefont {{Frebel}}, \citenamefont {{Cain}}, \citenamefont {{Placco}},
  \citenamefont {{Ji}}, \citenamefont {{Abate}}, \citenamefont {{Ezzeddine}},
  \citenamefont {{Karakas}}, \citenamefont {{Hansen}}, \citenamefont
  {{Sakari}}, \citenamefont {{Holmbeck}}, \citenamefont {{Santucci}},
  \citenamefont {{Casey}},\ and\ \citenamefont {{Beers}}}]{gull2018}%
  \BibitemOpen
  \bibfield  {author} {\bibinfo {author} {\bibfnamefont {M.}~\bibnamefont
  {{Gull}}}, \bibinfo {author} {\bibfnamefont {A.}~\bibnamefont {{Frebel}}},
  \bibinfo {author} {\bibfnamefont {M.~G.}\ \bibnamefont {{Cain}}}, \bibinfo
  {author} {\bibfnamefont {V.~M.}\ \bibnamefont {{Placco}}}, \bibinfo {author}
  {\bibfnamefont {A.~P.}\ \bibnamefont {{Ji}}}, \bibinfo {author}
  {\bibfnamefont {C.}~\bibnamefont {{Abate}}}, \bibinfo {author} {\bibfnamefont
  {R.}~\bibnamefont {{Ezzeddine}}}, \bibinfo {author} {\bibfnamefont {A.~I.}\
  \bibnamefont {{Karakas}}}, \bibinfo {author} {\bibfnamefont {T.~T.}\
  \bibnamefont {{Hansen}}}, \bibinfo {author} {\bibfnamefont {C.}~\bibnamefont
  {{Sakari}}}, \bibinfo {author} {\bibfnamefont {E.~M.}\ \bibnamefont
  {{Holmbeck}}}, \bibinfo {author} {\bibfnamefont {R.~M.}\ \bibnamefont
  {{Santucci}}}, \bibinfo {author} {\bibfnamefont {A.~R.}\ \bibnamefont
  {{Casey}}}, \ and\ \bibinfo {author} {\bibfnamefont {T.~C.}\ \bibnamefont
  {{Beers}}},\ }\href {\doibase 10.3847/1538-4357/aacbc3} {\bibfield  {journal}
  {\bibinfo  {journal} {\apj}\ }\textbf {\bibinfo {volume} {862}},\ \bibinfo
  {eid} {174} (\bibinfo {year} {2018})},\ \Eprint
  {http://arxiv.org/abs/1806.00645} {arXiv:1806.00645 [astro-ph.SR]}
  \BibitemShut {NoStop}%
\bibitem [{\citenamefont {{Sakari}}\ \emph
  {et~al.}(2018{\natexlab{b}})\citenamefont {{Sakari}}, \citenamefont
  {{Placco}}, \citenamefont {{Hansen}}, \citenamefont {{Holmbeck}},
  \citenamefont {{Beers}}, \citenamefont {{Frebel}}, \citenamefont
  {{Roederer}}, \citenamefont {{Venn}}, \citenamefont {{Wallerstein}},
  \citenamefont {{Davis}}, \citenamefont {{Farrell}},\ and\ \citenamefont
  {{Yong}}}]{sakari2018b}%
  \BibitemOpen
  \bibfield  {author} {\bibinfo {author} {\bibfnamefont {C.~M.}\ \bibnamefont
  {{Sakari}}}, \bibinfo {author} {\bibfnamefont {V.~M.}\ \bibnamefont
  {{Placco}}}, \bibinfo {author} {\bibfnamefont {T.}~\bibnamefont {{Hansen}}},
  \bibinfo {author} {\bibfnamefont {E.~M.}\ \bibnamefont {{Holmbeck}}},
  \bibinfo {author} {\bibfnamefont {T.~C.}\ \bibnamefont {{Beers}}}, \bibinfo
  {author} {\bibfnamefont {A.}~\bibnamefont {{Frebel}}}, \bibinfo {author}
  {\bibfnamefont {I.~U.}\ \bibnamefont {{Roederer}}}, \bibinfo {author}
  {\bibfnamefont {K.~A.}\ \bibnamefont {{Venn}}}, \bibinfo {author}
  {\bibfnamefont {G.}~\bibnamefont {{Wallerstein}}}, \bibinfo {author}
  {\bibfnamefont {C.~E.}\ \bibnamefont {{Davis}}}, \bibinfo {author}
  {\bibfnamefont {E.~M.}\ \bibnamefont {{Farrell}}}, \ and\ \bibinfo {author}
  {\bibfnamefont {D.}~\bibnamefont {{Yong}}},\ }\href {\doibase
  10.3847/2041-8213/aaa9b4} {\bibfield  {journal} {\bibinfo  {journal} {\apjl}\
  }\textbf {\bibinfo {volume} {854}},\ \bibinfo {eid} {L20} (\bibinfo {year}
  {2018}{\natexlab{b}})},\ \Eprint {http://arxiv.org/abs/1801.07727}
  {arXiv:1801.07727 [astro-ph.SR]} \BibitemShut {NoStop}%
\bibitem [{\citenamefont {{Ji}}\ and\ \citenamefont
  {{Frebel}}(2018)}]{jiap2018}%
  \BibitemOpen
  \bibfield  {author} {\bibinfo {author} {\bibfnamefont {A.~P.}\ \bibnamefont
  {{Ji}}}\ and\ \bibinfo {author} {\bibfnamefont {A.}~\bibnamefont
  {{Frebel}}},\ }\href {\doibase 10.3847/1538-4357/aab14a} {\bibfield
  {journal} {\bibinfo  {journal} {\apj}\ }\textbf {\bibinfo {volume} {856}},\
  \bibinfo {eid} {138} (\bibinfo {year} {2018})},\ \Eprint
  {http://arxiv.org/abs/1802.07272} {arXiv:1802.07272 [astro-ph.SR]}
  \BibitemShut {NoStop}%
\bibitem [{\citenamefont {{Beers}}\ and\ \citenamefont
  {{Christlieb}}(2005)}]{beers2005}%
  \BibitemOpen
  \bibfield  {author} {\bibinfo {author} {\bibfnamefont {T.~C.}\ \bibnamefont
  {{Beers}}}\ and\ \bibinfo {author} {\bibfnamefont {N.}~\bibnamefont
  {{Christlieb}}},\ }\href {\doibase 10.1146/annurev.astro.42.053102.134057}
  {\bibfield  {journal} {\bibinfo  {journal} {\araa}\ }\textbf {\bibinfo
  {volume} {43}},\ \bibinfo {pages} {531} (\bibinfo {year} {2005})}\BibitemShut
  {NoStop}%
\bibitem [{\citenamefont {{Beers}}\ \emph {et~al.}(1992)\citenamefont
  {{Beers}}, \citenamefont {{Preston}},\ and\ \citenamefont
  {{Shectman}}}]{beers1992}%
  \BibitemOpen
  \bibfield  {author} {\bibinfo {author} {\bibfnamefont {T.~C.}\ \bibnamefont
  {{Beers}}}, \bibinfo {author} {\bibfnamefont {G.~W.}\ \bibnamefont
  {{Preston}}}, \ and\ \bibinfo {author} {\bibfnamefont {S.~A.}\ \bibnamefont
  {{Shectman}}},\ }\href {\doibase 10.1086/116207} {\bibfield  {journal}
  {\bibinfo  {journal} {\aj}\ }\textbf {\bibinfo {volume} {103}},\ \bibinfo
  {pages} {1987} (\bibinfo {year} {1992})}\BibitemShut {NoStop}%
\bibitem [{\citenamefont {{Sneden}}\ \emph {et~al.}(2003)\citenamefont
  {{Sneden}}, \citenamefont {{Cowan}}, \citenamefont {{Lawler}}, \citenamefont
  {{Ivans}}, \citenamefont {{Burles}}, \citenamefont {{Beers}}, \citenamefont
  {{Primas}}, \citenamefont {{Hill}}, \citenamefont {{Truran}}, \citenamefont
  {{Fuller}}, \citenamefont {{Pfeiffer}},\ and\ \citenamefont
  {{Kratz}}}]{sneden2003}%
  \BibitemOpen
  \bibfield  {author} {\bibinfo {author} {\bibfnamefont {C.}~\bibnamefont
  {{Sneden}}}, \bibinfo {author} {\bibfnamefont {J.~J.}\ \bibnamefont
  {{Cowan}}}, \bibinfo {author} {\bibfnamefont {J.~E.}\ \bibnamefont
  {{Lawler}}}, \bibinfo {author} {\bibfnamefont {I.~I.}\ \bibnamefont
  {{Ivans}}}, \bibinfo {author} {\bibfnamefont {S.}~\bibnamefont {{Burles}}},
  \bibinfo {author} {\bibfnamefont {T.~C.}\ \bibnamefont {{Beers}}}, \bibinfo
  {author} {\bibfnamefont {F.}~\bibnamefont {{Primas}}}, \bibinfo {author}
  {\bibfnamefont {V.}~\bibnamefont {{Hill}}}, \bibinfo {author} {\bibfnamefont
  {J.~W.}\ \bibnamefont {{Truran}}}, \bibinfo {author} {\bibfnamefont {G.~M.}\
  \bibnamefont {{Fuller}}}, \bibinfo {author} {\bibfnamefont {B.}~\bibnamefont
  {{Pfeiffer}}}, \ and\ \bibinfo {author} {\bibfnamefont {K.-L.}\ \bibnamefont
  {{Kratz}}},\ }\href {\doibase 10.1086/375491} {\bibfield  {journal} {\bibinfo
   {journal} {\apj}\ }\textbf {\bibinfo {volume} {591}},\ \bibinfo {pages}
  {936} (\bibinfo {year} {2003})},\ \Eprint
  {http://arxiv.org/abs/astro-ph/0303542} {arXiv:astro-ph/0303542 [astro-ph]}
  \BibitemShut {NoStop}%
\bibitem [{\citenamefont {{Roederer}}\ \emph {et~al.}(2022)\citenamefont
  {{Roederer}}, \citenamefont {{Lawler}}, \citenamefont {{Den Hartog}},
  \citenamefont {{Placco}}, \citenamefont {{Surman}}, \citenamefont {{Beers}},
  \citenamefont {{Ezzeddine}}, \citenamefont {{Frebel}}, \citenamefont
  {{Hansen}}, \citenamefont {{Hattori}}, \citenamefont {{Holmbeck}},\ and\
  \citenamefont {{Sakari}}}]{Roederer2022}%
  \BibitemOpen
  \bibfield  {author} {\bibinfo {author} {\bibfnamefont {I.~U.}\ \bibnamefont
  {{Roederer}}}, \bibinfo {author} {\bibfnamefont {J.~E.}\ \bibnamefont
  {{Lawler}}}, \bibinfo {author} {\bibfnamefont {E.~A.}\ \bibnamefont {{Den
  Hartog}}}, \bibinfo {author} {\bibfnamefont {V.~M.}\ \bibnamefont
  {{Placco}}}, \bibinfo {author} {\bibfnamefont {R.}~\bibnamefont {{Surman}}},
  \bibinfo {author} {\bibfnamefont {T.~C.}\ \bibnamefont {{Beers}}}, \bibinfo
  {author} {\bibfnamefont {R.}~\bibnamefont {{Ezzeddine}}}, \bibinfo {author}
  {\bibfnamefont {A.}~\bibnamefont {{Frebel}}}, \bibinfo {author}
  {\bibfnamefont {T.~T.}\ \bibnamefont {{Hansen}}}, \bibinfo {author}
  {\bibfnamefont {K.}~\bibnamefont {{Hattori}}}, \bibinfo {author}
  {\bibfnamefont {E.~M.}\ \bibnamefont {{Holmbeck}}}, \ and\ \bibinfo {author}
  {\bibfnamefont {C.~M.}\ \bibnamefont {{Sakari}}},\ }\href@noop {} {\bibfield
  {journal} {\bibinfo  {journal} {arXiv e-prints}\ ,\ \bibinfo {eid}
  {arXiv:2205.03426}} (\bibinfo {year} {2022})},\ \Eprint
  {http://arxiv.org/abs/2205.03426} {arXiv:2205.03426 [astro-ph.SR]}
  \BibitemShut {NoStop}%
\bibitem [{\citenamefont {Diehl}\ \emph {et~al.}(2006)\citenamefont {Diehl}
  \emph {et~al.}}]{Diehl:2006cf}%
  \BibitemOpen
  \bibfield  {author} {\bibinfo {author} {\bibfnamefont {R.}~\bibnamefont
  {Diehl}} \emph {et~al.},\ }\href {\doibase 10.1038/nature04364} {\bibfield
  {journal} {\bibinfo  {journal} {Nature}\ }\textbf {\bibinfo {volume} {439}},\
  \bibinfo {pages} {45} (\bibinfo {year} {2006})},\ \Eprint
  {http://arxiv.org/abs/astro-ph/0601015} {arXiv:astro-ph/0601015} \BibitemShut
  {NoStop}%
\bibitem [{\citenamefont {{Wang}}\ \emph {et~al.}(2020)\citenamefont {{Wang}},
  \citenamefont {{Siegert}}, \citenamefont {{Dai}}, \citenamefont {{Diehl}},
  \citenamefont {{Greiner}}, \citenamefont {{Heger}}, \citenamefont {{Krause}},
  \citenamefont {{Lang}}, \citenamefont {{Pleintinger}},\ and\ \citenamefont
  {{Zhang}}}]{Wang:2020xx}%
  \BibitemOpen
  \bibfield  {author} {\bibinfo {author} {\bibfnamefont {W.}~\bibnamefont
  {{Wang}}}, \bibinfo {author} {\bibfnamefont {T.}~\bibnamefont {{Siegert}}},
  \bibinfo {author} {\bibfnamefont {Z.~G.}\ \bibnamefont {{Dai}}}, \bibinfo
  {author} {\bibfnamefont {R.}~\bibnamefont {{Diehl}}}, \bibinfo {author}
  {\bibfnamefont {J.}~\bibnamefont {{Greiner}}}, \bibinfo {author}
  {\bibfnamefont {A.}~\bibnamefont {{Heger}}}, \bibinfo {author} {\bibfnamefont
  {M.}~\bibnamefont {{Krause}}}, \bibinfo {author} {\bibfnamefont
  {M.}~\bibnamefont {{Lang}}}, \bibinfo {author} {\bibfnamefont {M.~M.~M.}\
  \bibnamefont {{Pleintinger}}}, \ and\ \bibinfo {author} {\bibfnamefont
  {X.~L.}\ \bibnamefont {{Zhang}}},\ }\href {\doibase 10.3847/1538-4357/ab6336}
  {\bibfield  {journal} {\bibinfo  {journal} {\apj}\ }\textbf {\bibinfo
  {volume} {889}},\ \bibinfo {eid} {169} (\bibinfo {year} {2020})},\ \Eprint
  {http://arxiv.org/abs/1912.07874} {arXiv:1912.07874 [astro-ph.HE]}
  \BibitemShut {NoStop}%
\bibitem [{\citenamefont {{Hudson}}\ \emph {et~al.}(1989)\citenamefont
  {{Hudson}}, \citenamefont {{Kennedy}}, \citenamefont {{Podosek}},\ and\
  \citenamefont {{Hohenberg}}}]{Hudson1989}%
  \BibitemOpen
  \bibfield  {author} {\bibinfo {author} {\bibfnamefont {G.~B.}\ \bibnamefont
  {{Hudson}}}, \bibinfo {author} {\bibfnamefont {B.~M.}\ \bibnamefont
  {{Kennedy}}}, \bibinfo {author} {\bibfnamefont {F.~A.}\ \bibnamefont
  {{Podosek}}}, \ and\ \bibinfo {author} {\bibfnamefont {C.~M.}\ \bibnamefont
  {{Hohenberg}}},\ }\href@noop {} {\bibfield  {journal} {\bibinfo  {journal}
  {Lunar and Planetary Science Conference Proceedings}\ }\textbf {\bibinfo
  {volume} {19}},\ \bibinfo {pages} {547} (\bibinfo {year} {1989})}\BibitemShut
  {NoStop}%
\bibitem [{\citenamefont {Hotokezaka}\ \emph {et~al.}(2015)\citenamefont
  {Hotokezaka}, \citenamefont {Piran},\ and\ \citenamefont
  {Paul}}]{Hotokezaka:2015zea}%
  \BibitemOpen
  \bibfield  {author} {\bibinfo {author} {\bibfnamefont {K.}~\bibnamefont
  {Hotokezaka}}, \bibinfo {author} {\bibfnamefont {T.}~\bibnamefont {Piran}}, \
  and\ \bibinfo {author} {\bibfnamefont {M.}~\bibnamefont {Paul}},\ }\href
  {\doibase 10.1038/nphys3574} {\bibfield  {journal} {\bibinfo  {journal}
  {Nature Phys.}\ }\textbf {\bibinfo {volume} {11}},\ \bibinfo {pages} {1042}
  (\bibinfo {year} {2015})},\ \Eprint {http://arxiv.org/abs/1510.00711}
  {arXiv:1510.00711 [astro-ph.HE]} \BibitemShut {NoStop}%
\bibitem [{\citenamefont {{Nishimura}}\ \emph {et~al.}(2006)\citenamefont
  {{Nishimura}}, \citenamefont {{Kotake}}, \citenamefont {{Hashimoto}},
  \citenamefont {{Yamada}}, \citenamefont {{Nishimura}}, \citenamefont
  {{Fujimoto}},\ and\ \citenamefont {{Sato}}}]{Nishimura2006}%
  \BibitemOpen
  \bibfield  {author} {\bibinfo {author} {\bibfnamefont {S.}~\bibnamefont
  {{Nishimura}}}, \bibinfo {author} {\bibfnamefont {K.}~\bibnamefont
  {{Kotake}}}, \bibinfo {author} {\bibfnamefont {M.-a.}\ \bibnamefont
  {{Hashimoto}}}, \bibinfo {author} {\bibfnamefont {S.}~\bibnamefont
  {{Yamada}}}, \bibinfo {author} {\bibfnamefont {N.}~\bibnamefont
  {{Nishimura}}}, \bibinfo {author} {\bibfnamefont {S.}~\bibnamefont
  {{Fujimoto}}}, \ and\ \bibinfo {author} {\bibfnamefont {K.}~\bibnamefont
  {{Sato}}},\ }\href {\doibase 10.1086/500786} {\bibfield  {journal} {\bibinfo
  {journal} {\apj}\ }\textbf {\bibinfo {volume} {642}},\ \bibinfo {pages} {410}
  (\bibinfo {year} {2006})},\ \Eprint {http://arxiv.org/abs/astro-ph/0504100}
  {arXiv:astro-ph/0504100 [astro-ph]} \BibitemShut {NoStop}%
\bibitem [{\citenamefont {{Winteler}}\ \emph {et~al.}(2012)\citenamefont
  {{Winteler}}, \citenamefont {{K{\"a}ppeli}}, \citenamefont {{Perego}},
  \citenamefont {{Arcones}}, \citenamefont {{Vasset}}, \citenamefont
  {{Nishimura}}, \citenamefont {{Liebend{\"o}rfer}},\ and\ \citenamefont
  {{Thielemann}}}]{Winteler+12}%
  \BibitemOpen
  \bibfield  {author} {\bibinfo {author} {\bibfnamefont {C.}~\bibnamefont
  {{Winteler}}}, \bibinfo {author} {\bibfnamefont {R.}~\bibnamefont
  {{K{\"a}ppeli}}}, \bibinfo {author} {\bibfnamefont {A.}~\bibnamefont
  {{Perego}}}, \bibinfo {author} {\bibfnamefont {A.}~\bibnamefont {{Arcones}}},
  \bibinfo {author} {\bibfnamefont {N.}~\bibnamefont {{Vasset}}}, \bibinfo
  {author} {\bibfnamefont {N.}~\bibnamefont {{Nishimura}}}, \bibinfo {author}
  {\bibfnamefont {M.}~\bibnamefont {{Liebend{\"o}rfer}}}, \ and\ \bibinfo
  {author} {\bibfnamefont {F.-K.}\ \bibnamefont {{Thielemann}}},\ }\href
  {\doibase 10.1088/2041-8205/750/1/L22} {\bibfield  {journal} {\bibinfo
  {journal} {\apjl}\ }\textbf {\bibinfo {volume} {750}},\ \bibinfo {eid} {L22}
  (\bibinfo {year} {2012})},\ \Eprint {http://arxiv.org/abs/1203.0616}
  {arXiv:1203.0616 [astro-ph.SR]} \BibitemShut {NoStop}%
\bibitem [{\citenamefont {{M{\"o}sta}}\ \emph {et~al.}(2018)\citenamefont
  {{M{\"o}sta}}, \citenamefont {{Roberts}}, \citenamefont {{Halevi}},
  \citenamefont {{Ott}}, \citenamefont {{Lippuner}}, \citenamefont {{Haas}},\
  and\ \citenamefont {{Schnetter}}}]{Mosta+17}%
  \BibitemOpen
  \bibfield  {author} {\bibinfo {author} {\bibfnamefont {P.}~\bibnamefont
  {{M{\"o}sta}}}, \bibinfo {author} {\bibfnamefont {L.~F.}\ \bibnamefont
  {{Roberts}}}, \bibinfo {author} {\bibfnamefont {G.}~\bibnamefont {{Halevi}}},
  \bibinfo {author} {\bibfnamefont {C.~D.}\ \bibnamefont {{Ott}}}, \bibinfo
  {author} {\bibfnamefont {J.}~\bibnamefont {{Lippuner}}}, \bibinfo {author}
  {\bibfnamefont {R.}~\bibnamefont {{Haas}}}, \ and\ \bibinfo {author}
  {\bibfnamefont {E.}~\bibnamefont {{Schnetter}}},\ }\href {\doibase
  10.3847/1538-4357/aad6ec} {\bibfield  {journal} {\bibinfo  {journal} {\apj}\
  }\textbf {\bibinfo {volume} {864}},\ \bibinfo {eid} {171} (\bibinfo {year}
  {2018})},\ \Eprint {http://arxiv.org/abs/1712.09370} {arXiv:1712.09370
  [astro-ph.HE]} \BibitemShut {NoStop}%
\end{thebibliography}
%

\end{document}